\documentclass[a4paper,12pt]{article}

\usepackage{color}
\usepackage[dvipsnames]{xcolor}
\usepackage{graphicx}
\usepackage{amsmath}
\usepackage{caption}
\usepackage{graphicx}
\usepackage{subfigure}
\usepackage[separate-uncertainty=true]{siunitx}
\usepackage{booktabs}
\usepackage{tabularx}
\usepackage{multirow}
\usepackage{multicol}
\usepackage{transparent}
\usepackage{placeins}
\usepackage{braket}
\usepackage{slashed}
\usepackage{relsize}

\pdfoutput=1 

\usepackage{jheppub} 

\hypersetup{%
  colorlinks=true, linktocpage=true, pdfstartpage=1, pdfstartview=FitV,
  breaklinks=true, pageanchor=true,
  pdfpagemode=UseNone,
  plainpages=false, bookmarksnumbered, bookmarksopen=true, bookmarksopenlevel=1,
  hypertexnames=true, pdfhighlight=/O,
  urlcolor=RoyalBlue, linkcolor=ForestGreen, citecolor=RoyalBlue,
  pdftitle={QCD corrections to charged-current decays with Heavy Sterile Neutrinos in initial or final state and their impact on $\tau$ decays},
  pdfauthor={Tim Kretz, Ulrich Nierste},
  pdfsubject={},
  pdfkeywords={},
  pdfcreator={pdfLaTeX},
  pdfproducer={LaTeX with hyperref}
}

\usepackage[normalem]{ulem}

\newcommand{\lf}{\left(}
\newcommand{\rt}{\right)}
\newcommand{\dx}{\mathrm{d}}

\newcommand{\nn}{\nonumber\\}
\newcommand{\lt}{\left}
\newcommand{\rrt}{\right}

\newcommand{\eq}[1]{Eq.~(\ref{#1})}

\newcommand{\eqsto}[2]{Eqs.~(\ref{#1}--\ref{#2})}

\newcommand{\un}[1]{{#1}}
\newcommand{\tk}[1]{{{#1}}}
\newcommand{\kt}[1]{{{#1}}}

\preprint{P3H–25–103, TTP25-049}

\author[a]{Tim Kretz,}
\author[a]{Ulrich Nierste}

\affiliation{
$^a$ Institute for Theoretical Particle Physics, Karlsruhe Institute of Technology (KIT),\\
  Wolfgang-Gaede-Str. 1, D-76131 Karlsruhe, Germany}

\emailAdd{tim.kretz@kit.edu}
\emailAdd{ulrich.nierste@kit.edu}

\title{QCD corrections to charged-current decays with Heavy Sterile Neutrinos in initial or final state and their impact on $\tau$ decays}

\abstract{Searches for a Heavy Sterile Neutrino $N$ profit from precise predictions of inclusive decay rates, 
which enter predictions for branching fractions and lifetime. 
Once the decay channels into semi-hadronic final states are open, a reliable calculation of  inclusive decay rates
is only possible if $N$ is heavy enough to permit a 
perturbative calculation, in analogy to the well-known case of semi-hadronic $\tau$ lepton decays.  
We adopt the popular scenario in which  $N$ only interacts with the SM particles through  
$N$-$\nu_\ell$ mixing, where $\ell=e,\mu,\tau$. 
Using literature results for $W$ boson correlators calculated to fourth order in the strong coupling 
$\alpha_s$, we study the quality of the perturbation series for $N\to \ell +\mbox{hadrons}$ to determine 
the mass ranges for which inclusive decay widths can be predicted in a robust way. We present novel analytic results 
for the decay rate of $N\to \tau +\mbox{hadrons}$ in terms of $m_\tau/m_N$.   
Our expressions equally apply to  
$\tau \to N +\mbox{hadrons}$, the width of which is found to be perturbatively calculable for
$m_N\lesssim 600\,$MeV. Applying our result to the $\tau$ lifetime, we 
determine the allowed parameter space 
for the $N$-$\nu_\tau$ mixing angle $\theta$ and $m_N$. We find $|\sin\theta| \leq 0.2 $ for $m_N=600\,$MeV and weaker bounds  for a 
lighter $N$. In the mass region $m_N\geq m_{\tau}$ we find constraints from the dependence of $\tau$   
decay rates on $\cos\theta$, \un{with $|\sin\theta|\leq 0.12$ inferred from the $\tau$ lifetime.}
Combining $\tau \to \pi^- \nu_\tau$ and $\tau \to K^- \nu_\tau$ data 
gives \tk{$|\sin\theta| =\un{\left(9.1_{-7.8}^{+3.7} \right)} \cdot 10^{-2}$ at $1\sigma$} while  $N$-$\nu_\tau$ mixing does not improve the 
agreement between theory and data for $\tau \to \ell \bar\nu_\ell \nu_\tau $. 
We find current data for  the decay rate $\Gamma(\tau \rightarrow \ell+\mbox{nothing})$ about 
1$\sigma$ above the SM prediction for $\Gamma(\tau \rightarrow \ell \bar\nu_\ell \nu_\tau)$, which leads to useful   
constraints on $\Gamma(\tau \rightarrow \ell X_{\mathrm{dark}})$ or $\Gamma(\tau \rightarrow \ell X_{\mathrm{dark}} X_{\mathrm{dark}})$ 
with dark-sector particles $X_{\mathrm{dark}}$ and might stimulate additional experimental effort on $\tau \rightarrow \ell+\mbox{nothing}$. 
}

\begin{document}

\maketitle
\flushbottom
\thispagestyle{empty}


\newpage
\setcounter{page}{1}

\section{Introduction}\label{sec:introduction}
Heavy Sterile Neutrinos (HSNs) (a.k.a.\ Heavy Neutral Leptons) are a
widely-studied extension to the Standard Model (SM), postulated in
theories of Dark Matter \cite{Asaka_2005,Asaka_2005_2}, leptogenesis \cite{Fukugita:1986hr,Davidson:2002qv}, \kt{or neutrino masses
\cite{Yanagida:1979as,Minkowski:1977sc,Gell-Mann:1979vob,Glashow:1979nm,Mohapatra:1979ia}} (see Ref.~\cite{Bondarenko_2018} for an overview). In standard
scenarios it is assumed that HSNs interact with SM particles only
through mixing with the active neutrinos $\nu_e$, $\nu_\mu$, and
$\nu_\tau$, so that different observables are highly correlated. In
particular, $W$-mediated weak decays involving an HSN $N$ and a charged
lepton $\ell$ in the initial or final state all involve the same element
$V_{N\ell}$ of the matrix describing the mixing of $N$ with the active
neutrinos. In the most simple case one assumes that a given HSN $N$ only
mixes with one active neutrino $\nu_\ell$, so that $V_{N\ell}$ reduces
to $V_{N\ell}=\sin \theta$ with a neutrino mixing angle $\theta$. It is
therefore desirable to study as many different decays as possible to
first discover HSNs and subsequently corroborate or falsify the
active-sterile mixing concept. In the former case one will aim at the
precise determination of $V_{N\ell}$ for $\ell=e,\mu,\tau$. In the latter case one
will find apparent different values for $V_{N\ell}$ in different decays
and turn the attention to less minimal models featuring new interactions
between $N$ and SM fermions, possibly mediated by new Higgs bosons or
leptoquarks \cite{Robinson:2018gza,Bernlochner:2024xiz}.

The search for HSNs produced at colliders and the interpretation of
experimental limits requires accurate predictions of branching
fractions. In addition, it is important to know the $N$ lifetime to
understand whether $N$ decays outside or within the detector.
Especially interesting is the parameter space for which $N$ decays with
a displaced decay vertex, which corresponds to an ideal, background-free
signal. To predict branching fractions or lifetimes one must calculate
the total decay width $\Gamma_{\rm tot}(N)$. In this paper we consider
semi-hadronic decays of $N$ and present precise predictions for $\Gamma (N\to \ell + \mbox{hadrons})$ for the case that the HSN mass $m_N$ is large enough that accurate perturbative
QCD predictions are possible. To this end we exploit literature results
on the correlator of two charged quark currents to five-loop order of
QCD \cite{Baikov:2008jh} and add the appropriate phase-space integration
to predict $\Gamma (N\to \ell + \mbox{hadrons})$. The main purpose of
this analysis is the determination of the range for $m_N$ for which the
perturbative calculation is applicable, by requiring that with higher orders
in $\alpha_s$ the size of the perturbative correction and 
the dependence on the unphysical renormalization scale $\mu$ decrease. 
Specifically, if the invariant mass of the hadronic system governed by
$m_N-m_\ell$ is too small, perturbation theory breaks down, and the same
is true for certain larger values of $m_N$ which are closely above the masses of
the $D$ or $B$ mesons. For $\ell=e$ or $\mu$ we can simply use the
formulae for $\tau \to \nu_\tau + \mbox{hadrons}$ decays from the
literature \cite{Baikov:2008jh,Beneke_2008}, but for
$\Gamma (N\to \tau + \mbox{hadrons})$ we  perform a  new
calculation to account for $m_\tau\neq 0$ and find closed results
in terms of $m_\tau/m_N$ with polylogarithms (up to $\alpha_s^3$) or
a series representation (at order  $\alpha_s^4$). With this result we can further
predict  $\Gamma (\tau \to N + \mbox{hadrons})$ for the case that $m_N$
is too large to be neglected and analyse the impact of $N$--$\nu_\tau$
mixing on $\tau$ lifetime and branching fractions. 
\un{While our calculations are complete for $\Gamma (\tau \to N + \mbox{hadrons})$ and the impact of these rates on 
the $\tau$ lifetime, the calculated $N \to \ell + \mbox{hadrons}$ decay rates do not fully account for the 
$N$ lifetime, which receives further important contributions from $Z$-mediated $N \to \nu_\ell + \mbox{hadrons}$
decay modes. These calculations are relegated to a future publication.} 

Our paper is organised as follows: In Sec.~\ref{sec:preliminaries} we recapitulate the known results entering the calculation of the inclusive hadronic HSN decay width, in Sec.~\ref{sec:qcdcorr} we show the details of the actual calculation followed by a discussion of the phenomenology of our results in Sec.~\ref{sec:numana}. We conclude in Sec.~\ref{sec:conc}.

\section{Preliminaries}\label{sec:preliminaries}
\subsection{Sterile neutrino decays at tree level}
Several decays of HSN or of leptons and hadrons into HSN have been calculated at tree level
\cite{PhysRevD.94.053001,Johnson_1997,Gribanov_2001,Gorbunov_2007,
  Atre_2009,Helo_2011}. Here we briefly collect the 
tree-level results relevant to our study and introduce our notation.

An HSN $N$ can decay via a $W$ boson into a charged lepton and a 
quark-antiquark pair $(u^\prime,\bar d^\prime)$. Here, we only consider 
cases in which $N$ is heavy enough to permit $N\to \ell u\bar d$ and
$N\to \ell u\bar s$ decays, but for large enough values of $m_N$ also
final states with $u^\prime=c$ or $ d^\prime=b$ can be accessible. The
decay width reads  
\begin{align}
  \Gamma(N \rightarrow \ell^-
  u' \bar{d}') &=N_c \frac{G_F^2 m_N^5 |V_{N\ell}|^2 |V_{u^\prime d^\prime}|^2}{192\pi^3} \nonumber\\ 
 &\hspace{-15mm}\times 12 \int\limits_{(x_{u'}+x_{d'})^2}^{(1-x_{\ell})^2}
\frac{\mathrm{d}\, x}{x} (1+x_{\ell}^2-x) (x-x_{u'}^2-x_{d'}^2) \sqrt{\lambda(1,x,x_{\ell}^2) \lambda(x,x_{u'}^2,x_{d'}^2)} , \label{eq:chargedDecay}
\end{align}
here $G_F$ is the Fermi constant, $m_N$ is the mass of the sterile
neutrino, $V_{N\ell}$ parametrises the mixing between $N$ and
$\nu_\ell$, $V_{u^\prime d^\prime}$ is the relevant CKM matrix element,
$N_c=3$ is the number of colours, $x_i=m_i/m_N$, and
$\lambda(a,b,c)=(a-b-c)^2 - 4 bc$ is the Källén function.

\begin{figure}[t]
\centering
\subfigure[charged current \label{fig:CClepton}]{\includegraphics[width=0.45\textwidth]{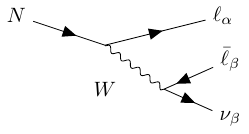}}
\subfigure[neutral current \label{fig:NClepton}]{\includegraphics[width=0.45\textwidth]{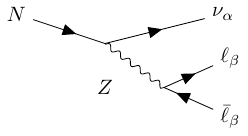}}
\caption{Tree diagrams for an HSN decay. For $\alpha=\beta$ both
  diagrams interfere. Diagrams were drawn with the help of TikZ-Feynman \cite{Ellis:2016jkw}. \label{fig:HSNlepton}}
~\\[-3mm]\hrule
\end{figure}
\begin{figure}[t]
\centering
\includegraphics[width=0.65\textwidth]{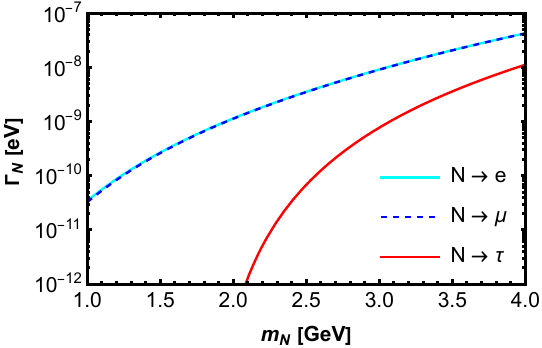}
\caption{Decay width of an HSN into leptons. Here the notation
  $N \rightarrow \ell$ means the plotted line is the sum of all partial
  widths associated with the tagging lepton i.e.
  $\Gamma(N \rightarrow e) = \sum_\ell \Gamma(N \rightarrow e \bar{\ell}
  \nu_\ell)$. For all three channels the mixing angle is chosen as
  $V_{N\ell} = 10^{-3}$.\label{fig:leptonwidth}}
~\\[-3mm]\hrule
\end{figure}

 
Decay rates of an HSN into leptons
$N \rightarrow \ell_i \bar{\ell}_j \nu_j $ are found from
Eq.~(\ref{eq:chargedDecay}) by setting $N_c=1$ and $V_{u^\prime d^\prime}=1$:
\begin{align}
  \lefteqn{\Gamma(N \rightarrow \ell_\alpha
   \bar{\ell}_\beta \nu_\beta)} \nonumber \\
  &=\frac{G_F^2 m_N^5 |V_{N\ell}|^2 }{192 \pi^3} 
    \times 12 \int\limits_{x_{\beta}^2}^{(1-x_{\alpha})^2}
    \frac{\mathrm{d}\, x}{x} (1+x_{\alpha}^2-x) (x-x_{\beta}^2)
    \sqrt{\lambda(1,x,x_{\alpha}^2) \lambda(x,x_{\beta}^2,0)} ,
    \label{eq:chargedDecaylepton}
\end{align}
where $x_\alpha=m_{\ell_\alpha}/m_N$. This relation only holds for
$\alpha \neq \beta$. We dub the lepton directly attached to the HSN the
tagging lepton (see Fig. \ref{fig:HSNlepton}). 
For $\alpha=\beta$ there are additional contributions to the leptonic
width due to interference with the neutral current decay. The decay
width becomes
\begin{align}
 \Gamma(&N \rightarrow  \nu_\alpha \ell_\alpha
  \bar{\ell}_\alpha) =  \\
  & \frac{G_F^2 m_N^5 |V_{N\ell}|^2}{192 \pi^3} 
     \times \bigg[ \sqrt{1-4x_\ell^2} \bigg( \bigg( \frac{g_V^2}{2} +
    g_V  \bigg) (1-10x_\ell^2+18x_\ell^4-36x_\ell^6) \nonumber\\
  & \hspace{1cm} + \bigg( \frac{g_A^2}{2} + g_A \bigg)
    (1-18x_\ell^2-22x_\ell^4+12x_\ell^6) +1 -14 x_\ell^2
    - 2 x_\ell^4 - 12 x_\ell^6\bigg) \nonumber\\
  & \quad - 24 \bigg( \bigg(g_V^2 +2 g_V\bigg) x_\ell^6 (2-3x_\ell^2)
    \nonumber\\
  &\hspace{1cm} +  \bigg(g_A^2 +2g_A\bigg) x_\ell^4
    (2-2x_\ell^2+x_\ell^4) + x_\ell^4(2 - 2 x_\ell^4) \bigg)
    \ln\bigg( \frac{2x_\ell}{1+\sqrt{1-4x_\ell^2}} \bigg)  \bigg], \nonumber
\end{align}
here $g_A = -1/2$, $g_V = -1/2 + 2 s_w^2$ with $s_w = \sin\theta_w$ and
the Weinberg angle $\theta_w$.
Our results agree with those of Bodarenko et al. in
Ref.~\cite{Bondarenko_2018}. In Fig. \ref{fig:leptonwidth} we show the
decay widths as a function of $m_N$. The widths for $e$ or $\mu$ as
tagging leptons are almost indistinguishable. 

\subsection{QCD correlators}
The QCD dynamics of decays of HSNs into hadrons is governed by the
decays of virtual electroweak gauge bosons into hadrons. The decays of
$W$ and $Z$ bosons into hadrons have been calculated up to
$\mathcal{O}(\alpha_s^4)$
\cite{PhysRevLett.108.222003,Baikov:2008jh}. Here we use these results
to calculate the charged-current contribution to the decay of an HSN.

\begin{figure}[t]
\centering
\subfigure[Sample non-singlet diagram to $\mathcal{O}(\alpha_S^2)$. \label{fig:nonsingelt}]{\includegraphics[width=0.5\textwidth]{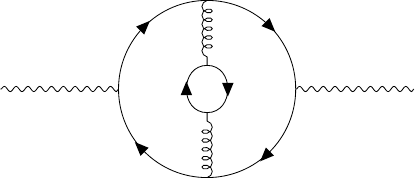}}
\subfigure[One possible cut of the loop diagram in Fig. \ref{fig:nonsingelt}. This corresponds to the decay amplitude $W \rightarrow q \bar{q}$ at $\mathcal{O}(\alpha_s^2)$ interfering with the tree-level amplitude. \label{fig:cut}]{\includegraphics[width=0.8\textwidth]{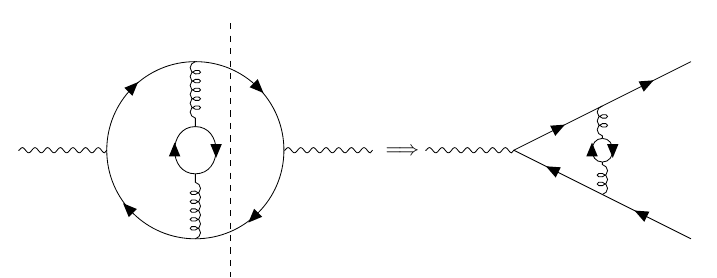}}
\caption{$W$ boson self-energy diagram contributing to the correlator.}
~\\[-3mm]\hrule
\label{fig:allcut}
\end{figure}

The optical theorem relates any inclusive decay width to the imaginary
part of the self-energy of the decaying particle. The non-trivial part
of the $W$ or $Z$ boson self-energy is the correlator of two charged quark
currents, namely
the time ordered product of the vector and axial vector currents of the
quarks $j^{V/A}_{\mu, \, ij} = \bar{q}_i \gamma_{\mu} (\gamma_5)
q_j$. It may be parametrized as follows \cite{BRAATEN1992581,Chetyrkin_1998}
\begin{align}
  \Pi^{V/A}_{\mu\nu, \, ij}(q,m_i,m_j,\mu,\alpha_s)
  &= i \int \dx x e^{iqx} \braket{0|
     \hat{T}\{ j^{V/A}_{\mu, \, ij}(x) j^{V/A \, \dagger}_{\nu, \, ij}(0)  \}|0}  \nonumber\\
  &= g_{\mu\nu} \Pi^{[1]}_{ij, \, V/A}(q^2) + q_\mu q_\nu
    \Pi^{[2]}_{ij, \, V/A}(q^2) \nonumber\\
  &= (-g_{\mu\nu} q^2 + q_\mu q_\nu) \Pi^{(1)}_{ij, \, V/A}(q^2)
    + q_\mu q_\nu \Pi^{(0)}_{ij, \, V/A}(q^2). \label{eq:pva}
\end{align}
Here $m_{i,j}$ are quark masses and $q$ is the momentum of the gauge
boson. In our case of the hadronic charged current $q_i$ and $q_j$ are
up-type and down-type quarks, respectively.  In the last line above the
correlators have been decomposed into their transverse and longitudinal
components
\begin{equation}
  \Pi^{(1)}_{ij, \, V/A}(q^2) = - \frac{\Pi^{[1]}_{ij, \,
      V/A}(q^2)}{q^2},
  \quad \Pi^{(0)}_{ij, \, V/A}(q^2) = \Pi^{[2]}_{ij, \, V/A}(q^2)
  + \frac{\Pi^{[1]}_{ij, \, V/A}(q^2)}{q^2}.
\end{equation}
In studies of $\tau$ decays one commonly adopts the definition 
\cite{BRAATEN1992581}
\begin{equation}
  \Pi^{(1+0)}_{ij, \, V/A}(q^2)\equiv
  \Pi^{(1)}_{ij, \, V/A}(q^2) +\Pi^{(0)}_{ij, \, V/A}(q^2).
\end{equation}
The correlators $\Pi^{[1]}$ and $\Pi^{[2]}$ are related to each other
via a Ward identity \cite{Becchi:1980vz,Pich_2021}
\begin{equation}
  q^\mu q^\nu \Pi^{V/A}_{\mu\nu, \, ij} = q^4 \Pi^{(0)}_{ij, \, V/A} = (m_j \mp m_i)^2 \Pi_{ij}^{S/P} + (m_j \mp m_i) \braket{0| \bar{q}_j q_j \mp \bar{q}_i q_i |0}.
\label{eq:qcs}
\end{equation}
This identity connects the longitudinal part of the $V$ or $A$
correlator to the scalar or pseudo-scalar correlators $\Pi_{ij}^{S/P}$,
respectively, where
\begin{equation}
\Pi^{S/P}_{ij}(q) = i \int \dx x e^{iqx} \braket{0|\hat{T}\{ j^{S/P}_{ij}(x) j^{S/P \, \dagger}_{ij}(0)  \}|0}, \qquad j^{S/P}_{ij} = \bar{q}_i (\gamma_5) q_j,
\end{equation}
and the last term in \eq{eq:qcs} is the quark condensate. Thus in the
chiral limit
\begin{equation}
\Pi^{(0)}_{ij, \, V/A} = 0.
\end{equation}
For the $W$ boson there are only contributions to the correlator by so-called non-singlet diagrams as seen in Fig. \ref{fig:nonsingelt}. 
Non-singlet diagrams only have cuts through at least two fermion lines. Cuts {indicate} individual contributions to the imaginary part of a loop diagram with the particles on the cut line being on-shell (see. Fig. \ref{fig:cut}). From now on we adopt the chiral limit with zero quark masses.  The
applicability of perturbative QCD requires $q^2 \gg m_i^2$, which is
certainly fulfilled for the three lightest quarks for \un{$m_N \gtrsim \mbox{a few times}\, \Lambda_{\rm QCD}$, 
which must be fulfilled in any perturbative QCD calculation}. 
If one of the final state quarks is $c$ or $b$, there is a range of $q^2$ for which the $c$
or $b$ quark mass is non-negligible and can be included in an expansion
in terms of $m_{c,b}^2/q^2$.  These mass corrections are beyond the
scope of this paper, i.e. for heavy hadrons our formulae are only valid for HSN decays with $q^2$ large enough that $m_c$ or $m_b$ can be set to zero i.e. $m_b^2 > q^2 \gg m_c^2$ or $q^2 \gg m_b^2$.

The correlators may be expressed as a series in 
\begin{equation}
a_\mu = \frac{\alpha_s(\mu)}{\pi}, \label{eq:alphas}
\end{equation}
where the coefficients are functions of the
logarithm $\ln(-s/\mu^2)$ with $s = q^2 $ the invariant mass of the hadronic state and the logarithm evaluated below the branch cut according to the Feynman prescription $s \rightarrow s +i \delta$, so that $\mathrm{Im}\,\ln(-s/\mu^2-i\delta) = \ln(s/\mu^2) - i\pi$. Since the longitudinal part of the correlator is proportional to the square of the quark mass it will not contribute in the chiral limit (see \eq{eq:qcs}). The sum of the longitudinal and transverse part of the correlator is given by \cite{Beneke_2008}
\begin{equation}
\Pi^{(1+0)}_{ij, \, V/A}(s) = -\frac{1}{12 \pi^2} \sum\limits_{n=0}^{\infty} a_\mu^n \sum\limits_{k=1}^{n+1} c_{n,k} \ln^k\bigg( \frac{-s}{\mu^2} \bigg), 
\label{eq:Wcorr}
\end{equation}
and the coefficients $c_{n,1}$ are given by \cite{Chetyrkin:1979bj,PhysRevLett.43.668,PhysRevLett.44.560,Gorishnii:1988bc,Surguladze:1990tg,Gorishnii:1990vf,Baikov:2008jh}
\begin{align}
c_{0,1}&=c_{1,1}=1 \label{eq:c01},\\
c_{2,1}&= - \bigg( \frac{11}{12} -\frac{2}{3} \zeta_3 \bigg) n_f + \frac{365}{24} - 11 \zeta_3  ,\\
c_{3,1}&=  \bigg( \frac{151}{162} - \frac{19}{27} \zeta_3 \bigg) n_f^2 - \bigg( \frac{7847}{216} - \frac{262}{9} \zeta_3 + \frac{25}{9} \zeta_5 \bigg) n_f \nonumber\\
& \hspace{1cm}+\frac{87029}{288} - \frac{1103}{4} \zeta_3 + \frac{275}{6} \zeta_5 ,\\
c_{4,1}&= \bigg(\frac{203}{324}\zeta_3+\frac{5}{18}\zeta_5-\frac{6131}{5832}\bigg) n_f^3  \nonumber\\
   &\hspace{1cm}+\bigg(-\frac{40655}{864}\zeta_3+\frac{5}{6} \zeta_3^2-\frac{260}{27} \zeta_5+\frac{1045381}{15552}\bigg) n_f^2 \nonumber\\
   &\hspace{1cm}  + \bigg(\frac{12205}{12} \zeta_3-55
   \zeta (3)^2+\frac{29675}{432} \zeta_5+\frac{665}{72} \zeta_7-\frac{13044007}{10368}\bigg) n_f \nonumber \\
   &\hspace{1cm}  -\frac{7315}{48} \zeta_7+\frac{65945}{288} \zeta_5+\frac{5445}{8} \zeta_3^2-\frac{5693495}{864} \zeta_3+\frac{144939499}{20736} .\label{eq:c41}
\end{align}
Here $n_f$ denotes the number of active flavours and $\zeta_n \equiv \zeta(n)$ are values of the Riemann $\zeta$-function. To reconstruct the
remaining coefficients $c_{n,k}$ we can use the renormalization group (RG)
equation. To this end, one considers the Adler function \cite{Adler:1974gd}
\begin{equation}
  D^{(1+0)}(s) = - s \frac{d}{d s} \Pi^{(1+0)}_{ij, \, V/A}(s),
\end{equation}
which is a physical object and $\mu$-independent. Employing
\begin{equation}
\mu^2 \frac{d}{d \mu^2} D^{(1+0)}_{ij, \, V/A}(s) = 0,
\end{equation}
one finds \cite{Beneke_2008} 
\begin{alignat}{2}
c_{2,2} &= -\frac{\beta_0}{2} c_{1,1} &\, & \label{eq:rgerel1}\\
c_{3,3} &= \frac{\beta_0^2}{3} c_{1,1} , &\qquad c_{3,2}&= -\frac{\beta_1}{2} c_{1,1} - \beta_0 c_{2,1}  \label{eq:rgerel2}\\
c_{4,4} &= - \frac{\beta_0^3}{4} c_{1,1}, &\qquad c_{4,3}&=\frac{5}{6}  \beta_1 \beta_0 c_{1,1} +  \beta_0^2 c_{2,1}, \quad \label{eq:rgerel3}\\
 & &\qquad  c_{4,2}&= - \frac{1}{2} \big( \beta_2 c_{1,1} + 2 \beta_1 c_{2,1} + 3 \beta_0 c_{3,1} \big), \nonumber
\end{alignat}
where $c_{n,k}=0$ if $k>n$ and $n \neq 0$ and the coefficients
$\beta_i$ of the QCD beta function are listed in Appendix~\ref{app:beta}.

\section{QCD Corrections to Charged Current HSN Decays}\label{sec:qcdcorr}
Using the correlator in Eq.~(\ref{eq:Wcorr}) the differential inclusive decay rate of a HSN via a charged current reads
\begin{align}
\frac{d \Gamma(N \rightarrow \ell X )}{ds} =& N_c \frac{G_F^2 m_N^5 |V_{N\ell}|^2 }{192 \pi^3} \times \frac{12 \pi}{m_N^2}  \bigg(1+x_\ell^2 - \frac{s}{m_N^2} \bigg) \sqrt{\lambda\bigg(1,\frac{s}{m_N^2},x_\ell^2\bigg)} \\
 \times  \bigg[ &\bigg(1+2 \frac{s}{m_N^2}+x_\ell^2 - \frac{4 x_\ell^2}{1+x_\ell^2 -\frac{s}{m_N^2}}\bigg) \mathrm{Im} \Pi^{(1+0)}(s) - 2 \frac{s}{m_N^2} \mathrm{Im} \Pi^{(0)}(s)  \bigg], \nonumber
\end{align}
integrating over $s$, or the dimensionless variable $x=s/m_N^2$, leads to \cite{BRAATEN1992581}
\begin{align}
\label{eq:incldecaywidth}
\Gamma(N \rightarrow \ell X ) =& N_c \frac{G_F^2 m_N^5 |V_{N\ell}|^2 }{192 \pi^3} \times 12 \pi \int\limits_0^{(1-x_\ell)^2} d x \, (1+x_\ell^2 -x) \sqrt{\lambda(1,x,x_\ell^2)} \\
 \times \bigg[& \bigg(1+2x+x_\ell^2 - \frac{4 x_\ell^2}{1+x_\ell^2 -x}\bigg) \mathrm{Im} \Pi^{(1+0)}(m_N^2 x) - 2x \mathrm{Im} \Pi^{(0)}(m_N^2 x)  \bigg], \nonumber
\end{align}
with $x_\ell = m_\ell/m_N$ and where we used 
\begin{align}
\Pi^{(J)} =& |V_{ud}|^2 \bigg( \Pi^{(J)}_{ud, \, V} + \Pi^{(J)}_{ud, \, A} \bigg) + |V_{us}|^2 \bigg(\Pi^{(J)}_{us, \, V} + \Pi^{(J)}_{us, \, A}  \bigg)+... \\
=& 2 (|V_{ud}|^2 + |V_{us}|^2 +...) \Pi^{(J)}_{V}, \nonumber
\end{align}
since the vector and axial vector contribution are equal in magnitude. Here the dots indicate to the inclusion of heavier mesons i.e. $D$ or $B$ mesons. Since $\Pi^{(0)} \sim m_q^2$ we neglect it. The integration region with small $x$ involves hadronic resonances while the integrand in Eq.~(\ref{eq:incldecaywidth}) is smooth. Eq.~(\ref{eq:incldecaywidth}) nevertheless reproduces the correct result as can be derived from analyticity properties of $\Pi^{(J)}$, see Appendix~\ref{app:contourintegration} for details.

Choosing $\mu = m_N$ the imaginary part of the correlator reads
\begin{align}
\mbox{Im}\,  \Pi^{(1+0)}_{V}(m_N^2 x) &= -\frac{1}{12 \pi^2} \sum\limits_{n=0}^{\infty} a_{m_N}^n \sum\limits_{k=1}^{n+1} c_{n,k} \mathrm{Im} \, \ln^k(-x)  \label{eq:imp} \\
&= \frac{1}{12 \pi} \bigg\{ 1 + a_{m_N} + a_{m_N}^2 (c_{2,1} + 2 c_{2,2} \ln(x)) \nonumber\\
&\hspace{2cm}+ a_{m_N}^3 (c_{3,1} + 2 c_{3,2} \ln(x) - (\pi^2 - 3\ln^2(x))c_{3,3})  \nonumber\\
&\hspace{2cm}+ a_{m_N}^4 (c_{4,1} + 2 c_{4,2} \ln(x) -(\pi^2 - 3\ln^2(x)) c_{4,3} \nonumber\\
&\hspace{2cm} - (4\pi^2 \ln(x) - 4 \ln^3(x)) c_{4,4}) \bigg\} \nonumber
\end{align}
and leads to integrals of the form 
\begin{equation}
I_k = \int\limits_0^{(1-x_\ell)^2} d x \, \bigg((1+x_\ell^2 -x)  (1+2x+x_\ell^2) - 4 x_\ell^2 \bigg) \sqrt{\lambda(1,x,x_\ell^2)} \ln^k( x). \label{eq:defik}
\end{equation}
Up to $k=2$ the integrals are analytically calculable in terms of 
dilogarithm and trilogarithm functions:
\begin{align}
I_0 &= \frac{1}{2} \bigg( 1 - 8 x_\ell^2 + 8 x_\ell^6 - x_\ell^8 - 12 x_\ell^4 \ln(x_\ell^2) \bigg) ,\label{eq:I0}\\
I_1 &= - \frac{19}{24} + \frac{13}{3} x_\ell^2 - 12 \zeta_2 x_\ell^4 - \frac{13}{3} x_\ell^6 + \frac{19}{24} x_\ell^8 \nonumber\\
&+ \bigg( -3 x_\ell^4 -4x_\ell^6 + \frac{1}{2} x_\ell^8 \bigg) \ln(x_\ell^2) \nonumber\\
&+ \bigg( 1 - 8x_\ell^2 +8x_\ell^6 - x_\ell^8 \bigg) \ln(1-x_\ell^2) + 12 x_\ell^4 \mathrm{Li}_2(x_\ell^2) , \label{eq:I1}\\
I_2 &= \frac{265}{144} - \frac{151}{18} x_\ell^2 - 12 (\zeta_2 + 2 \zeta_3) x_\ell^4 + \bigg( \frac{151}{18} -16 \zeta_2 \bigg) x_\ell^6 + \bigg( -\frac{265}{144} + 2 \zeta_2 \bigg) x_\ell^8 \nonumber\\
&+ \bigg( -\frac{19}{6} + \frac{52}{3} x_\ell^2 - 48 \zeta_2 x_\ell^4 - \frac{52}{3} x_\ell^6 + \frac{19}{6} x_\ell^8 + 48 x_\ell^4 \mathrm{Li}_2(x_\ell^2) \bigg) \ln(1-x_\ell^2) \nonumber\\
&+ (2-16 x_\ell^2 + 16x_\ell^6 -2 x_\ell^8) \ln^2(1-x_\ell^2) \nonumber\\
&+ \bigg( -x_\ell^2 + \frac{15}{2} x_\ell^4 +\frac{23}{3}x_\ell^6 -\frac{19}{12} x_\ell^8  \nonumber\\
& + (-1+8x_\ell^2-8x_\ell^6+x_\ell^8)\ln(1-x_\ell^2) + 24 x_\ell^4 \ln^2(1-x_\ell^2)- 12 x_\ell^4 \mathrm{Li}_2(x_\ell^2) \bigg) \ln(x_\ell^2)  \nonumber\\ 
&+ (-1 +8 x_\ell^2 +12 x_\ell^4 +8 x_\ell^6 - x_\ell^8) \mathrm{Li}_2(x_\ell^2) \nonumber\\
& + 24 x_\ell^4 \mathrm{Li}_3(x_\ell^2) + 48 x_\ell^4 \mathrm{Li}_3(1-x_\ell^2) .\label{eq:I2}
\end{align}
In Appendix \ref{app:intfit} we give approximations to these integrals for easier use.

We next derive a new analytical result holding as well for $k \geq 3$, defined in terms of a well converging sum. To do this we first use the following series representation of the square root of the K\"all\'en function, 
\begin{align}
  \frac{\dx^n}{\dx x^n} \sqrt{\lambda(1,x,x_\ell^2)} =&
  \sum\limits_{j=0}^{\lfloor \frac{n}{2} \rfloor} (-1)^{n-j}
  \bigg(- \frac{1}{2} \bigg)_{n-j} (2j-1)!!
  \begin{pmatrix}
 n \\ 2j
  \end{pmatrix} \\
& \hspace{1cm} \times \frac{\bigg(\lambda'(1,x,x_\ell^2)\bigg)^{n-2j}
    \bigg(\lambda''(1,x,x_\ell^2)\bigg)^{j}}{\bigg(
    \lambda(1,x,x_\ell^2)\bigg)^{\frac{2(n-j)-1}{2}}}, \nonumber
\end{align}
where $(a)_n = a (a+1)(a+2)\cdot ... \cdot (a+n-1)$ is the Pochhammer
symbol, $\lfloor j \rfloor$ is the floor function \un{(rounding a real number to the nearest 
smaller integer number)} and
\begin{align}
\lambda(1,x,x_\ell^2) &= \tk{x^2 - 2 x (1+x_\ell^2) + (1-x_\ell^2)^2} ,\\
\lambda'(1,x,x_\ell^2)&= \frac{\dx \lambda(1,x,x_\ell^2)}{\dx x} = \tk{2 x - 2(1+x_\ell^2)}, \\
\lambda''(1,x,x_\ell^2)&= \frac{\dx^2 \lambda(1,x,x_\ell^2)}{\dx x^2} = 2.
\end{align}
We write  
\begin{equation}
I_{k}(x_\ell^2,a) = \sum\limits_{n=0}^{\infty} \frac{1}{n!}A_n(x_\ell^2) B_{n,k}(x_\ell^2,a) \label{eq:defab} 
\end{equation}
with
\begin{align}
A_n(x_\ell^2)&\equiv \, \lt. \frac{\dx^n}{\dx x^n}\sqrt{\lambda(1,x,x_\ell^2)}\rrt|_{x=0} \nn
&\hspace{-0.75cm}=  \tk{\sum\limits_{j=0}^{\lfloor \frac{n}{2} \rfloor} (-1)^{j} \bigg(- \frac{1}{2} \bigg)_{n-j} (2j-1)!! \begin{pmatrix}
n \\ 2j
\end{pmatrix}  2^{n-j} (1+x_\ell^2)^{n-2j} (1-x_\ell^2)^{1-2(n-j)}, }
\end{align}  
and
\begin{align}
B_{n,k} (x_\ell^2,a) &\equiv \,
      \int\limits_{0}^{a} \! d x \,  \bigg(
                  (1+x_\ell^2-x)(1+2x+x_\ell^2) -4 x_\ell^2 \bigg) x^n
                     \ln^k(x) \nn
  &=\,  -2 I_{n+2,k}\lt( a \rrt)  + (1+x_\ell^2)
    I_{n+1,k}\lt( a \rrt)  + (1-x_\ell^2)^2
    I_{n,k}\lt( a \rrt) .  
\end{align}
The integrals in this expression are given by
\begin{align}
  I_{n,k}(a) &\equiv \, \int_0^a dx \, x^n \ln^k x \nn
  &=\, a^{n+1} \sum_{l=0}^k (-k)_l (n+1)^{-l-1} \ln^{k-l} a
    \label{eq:defi} ,
\end{align}
which completes the calculation of $I_k$ in \eq{eq:defik}.

Inserting \eq{eq:imp} into \eq{eq:incldecaywidth} and using Eq.~(\ref{eq:defik}) finally gives the full
semi-hadronic width for decays of sterile neutrinos lighter than the $D^{+}$
meson
\begin{align}
\Gamma(N  \rightarrow \ell X ) =& N_c \frac{G_F^2 m_N^5 |V_{N\ell}|^2 }{192 \pi^3} \cdot 2 (|V_{ud}|^2 + |V_{us}|^2) \label{eq:fullrate}\\
& \hspace{-2cm}\times  \bigg[ I_0(x_\ell^2,(1-x_\ell)^2) c_{0,1} + a_{m_N}  c_{1,1} I_0(x_\ell^2,(1-x_\ell)^2) \nonumber \\
& \hspace{-2cm} +a_{m_N}^2 \big[ c_{2,1} I_0(x_\ell^2,(1-x_\ell)^2) +2 c_{2,2} I_1(x_\ell^2,(1-x_\ell)^2) \big] \nonumber\\
&\hspace{-2cm} +a_{m_N}^3 \big[  c_{3,1} I_0(x_\ell^2,(1-x_\ell)^2) +2 c_{3,2} I_1(x_\ell^2,(1-x_\ell)^2) \nonumber\\
&- (\pi^2 I_0(x_\ell^2,(1-x_\ell)^2)-3  I_2(x_\ell^2,(1-x_\ell)^2) )c_{3,3} \big] \nonumber\\
& \hspace{-2cm} +a_{m_N}^4 \big[  c_{4,1} I_0(x_\ell^2,(1-x_\ell)^2)+2 c_{4,2} I_1(x_\ell^2,(1-x_\ell)^2) \nonumber\\
&-(\pi^2 I_0(x_\ell^2,(1-x_\ell)^2)-3 I_2(x_\ell^2,(1-x_\ell)^2))c_{4,3} \nonumber\\
&- (4\pi^2 I_1(x_\ell^2,(1-x_\ell)^2) -4 I_3(x_\ell^2,(1-x_\ell)^2) )c_{4,4} \big] \bigg]. \nonumber
\end{align}
Logarithms involving the renormalization scale may be
reconstructed by using the relation in Appendix \ref{app:beta}. Sterile neutrinos heavier than the $D^{+}$ meson also obey the relation
above once the corresponding CKM element as well as threshold effects
are included.

Additionally we also calculated the integral in Eq.~(\ref{eq:defik}) for an arbitrary upper limit of the hadronic invariant mass, to {permit the calculation of} the integrated spectrum of the decay
\begin{align}
\Gamma_{\mathrm{cut}}(N \rightarrow \ell X) &= \int\limits_{0}^{s_{\mathrm{max}}}  ds \, \frac{d\Gamma(N \rightarrow \ell X)}{d s}, \quad s_\mathrm{max} \in [\mathcal{O}(\SI{1}{\giga\electronvolt}), (m_N - m_\ell)^2], \nonumber\\
&= \int\limits_{0}^{x_{\mathrm{max}}}  dx \, \frac{d\Gamma(N \rightarrow \ell X)}{d x} 
\end{align}
where we defined $x_\mathrm{max} = s_\mathrm{max}/m_N^2$. $\Gamma_{\mathrm{cut}}(N \rightarrow \ell X)$ obeys Eq.~(\ref{eq:fullrate}) with $I_k = I_k(x_\ell^2,x_\mathrm{max})$. Up to $k=2$ we found a closed analytical form of the $I_k(x_\ell^2,x_\mathrm{max})$. The expressions are found in Appendix \ref{app:integratedspectrum}, but the calculation from Eq.~(\ref{eq:defab}) is easier even for $k \leq 2$.

\section{Numerical Analysis}\label{sec:numana}
\subsection{Quality of the perturbation series}
Perturbative QCD describes observables in an expansion of the strong coupling constant accurately if the relevant energy scale $\mu$ is large enough such that $\alpha_s$ is small. Any observable $O_\mu$ at the scale $\mu$ may be written as 
\begin{equation}
O_\mu = \sum\limits_n r_n \alpha_s(\mu)^n  = r_0 + r_1 \alpha_s(\mu) + r_2 \alpha_s(\mu)^2 + \mathcal{O}(\alpha_s(\mu)^3).
\end{equation}
However, at small scales ($\mu \sim \mathcal{O}(\SI{1}{\giga\electronvolt})$) the strong coupling constant becomes large and an expansion in $\alpha_s$ is no longer guaranteed to converge. For the $\tau$ lepton with a mass of $m_\tau = \SI{1.776}{\giga\electronvolt}$ it was only possible to correctly describe the {inclusive} hadronic decay {width} through the inclusion of five-loop corrections. Only at the five-loop level does the perturbative description become stable i.e. the $\mathcal{O}(\alpha_s^4)$ corrections lead to a significantly reduced dependence on the renormalization scale \cite{Baikov:2008jh}. In the case of a HSN with mass comparable to the $\tau$ the kinematics is, up to the mass of the lepton attached to the HSN, the same. Setting $m_\ell = 0 $ leads to the same numerical coefficients as in $\tau$ decay 
\begin{align}
\frac{\Gamma(N \rightarrow \ell X )}{|V_{ud}|^2 + |V_{us}|^2}=& N_c\frac{G_F^2 m_N^5 |V_{N\ell}|^2}{192 \pi^3} \\ & \times \bigg[ 1 + a_{m_N} + 5.202 a_{m_N}^2 + 26.366 a_{m_N}^3 +127.079 a_{m_N}^4  \bigg], \nonumber
\end{align}
perfectly agreeing with Ref. \cite{Baikov:2008jh}. For a HSN mass around $\SI{3}{\giga\electronvolt}$ the inclusive semi-hadronic decay with $\ell = \tau$ in the final state is kinematically accessible. The lepton mass ratio is then sizable, $x_\tau = 0.592$, and 
\begin{align}
\frac{\Gamma(N \rightarrow \tau X )}{|V_{ud}|^2 + |V_{us}|^2} =& N_c\frac{G_F^2 m_N^5 |V_{N\ell}|^2}{192 \pi^3} \\
& \times 0.071  \bigg[1 + a_{m_N} + 8.376 a_{m_N}^2 + 74.605 a_{m_N}^3 + 669.805 a_{m_N}^4\bigg], \nonumber
\end{align}
largely reducing the decay rate and making the convergence of the perturbative series worse.

\begin{figure}[t]
\centering
\includegraphics[width=0.65\textwidth]{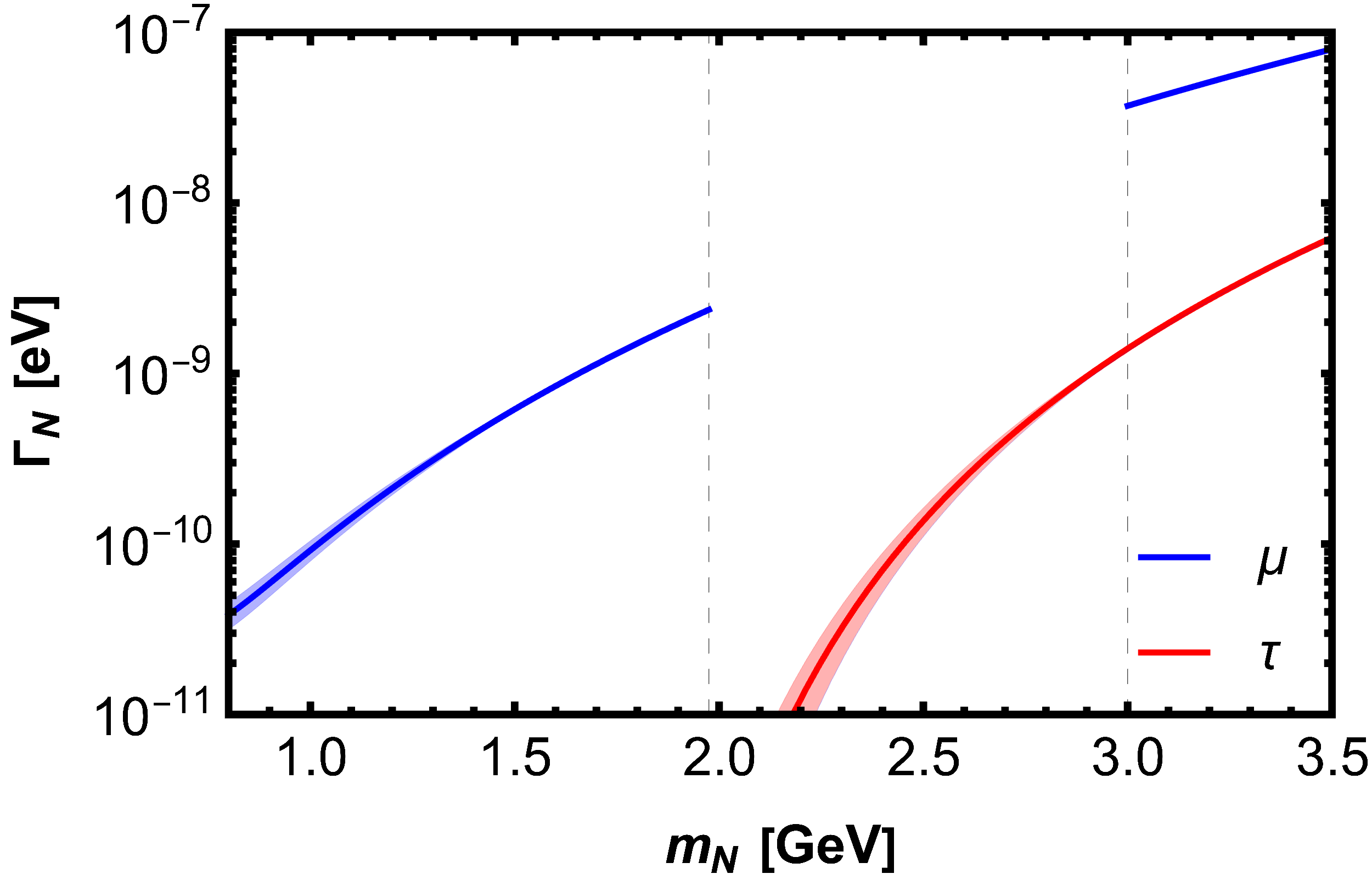}
\caption{Hadronic decay widths $\Gamma(N \rightarrow \ell X)$ for the tagging leptons $\ell = \mu, \tau$ with a mixing angle of $V_{N\ell} = 10^{-3}$. The width for $\ell=e$ looks like the $\ell = \mu$ width. The numerics of the running coupling were calculated with the help of \texttt{RunDec} \cite{Chetyrkin:2000yt,Herren:2017osy}. We do not show the region $m_N \in [m_\mu + m_D , \SI{3}{\giga\electronvolt}]$ where we expect large non-perturbative effects. The error band is taken as the difference between minimum and maximum of the width with respect to the renormalization scale $\sigma_\Gamma = (\max \, \Gamma_N(\mu) - \min \, \Gamma_N(\mu))/2$ for $ \SI{0.8}{\giga\electronvolt} \leq \mu \leq \SI{3.5}{\giga\electronvolt}$. We do not include the uncertainty {from our omission of} quark masses, which can be sizable, $\mathcal{O}(m_D^2/q^2) = \mathcal{O}(40\%)$ for $m_N = \SI{3}{\giga\electronvolt}$, beyond the charm production threshold.}
~\\[-3mm]\hrule
\label{fig:qcddecaywithtagginleps}
\end{figure}

\begin{figure}[t]
\centering
\subfigure[Inclusive HSN decay width with $\ell = \mu$. Here $V_{N\ell=\mu}=10^{-3}$. \label{fig:muonwidth}]{\includegraphics[width=0.65\textwidth]{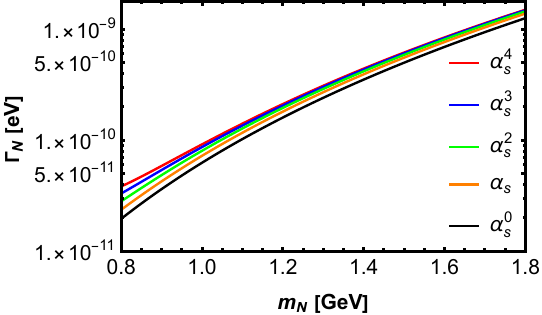}}
\subfigure[$\Gamma_N$ at different orders of $\alpha_s$, \kt{normalized} to the LO result. \label{fig:muonratio}]{\includegraphics[width=0.65\textwidth]{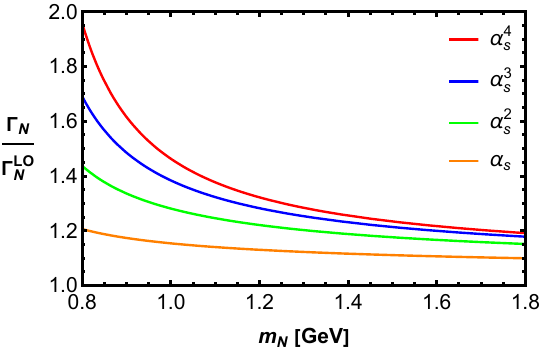}}
\caption{Inclusive decay width $\Gamma_N = \Gamma(N \rightarrow \mu X)$ {as a function of $m_N$}.}
~\\[-3mm]\hrule
\label{fig:muon}
\end{figure}

In Fig. \ref{fig:qcddecaywithtagginleps} we show the full decay width. The widths for $\ell = e$ and $\ell = \mu$ are indistinguishable. For $m_N \geq m_\mu + m_D$ the decay channel into charmed mesons opens up. For this reason we omit the region $m_N \in [m_\mu + m_D , \SI{3}{\giga\electronvolt}]$ in Fig. \ref{fig:qcddecaywithtagginleps} as here there are resonances of charmed mesons. They are dominated by non-perturbative effects making an accurate perturbative description in this region impossible. We show the $\ell = \mu $ rate again beyond $m_N \geq \SI{3}{\giga\electronvolt}$. We have computed the error band by taking as uncertainty the difference between the minimal and maximal width with respect to the renormalization scale i.e. $\sigma_\Gamma = (\max \, \Gamma_N(\mu) - \min \, \Gamma_N(\mu))/2$ for $ \SI{0.8}{\giga\electronvolt} \leq \mu \leq \SI{3.5}{\giga\electronvolt}$. Since we did not include charm mass corrections we expect an additional error of $\mathcal{O}(m_D^2/q^2) = \mathcal{O}(40\%)$ for $m_N = \SI{3}{\giga\electronvolt}$, which is not shown.
In Fig.~\ref{fig:muon} we plot the effect of each order in $\alpha_s$ on the total decay width for $\ell =\mu$. For small HSN masses, close to the kinematic threshold of multi-hadron production, we see that including more corrections in $\alpha_s$ does not improve the convergence. This is even more pronounced in Fig.~\ref{fig:muonratio} where the relative increase compared to the leading order is large for HSN masses in a window below $\sim \SI{1}{\giga\electronvolt}$. Beyond $\SI{1}{\giga\electronvolt}$ the perturbative behavior rapidly improves and higher order corrections {become smaller with higher orders in $\alpha_s$} in relation to the leading order.
We specifically analyzed the scale dependence of the widths by expressing $a_{m_N}$ in terms of logarithms of the renormalization scale $\mu$ and the HSN mass as seen in Appendix \ref{app:beta}.
\begin{figure}[t!]
\centering
\subfigure{\includegraphics[width=0.49\textwidth]{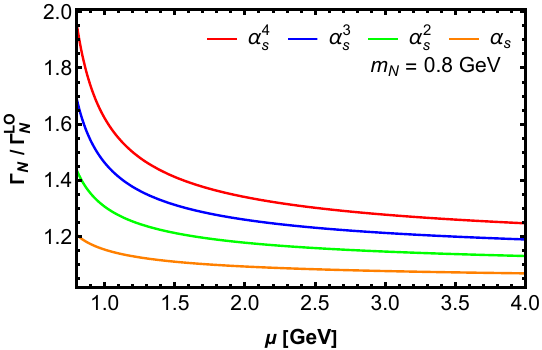}}
\subfigure{\includegraphics[width=0.49\textwidth]{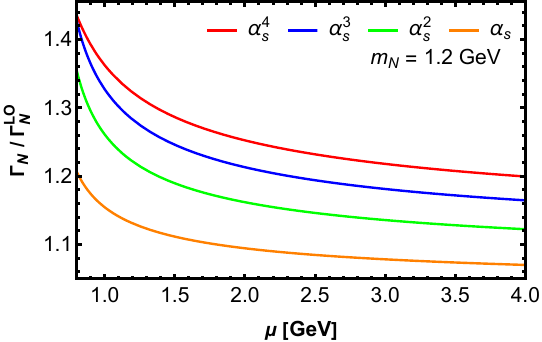}}
\subfigure{\includegraphics[width=0.49\textwidth]{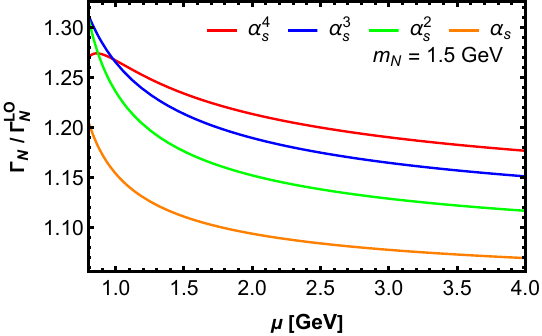}}
\subfigure{\includegraphics[width=0.49\textwidth]{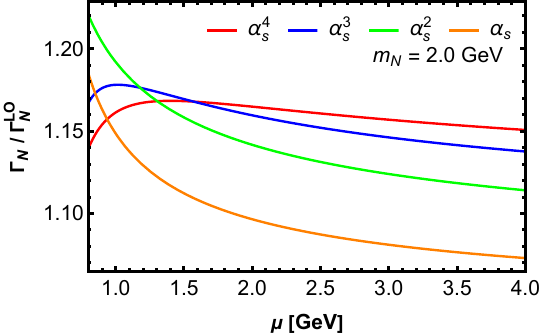}}
\caption{Scale dependence of the inclusive decay width $\Gamma_N = \Gamma(N \rightarrow \mu X)$ for different HSN masses.}
~\\[-3mm]\hrule
\label{fig:runningmuon}
\end{figure} 
In Fig. \ref{fig:runningmuon} the scale dependence for different HSN masses is shown. The scale dependence is very strong close to the kinematic threshold. Increasing the mass of the HSN the scale dependence reduces. We see that HSN masses around $\SI{1.5}{\giga\electronvolt}$, including the highest order corrections in $\alpha_s$ available, lead to a flat scale dependence.

\begin{figure}[t]
\centering
\subfigure[Inclusive HSN decay width with $\ell = \tau$. Here $V_{N\ell=\tau}=10^{-3}$. \label{fig:tauonwidth}]{\includegraphics[width=0.65\textwidth]{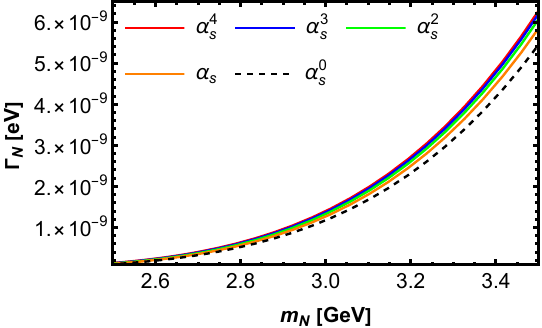}}
\centering
\subfigure[$\Gamma_N$ at different orders of $\alpha_s$, \kt{normalized} to the LO result. \label{fig:tauonratio}]{\includegraphics[width=0.65\textwidth]{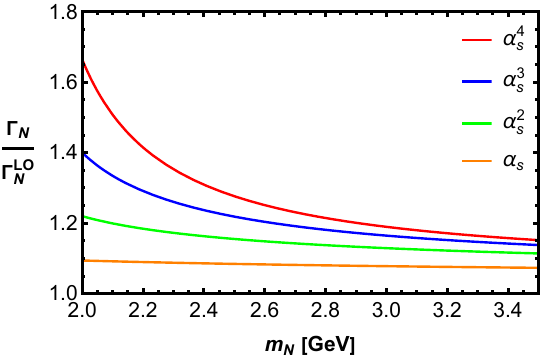}}
\caption{Inclusive decay width $\Gamma_N = \Gamma(N \rightarrow \tau X)$}
~\\[-3mm]\hrule
\label{fig:tauon}
\end{figure}

\begin{figure}[t!]
\centering
\subfigure{\includegraphics[width=0.49\textwidth]{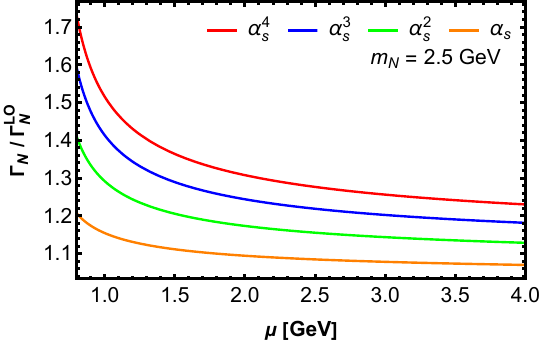}}
\subfigure{\includegraphics[width=0.49\textwidth]{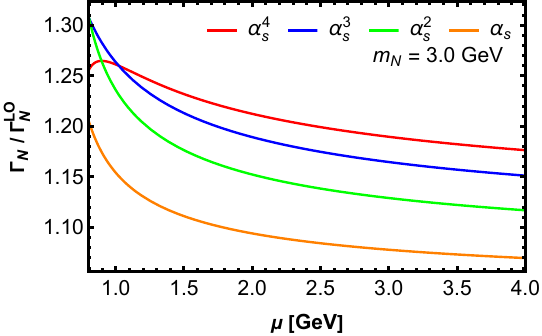}}
\subfigure{\includegraphics[width=0.49\textwidth]{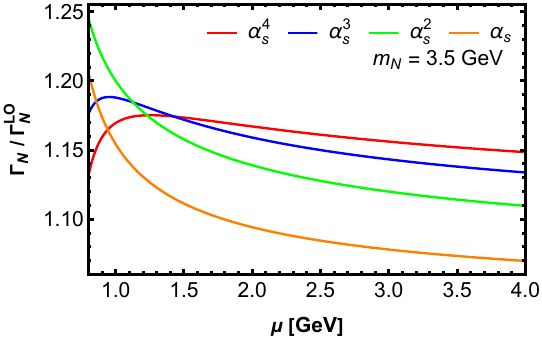}}
\subfigure{\includegraphics[width=0.49\textwidth]{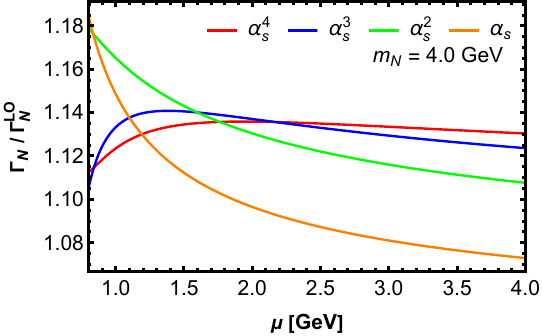}}
\caption{Scale dependence of the inclusive decay width $\Gamma_N = \Gamma(N \rightarrow \tau X)$ for different HSN masses.}
~\\[-3mm]\hrule
\label{fig:runningtau}
\end{figure}

In the $\ell = \tau$ final state lepton case the discussion is qualitatively similar to the $\ell=e$ and $\ell=\mu$ case. Looking only at the decay width in Fig. \ref{fig:tauonwidth} the effect of poor convergence does not seem to be as pronounced, however, in absolute values it is comparable to the decay into a muon as seen in Fig. \ref{fig:tauonratio}. As in the muon case, the quality of the perturbative series rapidly increases for larger HSN masses. 
In Fig. \ref{fig:runningtau} the scale dependence for different HSN masses is shown. The scale dependence reduces for larger masses. At a HSN mass of $\SI{3}{\giga\electronvolt}$ the $\mathcal{O}(\alpha_s^4)$ corrections reduce the scale dependence significantly.
For larger HSN masses already lower \kt{order} results show a satisfactory small scale dependence.

\subsection{Branching ratios of HSN}
Without the neutral current contributions mediated by the $Z$ boson one cannot calculate the total width, which is needed for branching ratio predictions. However ratios of branching ratios for the charged decays can be calculated
\begin{equation}
\kappa_{\ell}^{P} = \frac{Br(N \rightarrow \ell P)}{Br(N \rightarrow \ell X)} = \frac{\Gamma(N \rightarrow \ell P)}{\Gamma(N \rightarrow \ell X)}, \label{eq:kappa}
\end{equation}
here $P = \pi^+,K^+,D^+,...$ is some meson, $\Gamma(N \rightarrow \ell P)$ is the width of a HSN into a meson
\begin{equation}
\Gamma(N \rightarrow \ell P) = \frac{G_F^2 m_N^3 |V_{N\ell}|^2}{16\pi} |V_{u'd'}|^2 f_P^2 \bigg[(1 - x_\ell^2)^2 - x_P^2 (1+x_\ell^2) \bigg] \sqrt{\lambda(1,x_\ell^2,x_P^2)},
\end{equation}
where $f_P$ is the meson decay constant, $|V_{u'd'}|$ is the relevant CKM matrix element, and $x_i = m_i/m_N$, and $\Gamma(N \rightarrow \ell X)$ is the fully inclusive semi-hadronic decay rate as given in Eq.~(\ref{eq:fullrate}). For our predictions of $\kappa_\ell^P$ we assume $m_N > \SI{1.5}{\giga\electronvolt}$ for electron and muon tagging and $m_N > \SI{3}{\giga\electronvolt}$ for tau tagging. 
\begin{figure}[tp]
\centering
\subfigure[$\kappa_\mu^P$]{\includegraphics[width=0.65\textwidth]{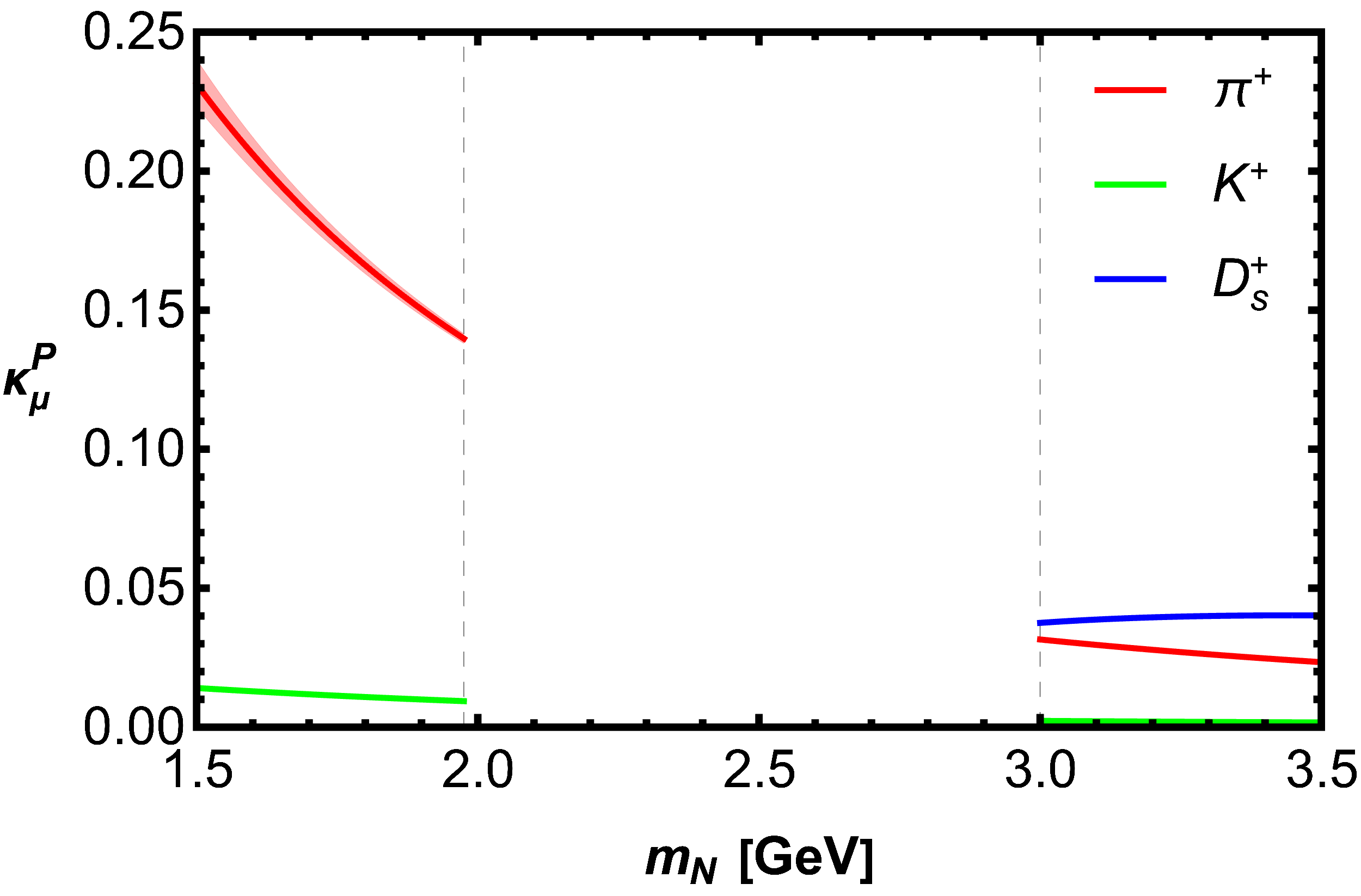}}
\subfigure[$\kappa_\tau^P$]{\includegraphics[width=0.65\textwidth]{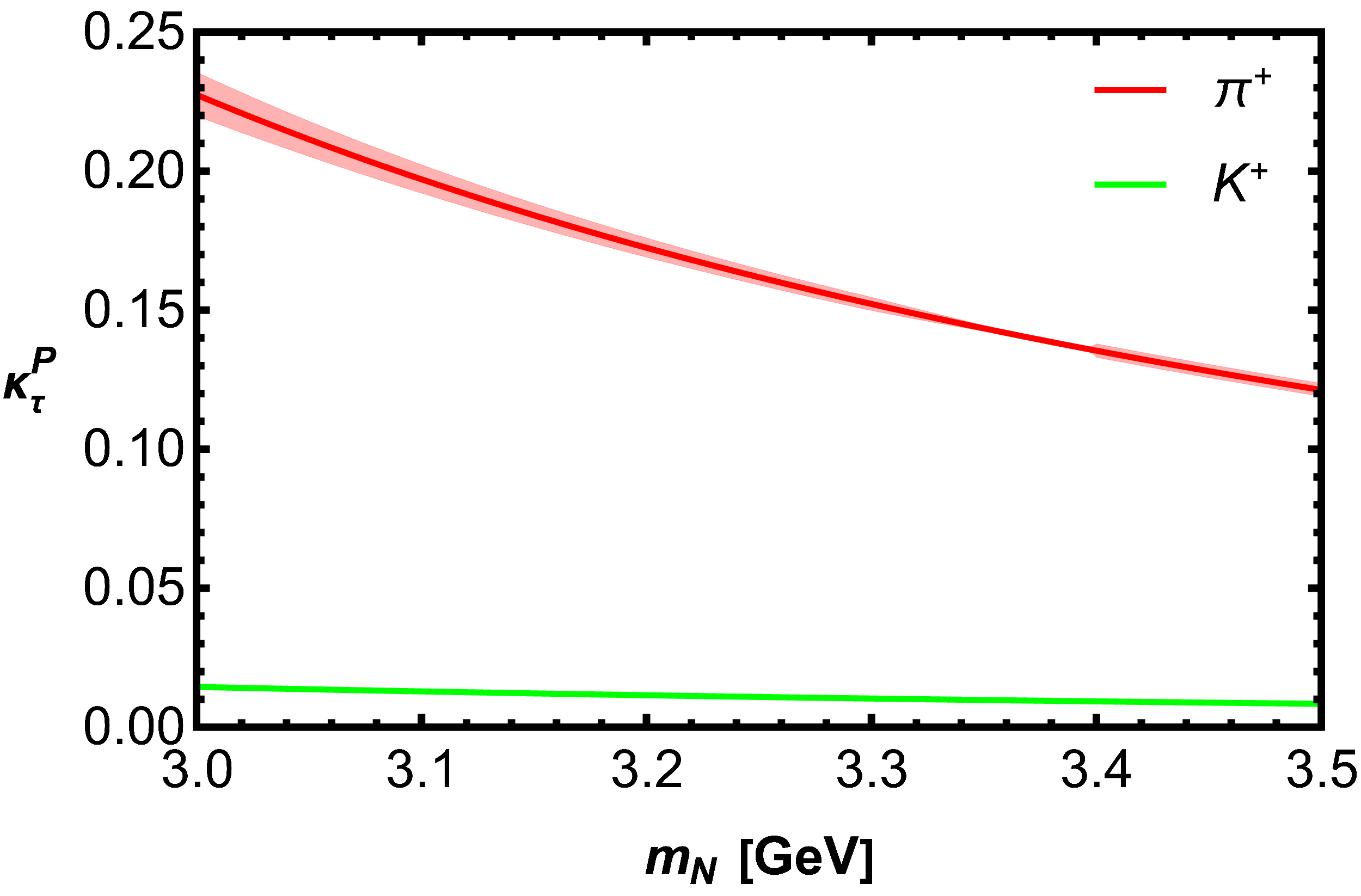}}
\caption{$N \rightarrow \ell P$ branching fractions (normalized to $Br(N \rightarrow \ell N)$) for $P=\pi^+, K^+, D_s^+$, see Eq.~(\ref{eq:kappa}). We use $f_D = \SI{212.0 \pm 0.7}{\mega\electronvolt}$, $f_{D_s} = \SI{249.9 \pm 0.5}{\mega\electronvolt}$ \cite{Bazavov:2017lyh,Carrasco:2014poa,FlavourLatticeAveragingGroupFLAG:2024oxs}, and $|V_{cd}| = 0.221 \pm 0.004$, and $|V_{cs}| = 0.975 \pm 0.006$ \cite{ParticleDataGroup:2024cfk}. The error bands are the same as in Fig.~\ref{fig:qcddecaywithtagginleps}.}
~\\[-3mm]\hrule
\label{fig:kappamuon}
\end{figure}
In Fig. \ref{fig:kappamuon} the $\kappa$-ratio for muon and tau tagging are shown. In both cases pions dominate the decay and only a few CKM-suppressed decays to kaons are predicted. For this reason we omit decays to $D$ mesons as they are more suppressed than those into kaons due to the smaller phase space. In addition we also show the impact of the inclusion of $D_s$ mesons. $D_s$ mesons are produced copiously beyond the charm threshold as $V_{cs} \approx 1$. We exclude the region $m_N \in [m_\mu +m_D, \SI{3}{\giga\electronvolt}]$ as here non-perturbative charm resonance effects dominate the width. As long as the decay channels to charmed final states are absent, at least $12\%$ of the charged-current decays of HSN are $N \rightarrow \ell \pi $ for all considered values of $m_N$.

\subsection{Constraints on $\theta_{N\tau}$ from the $\tau$ lifetime}
\label{sec:constraints}
The current world average of the $\tau$-lifetime
\begin{equation}
\tau_\tau^{\mathrm{exp.}} = \SI{290.29 \pm 0.53}{\femto\second} \quad \text{\cite{HeavyFlavorAveragingGroupHFLAV:2024ctg}}
\label{eq:tauexp}
\end{equation}
agrees with the SM theory prediction
\begin{equation}
\tau_\tau^{\mathrm{SM}} = \SI{288.59 \pm 2.31}{\femto\second} \quad \text{\cite{ParticleDataGroup:2024cfk,Erler:2002bu}.}
\label{eq:tausm}
\end{equation}

Whenever a sterile neutrino interaction is present it comes with a factor of the mixing angle $V_{N\tau} = \sin\theta_{N\tau}$. At the same time SM neutrino interactions receive a contribution involving $\cos\theta_{N\tau}$. Mixing is possible between one or more generations of SM neutrinos. If there is mixing with $\nu_e$ or $\nu_\mu$ this has an impact on the Fermi constant. A complete analysis including also mixing with $e$ or $\mu$ requires a full electroweak fit (see Ref.\cite{Lacker:2010zz}) and is beyond the scope of this work. Hence we only look at mixing with $\nu_\tau$, and de{fine} $\theta\equiv \theta_{N\tau}$. {Then} a sterile neutrino {will modify} the total width, {implying the following constraint on $\theta$:}
\begin{equation}
\tau_\tau^{\mathrm{exp.}} = \frac{1-\mathcal{B}^s_\tau}{\cos^2\theta \, (1-\mathcal{B}^s_\tau)\Gamma^{\mathrm{SM}} + \sin^2\theta \, \Gamma^{N}},
\label{eq:tautotalwidth}
\end{equation}
where $\mathcal{B}^s_\tau = 0.0292$ \cite{ParticleDataGroup:2024cfk} is the branching fraction of $\tau$ decays into strange final states, $(1-\mathcal{B}^s_\tau)\Gamma^{\mathrm{SM}}$ is the SM width involving leptonic and non-strange hadronic decays, and $\Gamma^{N}$ is the \tk{non-strange} sterile neutrino contribution, i.e.\ 
the $\tau \rightarrow N X$ \un{decay rate} \tk{with $X$ being a non-strange hadronic final state}. \un{$(1-\mathcal{B}^s_\tau)/\tau_\tau^{\mathrm{exp.}}$ is the experimentally determined total decay rate into 
non-strange particles; our approach follows Ref.~\cite{ParticleDataGroup:2024cfk}}. $\Gamma^{\mathrm{SM}}$ is calculated as $\Gamma^{\mathrm{SM}} = 1/\tau_\tau^{\mathrm{SM}}$ and $\Gamma^{N}$ follows from Eq.~(\ref{eq:fullrate}) by replacing $V_{N\ell} \rightarrow 1$, $V_{us} \rightarrow 0$, $m_N \rightarrow m_\tau$, and $I_k(x_\ell^2,(1-x_\ell)^2) \rightarrow I_k(m_N^2/m_\tau^2,(1-m_N/m_\tau)^2)$. For the strong coupling constant we use $\alpha_s^{n_f=3}(m_\tau) = 0.316$ which we obtain by using the weighted average $\alpha_s^{n_f=5}(M_Z) = 0.1182 $ of all measurements listed in the PDG \cite{ParticleDataGroup:2024cfk}, except the one using $\tau$ decay, and running $\alpha_s(M_Z)$ down to $m_\tau$. 
\begin{figure}[t]
\centering
\subfigure{\includegraphics[width=0.49\textwidth]{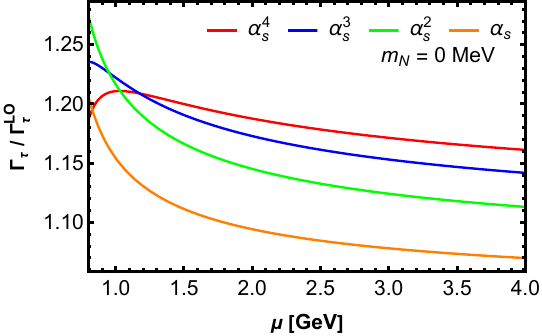}}
\subfigure{\includegraphics[width=0.49\textwidth]{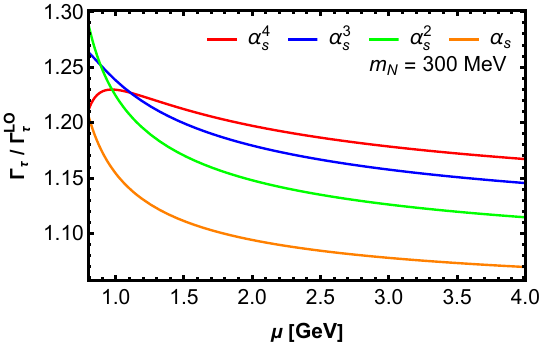}}
\subfigure{\includegraphics[width=0.49\textwidth]{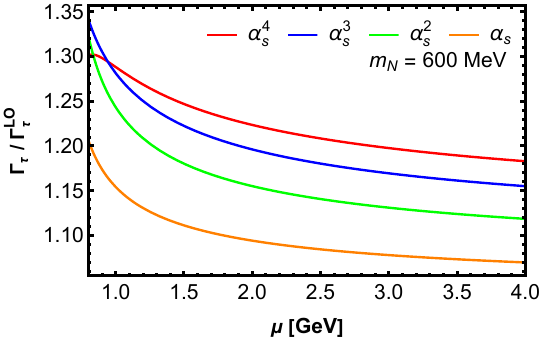}}
\subfigure{\includegraphics[width=0.49\textwidth]{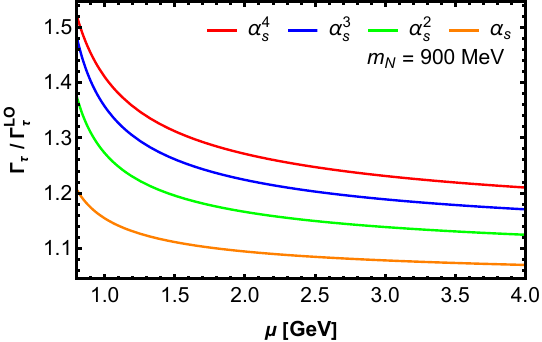}}
\caption{Scale dependence of the inclusive $\tau$ decay width with a final state HSN {for different HSN masses $m_N$}.}
~\\[-3mm]\hrule
\label{fig:taurunning}
\end{figure} 
In Fig. \ref{fig:taurunning} we show the renormalization scale dependence of the $\tau$ decay width. At $\mathcal{O}(\alpha_s^4)$ the scale dependence remains sufficiently flat up to HSN masses around $m_N =\SI{600}{\mega\electronvolt}$. For larger HSN masses the perturbative description becomes poorer.

\begin{figure}[t]
\centering
\includegraphics[width=0.65\textwidth]{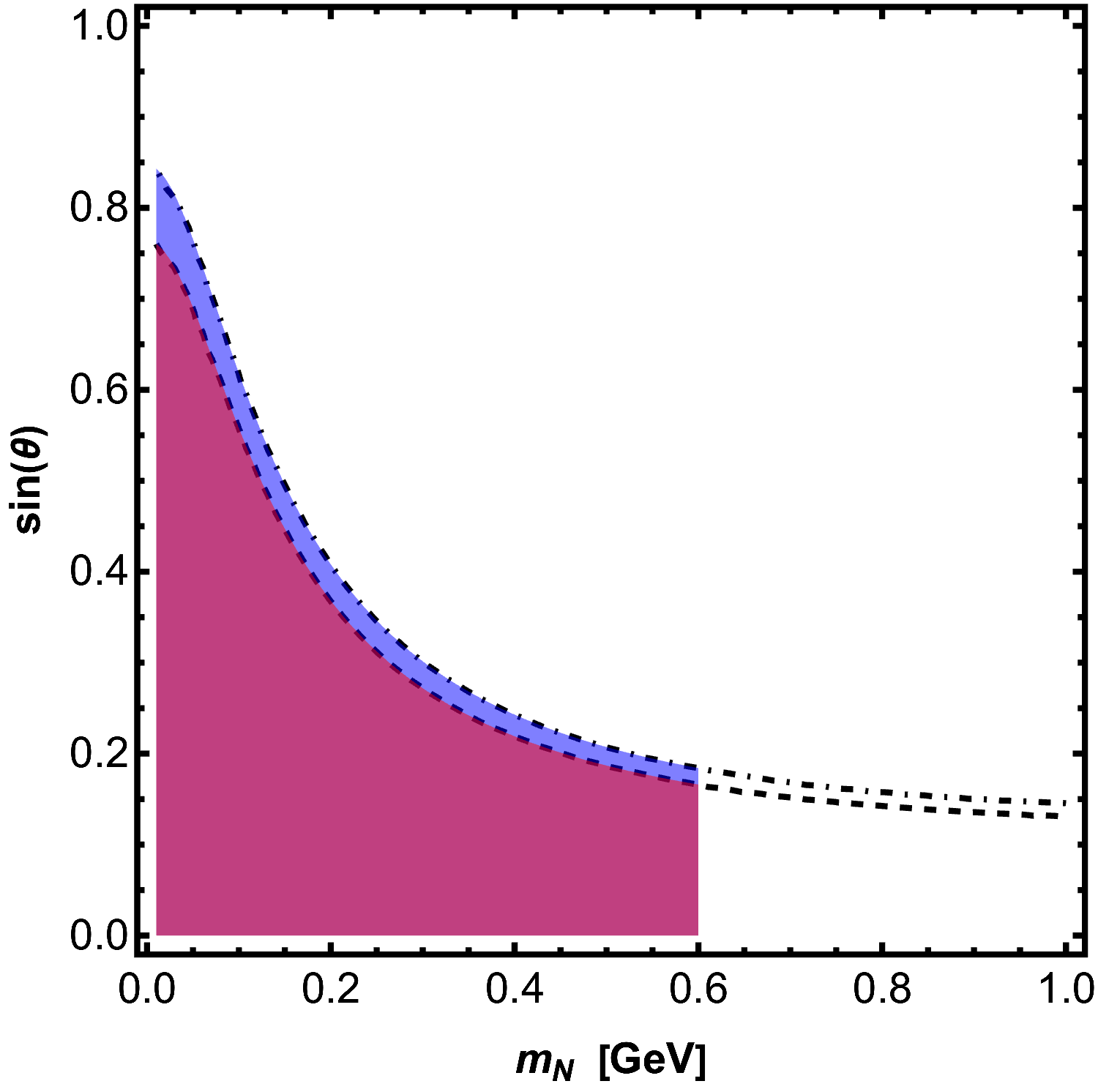}
\caption{Allowed parameter space deduced from the measured $\tau$ lifetime {in \eq{eq:tauexp} via \eq{eq:tautotalwidth}} at the $1\sigma$ (purple) and $3\sigma$ (purple and blue) levels. 
{For} $m_N\leq \SI{600}{\mega\electronvolt}$ we trust the perturbative calculation. The dashed lines for 
$m_N > \SI{600}{\mega\electronvolt}$ {indicate the perturbative result in a region where the} description is no longer reliable.}
~\\[-3mm]\hrule
\label{fig:taulifecont}
\end{figure}

Sterile neutrinos {solely} produced {through mixing with $\nu_\tau$ will always decrease } the $\tau$ width and thus {prolong} the predicted lifetime, {possibly} bringing it into tension with the experimental value. 
{This is so, because the decay rate to $\nu_\tau$ is decreased by a factor of $\cos^2 \theta_{N\tau}$ while the additional decay channel with $N$, proportional to  $\sin^2 \theta_{N\tau}$ involves phase-space suppression.}
Neglecting electroweak corrections in $\Gamma^N$ we {use \eq{eq:tautotalwidth} to determine}  the parameter space for $\theta$ and $m_N$ in agreement with the measured $\tau$ lifetime, see Fig.~\ref{fig:taulifecont}, which shows the $1\sigma$ region of the allowed mixing angle and HSN mass, up to $m_N \leq \SI{600}{\mega\electronvolt}$. {We further show the allowed region when the experimental value in \eq{eq:tauexp} is varied in the 3$\sigma$ range, while we do not triple the theoretical uncertainty.}
For $m_N > \SI{600}{\mega\electronvolt}$ we continue our $1\sigma$ and $3\sigma$ lines for illustration,  bearing in mind that the quality of the perturbative calculation diminishes significantly for $m_N \gtrsim \SI{600}{\mega\electronvolt}$, {while the perturbation series for the inclusive hadronic $\tau$ decay width converges well for $m_N \lesssim \SI{600}{\mega\electronvolt}$}.

Eq.~(\ref{eq:tautotalwidth}) holds for any value of $m_N$, in particular, it also holds for $m_N > m_\tau$. In this case $\Gamma_N = 0$ and Eq.~(\ref{eq:tautotalwidth}) becomes
\begin{equation}
    \sin^2\theta = 1 -\frac{\tau_\tau^{\mathrm{th.}}}{\tau_\tau^{\mathrm{exp.}}} = 0.00586 \pm 0.00816,\label{eq:noHSN}
\end{equation}
which leads to 
\begin{equation}
       |\sin\theta| \un{\leq 11.8} \cdot 10^{-2} \quad \mathrm{at} \,\, 1\sigma. \label{eq:mixinganglemnlarge}
\end{equation}

\subsection{Contributions on $\theta_{N\tau}$ for $m_N > m_\tau$ from $\tau$ branching fractions}
HSNs heavier than $\tau$ affect $\tau_\tau$ in Eq.~(\ref{eq:tautotalwidth}) indirectly through the cosine of the mixing angle. To study this scenario we look at the decays $\tau^- \rightarrow \pi^- \nu_\tau$, $\tau^- \rightarrow K^- \nu_\tau$, and $\tau \rightarrow \nu_\tau \ell \bar{\nu}_\ell$ \cite{Helo_2011,Kobach:2014hea}. {This will allow us to bypass the theoretical uncertainty of the total hadronic width feeding into $\tau_\tau$ in \eq{eq:tausm}.}   
The decay width{s} {of interest} read
\begin{align}
\Gamma(\tau \rightarrow P \nu_\tau) =& \cos^2\theta \Gamma_{\mathrm{SM}}(\tau \rightarrow P \nu_\tau) \\
=& \frac{G_F^2 m_\tau^3 \cos^2\theta}{16 \pi} |V_{uq}|^2 f_P^2 \bigg( 1-\frac{m_P^2}{m_\tau^2} \bigg)^2 \big(1+ \delta_{\mathrm{RC}}^{(P)}\big), \nonumber\\
\Gamma(\tau \rightarrow  \nu_\tau \ell \bar{\nu}_\ell)=&\cos^2\theta\Gamma_{\mathrm{SM}}(\tau \rightarrow  \nu_\tau \ell \bar{\nu}_\ell) = \frac{G_F^2 m_\tau^5 \cos^2\theta}{192 \pi^3} \\
\times &\bigg[  \big(1- 8 x_\ell^2 + 8x_\ell^6 -x_\ell^8 -12 x_\ell^4 \ln(x_\ell^2)\big) + \frac{\alpha(m_\tau)}{\pi} H_1 + \frac{\alpha^2(m_\tau)}{\pi^2} H_2 \bigg] ,\nonumber
\end{align}
where $x_\ell = m_\ell/m_\tau$, $P$ is either $K^-$ or $\pi^-$, $f_P$ is the meson decay constant, and $V_{uq}$ is the CKM matrix element. We have added the $\cos{\theta}$ factor to account for the effect of HSN mixing. Here the $\delta_{\mathrm{RC}}^{(P)}$ are the radiative QED corrections governed by $\alpha(m_\tau)$. {We use the values $\delta_{\mathrm{RC}}^{(\pi)} = 0.0194 \pm 0.0061$ and $\delta_{\mathrm{RC}}^{(K)} = 0.0204 \pm 0.0062$ presented in Ref.~\cite{Cirigliano:2021yto} which are based on Refs.~\cite{Rosner:2015wva,Arroyo-Urena:2021nil,Arroyo-Urena:2021dfe,Cirigliano:2007xi}}. The $H_i$ are the radiative QED corrections for leptonic decay, reading \cite{Kinoshita:1958ru,vanRitbergen:1999fi,Steinhauser:1999bx,Nir:1989rm,Pak:2008qt}
\begin{align}
H_1 &= \bigg( \frac{25}{8} - \frac{\pi^2}{2} \bigg) - (34+ 24 \ln x_\ell) x_\ell^2 +16 \pi^2 x_\ell^3 + \mathcal{O}(x_\ell^4) \\
H_2 &= \frac{156815}{5184} - \frac{518}{81} \pi^2 - \frac{895}{36} \zeta_3 +\frac{67}{720} \pi^4 + \frac{53}{6} \pi^2 \ln 2 \\
& \hspace{1cm} - (0.042 \pm 0.002) \tk{- \frac{5}{4} \pi^2 x_\ell }+ \mathcal{O}(x_\ell^2) \nonumber
\end{align}
We use the measured $\tau$ lifetime $\tau_\tau = \SI{290.29 \pm 0.53}{\femto\second}$ \cite{HeavyFlavorAveragingGroupHFLAV:2024ctg}, the branching ratios, decay constants, and parameters listed in Tab. \ref{tab:constants}. We have determined the pion decay constant from the ratio $f_{K^+}/f_{\pi^+}$ and $f_{K^+}$.

\begin{figure}[t!]
\centering
\includegraphics[width=0.75\textwidth]{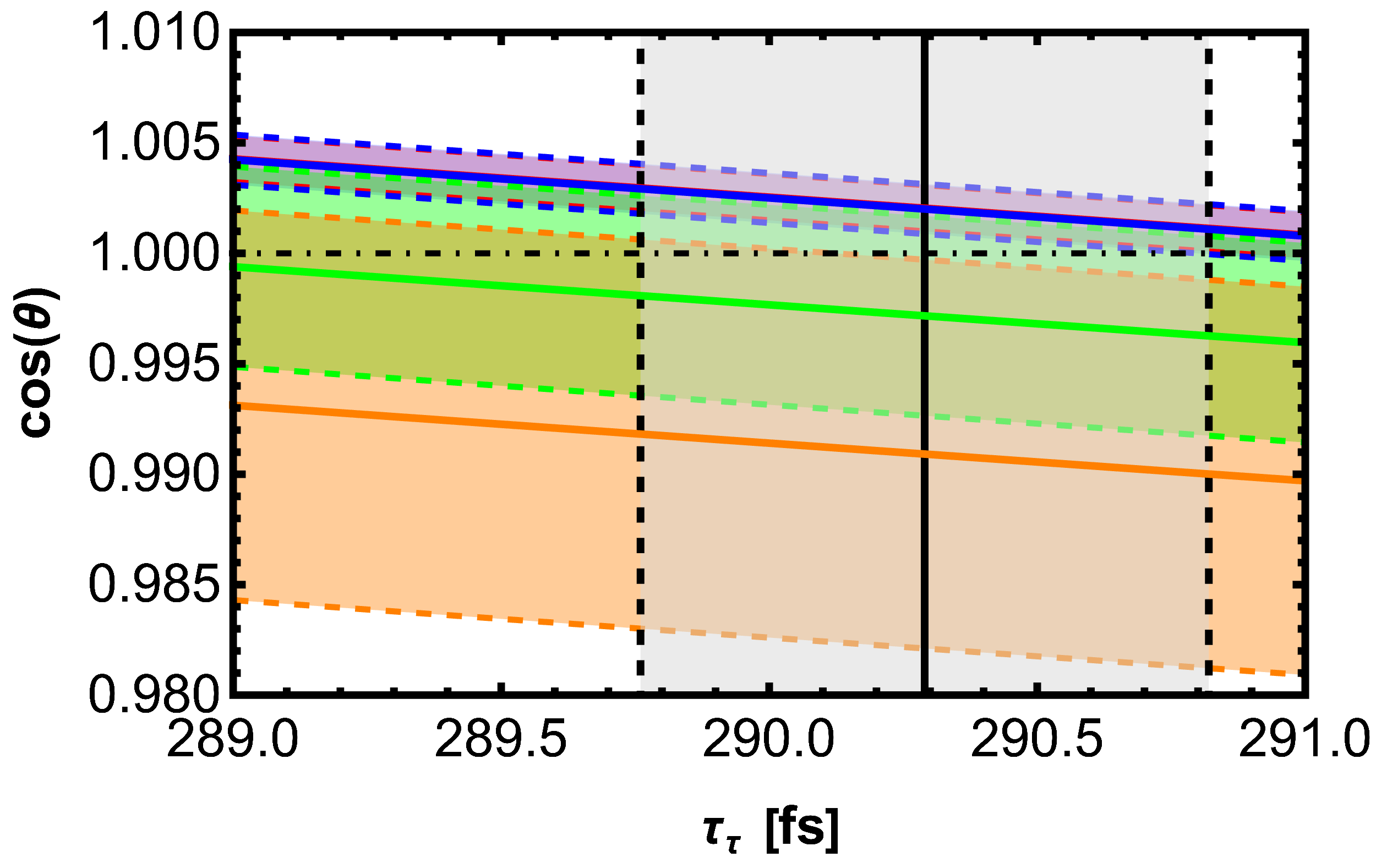}
\caption{Cosine of the mixing angle vs.\ the $\tau$ lifetime. The grey band is the $1\sigma$ band of the world average of the $\tau$ lifetime. The {solid} lines are the central values and the dashed lines {delimit} the $1\sigma$ band. The bands {from} $\tau \rightarrow \nu_\tau e \bar{\nu}_e$ (blue) and $\tau \rightarrow \nu_\tau \mu \bar{\nu}_\mu$ (red) overlap. {the constraints from} both $\tau \rightarrow \pi \nu_\tau$ (green) and $\tau \rightarrow K \nu_\tau$ (orange) fully {comply} with the measured $\tau$ lifetime. The area under the dot-dashed line at $\cos\theta = 1$ delimits the physically allowed region.}
~\\[-3mm]\hrule
\label{fig:costau}
\end{figure}

\begin{table}[tb]
\caption{Table of branching ratios, decay constants, and CKM elements used.}
\label{tab:constants}
\centering
\begin{tabular}{l|c|c}
 &  &  \\
\hline \hline
$\mathrm{Br}^{\mathrm{exp.}}(\tau \rightarrow \nu_\tau e \bar{\nu}_e)$ & $0.1785 \pm 0.0004$ & HFLAV \cite{HeavyFlavorAveragingGroupHFLAV:2024ctg} \\
$\mathrm{Br}^{\mathrm{exp.}}(\tau \rightarrow \nu_\tau \mu \bar{\nu}_\mu)$ & $0.17366 \pm 0.00036$ & HFLAV \cite{HeavyFlavorAveragingGroupHFLAV:2024ctg}\\
$\mathrm{Br}^{\mathrm{exp.}}(\tau \rightarrow \pi \nu_\tau )$ & $0.1082 \pm 0.0005$ & HFLAV \cite{HeavyFlavorAveragingGroupHFLAV:2024ctg}\\
$\mathrm{Br}^{\mathrm{exp.}}(\tau \rightarrow K \nu_\tau   )$ & $(0.697 \pm 0.010) \cdot 10^{-2}$ & HFLAV \cite{HeavyFlavorAveragingGroupHFLAV:2024ctg}\\
\hline
$f_{K^+}/f_{\pi^+}$  & $1.1934 \pm 0.0019$ & FLAG \cite{FlavourLatticeAveragingGroupFLAG:2024oxs,Bazavov:2017lyh,Miller:2020xhy,Dowdall:2013rya,Carrasco:2014poa,ExtendedTwistedMass:2021qui}\\
$f_{K^+}$    & $\SI{155.7 \pm 0.3}{\mega\electronvolt}$ & FLAG \cite{Dowdall:2013rya,Carrasco:2014poa,Bazavov:2014wgs,ExtendedTwistedMass:2021qui,FlavourLatticeAveragingGroupFLAG:2024oxs}\\
$f_{\pi^+} $ & $\SI{130.5 \pm 0.3}{\mega\electronvolt}$ &  \\
$|V_{ud}|$ &   $0.97367 \pm 0.00032$   & PDG \cite{ParticleDataGroup:2024cfk}\\
$|V_{us}|$ &   $0.22431 \pm 0.00085$   & PDG \cite{ParticleDataGroup:2024cfk}\\
$1/\alpha(m_\tau)$ & $133.50 \pm 0.02 $ & \cite{Erler:1998sy,Erler:2002mv}
\end{tabular}
~\\[1.5mm]\hrule
\end{table}
This leads to the following mixing angles allowed by the experimental bounds for the leptonic decays
\begin{align}
\bigg(\frac{\mathrm{Br}^{\mathrm{exp.}}(\tau \rightarrow \nu_\tau e \bar{\nu}_e)}{\tau_\tau^{\mathrm{exp.}}} \cdot \frac{1}{\Gamma_{\mathrm{SM}}(\tau \rightarrow \nu_\tau e \bar{\nu}_e))}\bigg)^{1/2} = \cos\theta =& 1.00201 \pm 0.00145 \\
\bigg(\frac{\mathrm{Br}^{\mathrm{exp.}}(\tau \rightarrow \nu_\tau \mu \bar{\nu}_\mu)}{\tau_\tau^{\mathrm{exp.}}} \cdot \frac{1}{\Gamma_{\mathrm{SM}}(\tau \rightarrow \nu_\tau \mu \bar{\nu}_\mu))}\bigg)^{1/2} = \cos\theta =& 1.00205 \pm 0.00139 
\end{align}
and for the semi-hadronic decays
\begin{align}
\bigg(\frac{\mathrm{Br}^{\mathrm{exp.}}(\tau \rightarrow \pi \nu_\tau)}{\tau_\tau^{\mathrm{exp.}}} \cdot \frac{1}{\Gamma_{\mathrm{SM}}(\tau \rightarrow \pi \nu_\tau )}\bigg)^{1/2} = \cos\theta =& 0.99717 \pm 0.00462 \\
\bigg(\frac{\mathrm{Br}^{\mathrm{exp.}}(\tau \rightarrow K \nu_\tau)}{\tau_\tau^{\mathrm{exp.}}} \cdot \frac{1}{\Gamma_{\mathrm{SM}}(\tau \rightarrow K \nu_\tau))}\bigg)^{1/2} = \cos\theta =& 0.99091 \pm 0.00884
\end{align}
here we roughly estimated the errors by assuming they are all uncorrelated and distributed Gaussian. The leptonic decays all prefer unphysical values with $\cos\theta > 1$ \tk{at the $1\sigma$ level}, the semi-hadronic decays, however, permit non-zero mixing angles \tk{at the $1\sigma$} level. For the decay $\tau \rightarrow \pi \nu_\tau$ we find the mixing angle \tk{at $1\sigma$}
\begin{equation}
\tau \rightarrow \pi \nu_\tau: \qquad |\sin\theta| \un{\leq 12.2} \cdot 10^{-2} \quad \mathrm{at }\,\, 1\sigma \label{eq:taupi}
\end{equation}
and for $\tau \rightarrow K \nu_\tau$ the mixing angle 
\begin{equation}
\tau \rightarrow K \nu_\tau: \qquad |\sin\theta| = \un{\left(13.5_{-11.3}^{+5.4} \right)}\cdot 10^{-2} \quad \mathrm{at }\,\, 1\sigma.\label{eq:taukaon}
\end{equation}
In Fig.~\ref{fig:costau} we show the $1 \sigma$ band of the cosines of the mixing angle as a function of the $\tau$ lifetime. Combining the semi-hadronic and leptonic decays yields $\cos\theta = 1.00168 \pm 0.00098$, while combining only the two semi-hadronic decays leads to $\cos\theta = 0.99582 \pm 0.00410$ 
corresponding to
\begin{equation}
\tau \rightarrow P \nu_\tau, \quad P = \pi,K \,\, \mathrm{combined}:  \qquad |\sin\theta| =\un{\left(9.1_{-7.8}^{+3.7} \right)} \cdot 10^{-2} \quad \mathrm{at }\,\, 1\sigma.\label{eq:taupikaoncombined}
\end{equation}
Taking the weighted average of the leptonic decays leads to $\cos\theta = 1.00203 \pm 0.00100$. These decays 
permit $\cos\theta \geq 0.99901$ at the $3\sigma$ level  leading to:
\begin{equation}
  \tau \to \nu_\tau \ell \bar{\nu}_\ell \quad \ell=e,\mu \,\, \mathrm{combined}: \qquad 
   |\sin\theta| \un{\leq 4.4} \cdot 10^{-2} \quad \mathrm{at }\,\, 3\sigma.
  \label{eq:lept3sig}
\end{equation}

\subsection{Detecting HSN for $m_N < m_\tau$ from spectra}
\begin{figure}[t!]
\centering
\includegraphics[width=0.65\textwidth]{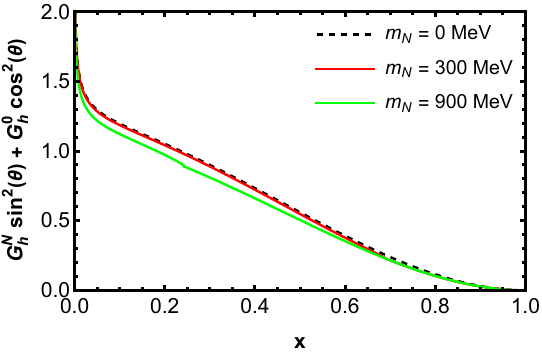}
\caption{{The plot shows} $G_h^N \sin^2\theta + G_h^0 \cos^2\theta = G(s/m_\tau^2,x_N)  \sin^2\theta +  G(s/m_\tau^2,0) \cos^2\theta $, the mixing angle  is {chosen to give} $\sin\theta = 0.32$.}
~\\[-3mm]\hrule
\label{fig:hadspec}
\end{figure}

HSNs light enough to be produced in $\tau$ decays are difficult to detect in branching ratios, because
\begin{equation}
\Gamma(\tau \rightarrow \nu_\tau Y ) + \Gamma(\tau \rightarrow N Y ) = \Gamma_{\mathrm{SM}}(\tau \rightarrow \nu_\tau Y ) + \mathcal{O}\bigg( \frac{m_N^2}{m_\tau^2} \cdot  \sin^2\theta\bigg)
\end{equation}
for $m_N < m_\tau$ and where $Y$ is any final state. To this end one better employs precise measurements of the charged lepton energy spectrum 
\begin{align}
\frac{d \Gamma(\tau \rightarrow \nu \ell \bar{\nu}_\ell)}{d E_\ell} = \frac{G_F^2 m_\tau^4 V_{\nu}^2}{2 \pi^3}   F(x_E,x_\ell,x_\nu), 
\end{align}
here $\nu = \nu_\tau , N$ is either a SM or a sterile neutrino, $x_E = E_\ell/m_\tau$, $x_i = m_i/m_\tau$, $V_\nu = \cos\theta$ for SM and $V_\nu = \sin\theta$ for sterile neutrinos, and 
\begin{align}
F(x_E,x_\ell,x_\nu) &= \frac{\sqrt{x_E^2-x_\ell^2} \big( 1- 2 x_E +x_\ell^2 - x_\nu^2 \big)^2}{6 \big( 1 - 2 x_E +x_\ell^2 \big)^3} \\
& \hspace{-2cm} \times \bigg[ 8 x_E^3 - 2 x_E^2 (5 + 5 x_\ell^2 + x_\nu^2) \nonumber\\
&+ x_E (3 + 10 x_\ell^2 + 3 x_\ell^4 + 3 x_\nu^2 (1 + x_\ell^2)) - 2x_\ell^2 (1 + x_\ell^2 +2 x_\nu^2) \bigg]. \nonumber
\end{align}
Since the experiment cannot distinguish between sterile and SM neutrino the measured lepton energy is
\begin{align}
\frac{d\Gamma^{\mathrm{exp.}}_{\mathrm{lep}}}{d E_\ell} =& \frac{d \Gamma(\tau \rightarrow \nu_\tau \ell \bar{\nu}_\ell)}{d E_\ell} + \frac{d\Gamma(\tau \rightarrow N \ell \bar{\nu}_\ell)}{d E_\ell} \\
 =& \frac{G_F^2 m_\tau^4 V_{\nu}^2}{2 \pi^3} \bigg[ F(x_E,x_\ell,x_N)  \sin^2\theta +  F(x_E,x_\ell,0) \cos^2\theta \bigg]. \nonumber
\end{align}
\begin{figure}[t!]
\centering
\includegraphics[width=0.65\textwidth]{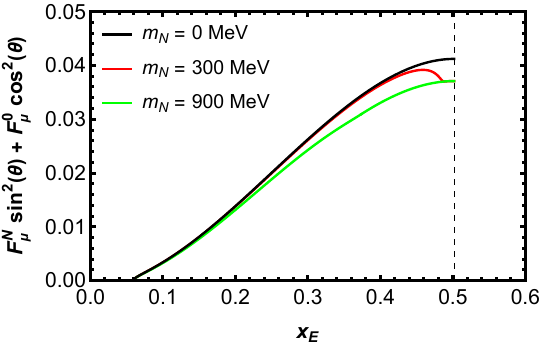}
\caption{Here we {show} $F_\mu^N \sin^2\theta + F_\mu^0 \cos^2\theta = F(E_\mu/m_\tau,x_\mu,x_N)  \sin^2\theta +  F(E_\mu/m_\tau,x_\mu,0) \cos^2\theta $ {for} $\ell = \mu$, the black line corresponds to the SM case $m_N=0$, the mixing angle assumed here corresponds to $\sin\theta = 0.32$.}
~\\[-3mm]\hrule
\label{fig:leptspec}
\end{figure}
The hadronic spectrum is derived analogously,
\begin{align}
\frac{d\Gamma(\tau \rightarrow \nu X)}{ds} = N_c \frac{G_F^2 m_\tau^3 V_\nu^2 \cdot 2 (|V_{ud}|^2+|V_{us}|^2)}{192 \pi^3} S(m_\tau,m_Z)  G\big( x , x_\nu\big), 
\end{align}
where $S(m_\tau,m_Z) = 1.01907$ \cite{PhysRevLett.61.1815,Erler:2002mv} accounts for electroweak corrections, $x = s/m_\tau^2$, $x_\nu = m_\nu/m_\tau$, and
\begin{align}
G\big( x , x_\nu\big) &= \bigg[ \bigg(1+x_\nu^2 - x \bigg) \bigg(1+2 x + x_\nu^2\bigg) - 4 x_\nu^2 \bigg] \\ & \hspace{0.5cm} \times  \sqrt{\lambda\big( 1, x , x_\nu^2 \big)} \cdot 12 \pi \mathrm{Im} \, \Pi^{1+0}_V(m_\tau^2 x). \nonumber
\end{align}
As for the leptonic decay the measured quantity is the sum of SM and sterile neutrino contributions
\begin{align}
\frac{d\Gamma^{\mathrm{exp.}}_{\mathrm{had}}}{d {s}} =& \frac{d \Gamma(\tau \rightarrow \nu_\tau X)}{d s} + \frac{d\Gamma(\tau \rightarrow N X)}{d s} \\
=& N_c \frac{G_F^2 m_\tau^3 V_\nu^2 \cdot 2 (|V_{ud}|^2+|V_{us}|^2)}{192 \pi^3} S(m_\tau,m_Z)  \nonumber\\
&  \times \bigg[ G(x,x_N)  \sin^2\theta +  G(x,0) \cos^2\theta\bigg]. \nonumber
\end{align}
In Figs.\ \ref{fig:leptspec} and \ref{fig:hadspec} we show the spectrum for the leptonic decay with $\ell = \mu$ in the final state and the  hadronic spectrum, {respectively}, in both cases with $\sin\theta = 0.32$. A HSN with a mass of $m_N = \SI{300}{\mega\electronvolt}$ leads to a significantly modified spectrum.
{The perturbative calculation of the hadronic spectrum can be compared to data for large $s\gtrsim 1\,$GeV. 
In addition, one can compare integrated spectra from $s=0$ to a 
chosen $s_{\rm max} \gtrsim 1\,$GeV, exploiting quark-hadron duality as described in Appendix~\ref{app:contourintegration}.}

\subsection{Comparisons}
It is interesting to compare the allowed values of the mixing angle with other experimental exclusions. \kt{Both ATLAS \cite{ATLAS:2024fcs,ATLAS:2025uah} and CMS \cite{2024CMS,CMS:2024ake,CMS:2024xdq} have} published upper and lower limits on the mixing angle. \kt{ATLAS performed an analysis for the HSN exclusively mixing with the $\tau$ neutrino and is less competitive than CMS. The CMS experiment specifically looked at $\nu_\tau$-$N$ mixing and is sensitive to HSN masses $\SI{1}{\giga\electronvolt} \leq m_N \lesssim \SI{2.2}{\giga\electronvolt}$ and $m_N \geq \SI{3}{\giga\electronvolt}$.}

Below $m_N < \SI{3}{\giga\electronvolt}$ there are exclusion limits for $N$-$\nu_\tau$ mixing from DELPHI \cite{DELPHI:1996qcc}, ArgoNeuT \cite{ArgoNeuT:2021clc}, BABAR \cite{BaBar:2022cqj}, and Belle \cite{Belle:2024wyk}. In the mass range $\SI{0.4}{\giga\electronvolt} \leq m_N \leq \SI{1.3}{\giga\electronvolt}$ ArgoNeuT and BABAR provide the strongest bounds. \kt{Beyond $m_N \geq \SI{1.3}{\giga\electronvolt}$ DELPHI and CMS provide the strongest bounds, with the DELPHI bound being strongest for $m_N \gtrsim \SI{1.9}{\giga\electronvolt}$.} {Furthermore,} there {are} recasts of {data} from the CHARM \cite{Boiarska:2021yho} and WA66 experiments \cite{Barouki:2022bkt}. {While} originally these experiments were not understood to be sensitive to $N$-$\nu_\tau$ mixing for HSN with masses $m_N \approx \SI{1}{\giga\electronvolt}$, 
the recasts provide {the} tightest constraints up to $m_N = \SI{1.6}{\giga\electronvolt}$. 
\kt{Only the DELPHI experiment is sensitive {to} masses in the range $ \SI{2.2}{\giga\electronvolt} \lesssim m_N < \SI{3}{\giga\electronvolt} $.}

For $\mathbf{m_N < m_\tau}$ we find 
the allowed range ${|}V_{N\tau}{|} = {|}\sin\theta{|} \lesssim 0.2 $ from the $\tau$ lifetime for $m_N \approx \SI{600}{\mega\electronvolt}$.  
In this mass regime our bound from $\tau$ decay is not competitive, the strongest bound {stems from} the BABAR analysis, 
{which finds {${|}\sin\theta{|} \lesssim 2.49 \cdot 10^{-2}$} \cite{BaBar:2022cqj},}
and the CHARM and WA66 recasts, ${|}\sin\theta{|} \lesssim 1.73 \cdot 10^{-3}$ \cite{Boiarska:2021yho,Barouki:2022bkt}, {which, however, are more model-dependent as explained below.}

For $\mathbf{m_N > m_\tau}$ we find the mixing angle in Eq.~(\ref{eq:mixinganglemnlarge}) from the $\tau$ lifetime. The semi-hadronic $\tau$ decays exclude mixing angles for HSN masses $m_N > m_\tau - m_P$ where $P=\pi,K$ and lead to Eq.~(\ref{eq:taupi}) for $\tau \rightarrow \pi \nu_\tau$, and Eq.~(\ref{eq:taukaon}) for $\tau \rightarrow K \nu_\tau$. In combination, the semi-hadronic decays lead to Eq.~(\ref{eq:taupikaoncombined}), {valid} 
for $m_N > m_\tau - m_P$. Kobach and Dobbs (2015) \cite{Kobach:2014hea} {have} updated the bound of Ref.~\cite{Helo_2011}. They used leptonic $\tau$ decays to constrain the mixing angle for $m_N>m_\tau$ and {deriv}ed the bound ${|}\sin\theta^{\mathrm{KD}} {|}\lesssim 7.07 \cdot 10^{-2}$ at the $\sim 2\sigma$ level. This is weaker than our bound obtained from leptonic $\tau$ decays given in Eq.~(\ref{eq:lept3sig}) at the $3\sigma$ level. {While} our bounds obtained 
from semi-hadronic $\tau$ decays (Eq.~(\ref{eq:taupikaoncombined})) and the $\tau$ lifetime (Eq.~(\ref{eq:mixinganglemnlarge})) 
{are weaker than the constraint from leptonic decays, they are less model-dependent as explained below. \tk{It is interesting to note that $\theta=0$ is ruled out by {1}$\sigma$ in Eq.~(\ref{eq:taupikaoncombined})}. Bearing in mind that our error estimates do not consider correlations, these numbers may well indicate that 
better data may lead to  \un{evidence} for $\theta\neq 0$ in the foreseeable future.}   

\kt{The experiment most sensitive in the mass range $\SI{1.9}{\giga\electronvolt} \lesssim m_N \leq \SI{3}{\giga\electronvolt}$ is the DELPHI experiment,} {determin}ing $\sin^2\theta^{\mathrm{DELPHI}} \lesssim 6 \cdot 10^{-5} \Rightarrow {|}\sin\theta^{\mathrm{DELPHI}} {|}\lesssim 7.75 \cdot 10^{-3}$ for $m_N \approx \SI{2}{\giga\electronvolt}$; the bound 
{becomes weaker} for smaller HSN masses \cite{DELPHI:1996qcc}. The DELPHI experiment {studied} $Z$ decays and {was therefore} sensitive to HSN and SM neutrinos of all flavors. However, they {searched for} the HSN through charged-current and neutral-current 
decays of HSN {and derived the bound} ${|}\sin\theta^{\mathrm{DELPHI}}{|} \lesssim 1.55 \cdot 10^{-2}$ {from the non-observation 
of HSN decays into $\tau$ leptons. The DELPHI bound} is  tighter than ours, {but is also more model-dependent:
While both our and the DELPHI study assume that $N$ only interacts with SM particles through $N$-$\nu_\tau$ mixing, 
we do not assume anything on other interactions of $N$. A heavy $N$ could be a mediator to a dark sector and possibly decay into  lighter sterile neutrinos and/or dark bosons. In such scenarios the branching ratios of the decay modes studied by DELPHI will
decrease and the bounds on ${|}\theta{|}$ will become weaker and the same remark applies to the recasts of CHARM and WA66 data. For the same reason we also put less emphasis on the bounds on 
${|}\theta{|}$ in \eq{eq:lept3sig} found from leptonic decays. Unlike decays into final states with a single neutrino, 
leptonic $\tau$ decays can be contaminated by decays into final states containing one or more massive dark bosons. Since the central values of the measured leptonic decay rates are higher than the SM predictions, we study this scenario in the next subsection.} 

Recently it was pointed out that indeed the LHC might already be capable of probing mixing angles for mixing of sterile neutrinos with $\tau$ neutrinos far more efficiently \cite{Tireli:2025pno}. A reinterpretation of the existing LHC data would be highly interesting.

\subsection{BSM scenarios with other light invisible particles}
Leptonic $\tau$ decays $\tau \rightarrow \nu_\tau \ell \bar\nu_\ell$ are interesting as there are two neutrinos produced in the final state: What is studied is $\tau \rightarrow \ell + E_\mathrm{miss}$ and the squared missing mass is much larger than zero, so that there can be contributions from $\Gamma(\tau \rightarrow \ell X_{\mathrm{dark}})$ or $\Gamma(\tau \rightarrow \ell X_{\mathrm{dark}} X_{\mathrm{dark}})$ with massive dark particles $X_{\mathrm{dark}}$. {A natural candidate for an invisible boson emitted in a $\tau$ decay is a majoron coupling to 
$\nu_\tau$ \cite{Chikashige:1980ui,Gelmini:1980re,Barenboim:2020dmg}.}

We write $\Gamma_{\mathrm{NP}}$ for the dark contributions i.e. $\Gamma(\tau \rightarrow \ell + E_\mathrm{miss}) = \Gamma(\tau \rightarrow \nu_\tau \ell \bar\nu_\ell) + \Gamma_\mathrm{NP}$. This constrains the new physics contribution to 
\begin{align}
\Gamma_\mathrm{NP} &= \\
 &\frac{\mathrm{Br}^{\mathrm{exp.}}(\tau \rightarrow \nu_\tau e \bar\nu_e)}{\mathrm{Br}^{\mathrm{exp.}}(\tau \rightarrow \pi \nu_\tau )} \Gamma_{\mathrm{SM}}(\tau \rightarrow \pi \nu_\tau ) - \Gamma_{\mathrm{SM}}(\tau \rightarrow \nu_\tau e \bar\nu_e) =  \SI{3.92 \pm 3.81}{\micro\electronvolt} \nonumber \\
\Gamma_\mathrm{NP} &= \\
 &\frac{\mathrm{Br}^{\mathrm{exp.}}(\tau \rightarrow \nu_\tau \mu \bar\nu_\mu)}{\mathrm{Br}^{\mathrm{exp.}}(\tau \rightarrow \pi \nu_\tau )} \Gamma_{\mathrm{SM}}(\tau \rightarrow \pi \nu_\tau ) - \Gamma_{\mathrm{SM}}(\tau \rightarrow \nu_\tau \mu \bar\nu_\mu) =  \SI{3.85 \pm 3.69}{\micro\electronvolt}, \nonumber
\end{align}
where we set $\cos\theta = 1$ {because of the tight constraints on $|\sin\theta|$ found above}.

\section{Conclusions}\label{sec:conc}
In this paper, we have calculated the inclusive semi-hadronic charged-current decay width of HSNs up to $\mathcal{O}(\alpha_s^4)$ in the strong coupling constant by using the known results of the electroweak gauge boson correlator which has been calculated up to the five-loop level. {Decay rates into massless leptons can be trivially 
obtained from SM $\tau$ decay rates, while $N\to \tau +\mbox{hadrons}$ decays involve new phase-space 
integrals. We derive analytic expressions in terms polylogarithms (up to $\mathcal{O}(\alpha_s^3)$) and a series representation (valid for all calculated orders) for these rates.  } 
We then calculated the perturbative series and analyzed its stability  {in order to determine the sterile neutrino mass range for which the perturbative calculation is valid and robust theory predictions are possible. Reliable predictions for the semi-hadronic width are found for $m_N \geq \SI{1.5}{\giga\electronvolt}$ for $\ell = e, \mu$ in the final state. For the final state with $\ell=\tau$ we find stability for $m_N \geq \SI{3}{\giga\electronvolt}$. Our 
result can also be used for $\tau \to N +\mbox{hadrons}$ in the mass range $m_N\lesssim 600\,\mbox{MeV}$.}
In $\tau$ decays the perturbative series is only stable {after including} the $\mathcal{O}(\alpha_s^4)$ corrections.  

{We have further} studied the impact of {an $N$-$\nu_\tau$ mixing angle $\theta$} on the $\tau$ lifetime. If  $N$ is light enough to be produced in $\tau$ decay, the SM {decay amplitude is modified by a factor of $\cos\theta$ and there is an additional $\tau \rightarrow N + W^*$ mode with amplitude proportional to $\sin\theta$.} We find the parameter space of $m_N$ and $\theta$ constrained by the $\tau$ lifetime for HSN masses $m_N \lesssim \SI{600}{\mega\electronvolt}$. 
{Furthermore, we present expressions for the lepton energy spectrum in $\tau\to \ell \nu_\ell N$ and the 
hadron invariant mass spectrum in $\tau \to N+\mbox{hadrons}$, which may be used to detect HSN effects in the future.} 

HSNs heavier than the $\tau$ lepton  may still {leave imprints on observables}  through the cosine of the mixing angle. \tk{The lifetime analysis leads to $|\sin\theta| \leq 11.8 \cdot 10^{-2}$ at $1\sigma$ in this case.}
For the four decays $\tau \rightarrow \nu_\tau \ell \bar{\nu}_\ell$ with $\ell = e,\mu$ and $\tau \rightarrow P \nu_\tau$ with $P=\pi^-,K^-$ we have compared the measured widths with the theoretical predictions. Only the data on the semi-{hadronic} decays permit a non-zero mixing angle and {we derive the constraints} \tk{$|\sin\theta| \leq  12.2 \cdot 10^{-2}$ at $1\sigma$ from $\tau \rightarrow \pi \nu_\tau$ and $|\sin\theta| =\left(13.5_{-11.3}^{+5.4} \right)\cdot 10^{-2}$ at $1\sigma$ from $\tau \rightarrow K \nu_\tau$, {respectively}. Combining both semi-hadronic decays leads to $|\sin\theta| = \left(9.1_{-7.8}^{+3.7} \right) \cdot 10^{-2} $ at $1\sigma$.} {Our results are complementary to those from CMS, which constrain $\theta$ for HSN masses above 3$\,$GeV, {and less model dependent \kt{than} those of DELPHI below $\SI{3}{\giga\electronvolt}$}.}
{In another application of our calculation we have derived relative branching fractions between certain exclusive decays and the inclusive semi-hadronic decay and determined the fraction of $N\to \pi^+ \ell^-$ decays of all 
$N \to \ell^- +\mbox{hadrons}$ decays.} 

{Finally we have derived a bound on the $\tau$ decay width into final states containing one or more new invisible particles, for example majorons. These decays contaminate $\tau \to \ell \bar\nu_\ell \nu_\tau$, whose decay rate is indeed a bit larger than the SM prediction. This finding calls for more precise experimental studies of the leptonic 
$\tau$ decays.}

\section*{Acknowledgements}
This research was supported by the Deutsche Forschungsgemeinschaft (DFG, German Research Foundation) under grant 396021762 - TRR 257 for the Collaborative Research Center \emph{Particle Physics Phenomenology after the Higgs Discovery (P3H)}.

\newpage
\appendix
\section{QCD $\beta$-Function}
\label{app:beta}
We use the $\beta$-function with the following normalization
\begin{equation}
\frac{d a_\mu}{d \ln \mu^2} = \beta(a_\mu) = - a_\mu^2 \bigg[ \beta_0 + \beta_1 a_\mu + \beta_2 a_\mu^2 + ... \bigg], \label{eq:runningalphas}
\end{equation}
where $a_\mu = a(\mu)= \alpha_s(\mu)/\pi$ and with the coefficients $\beta_i$ \cite{Czakon_2005,Chetyrkin:2016uhw}:
\begin{align}
\beta_0 &= \frac{1}{4}   \lf 11 - \frac{2}{3} n_f \rt, \\
\beta_1 &= \frac{1}{16}  \lf 102 - \frac{38}{3} n_f  \rt, \\
\beta_2 &= \frac{1}{64}  \lf \frac{2857}{2} - \frac{5033}{18} n_f + \frac{325}{54} n_f^2  \rt.
\end{align}
Eq.~(\ref{eq:runningalphas}) leads to 
\begin{align}
a(\mu) =& a(\mu_0) - a(\mu_0)^2 \beta_0 \ell_{\mu\mu_0} + a(\mu_0)^3 \bigg[ -\beta_1 \ell_{\mu\mu_0} + \beta_0^2 \ell_{\mu\mu_0}^2 \bigg] \nonumber\\
&\hspace{1cm}+ a(\mu_0)^4 \bigg[ -\beta_2 \ell_{\mu\mu_0} + \frac{5}{2} \beta_0 \beta_1 \ell_{\mu\mu_0}^2 - \beta_0^3 \ell_{\mu\mu_0}^3  \bigg]  + \mathcal{O}(a(\mu_0)^5),
\end{align}
where $\ell_{\mu\mu_0}=\ln\bigg( \frac{\mu^2}{\mu_0^2} \bigg)$.

\section{Contour Integration}
\label{app:contourintegration}
\begin{figure}[t]
\centering
\includegraphics[width=0.5\textwidth]{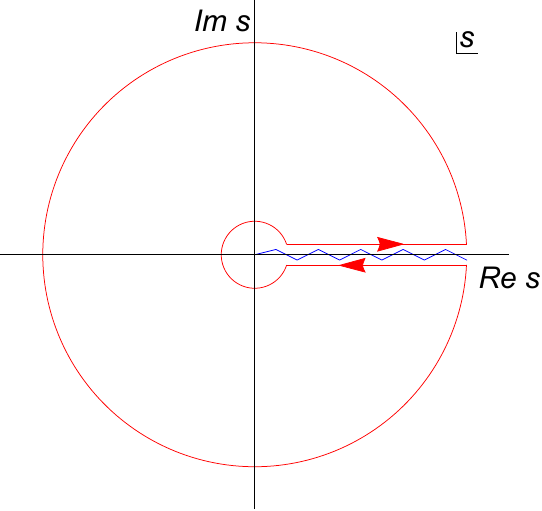}
\caption{Contour integration permitting to replace the integral over the resonance region 
by the integral over the large circle, which is far away from resonances. Thus, while the differential spectrum is not perturbatively calculable in that region, the integrated spectrum is found correctly in this way.}
\label{fig:contour}
\end{figure}

In the framework of perturbative calculations it is possible to calculate scattering or decay processes into quarks. However, in an experiment only hadrons are detectable and not quarks. This poses a problem as hadrons at different masses would appear as resonances in e.g. cross-sections. These resonances however are not accounted for by perturbative calculations using quarks. This conundrum was resolved by Poggio, Quinn and Weinberg \cite{Poggio:1975af}; they showed that a perturbative calculation smeared over a suitable energy range correctly averages over the resonances. 
This approach was then refined by showing that it is possible to transform the integration over the suitable energy range into an integration around a suitable complex contour \cite{PhysRevD.15.755,PhysRevD.16.703}.
This was used for $\tau$ decays \cite{BRAATEN1992581,PhysRevLett.60.1606}, the integral in 
Eq.~(\ref{eq:incldecaywidth}) can be rewritten in terms of a contour integration around a keyhole contour (see Fig. \ref{fig:contour}). Thus the decay width is proportional to the discontinuity over the branch cut along the positive real axis.
In this way the integral may be rewritten in the following way (setting the lepton mass to zero for brevity)
\begin{align*}
\Gamma(N \rightarrow X \ell ) \sim & 12 \pi \int\limits_0^{m_N^2} d s \, \bigg(1 - \frac{s}{m_N^2} \bigg)^2 \bigg[ \bigg( 1+2 \frac{s}{m_N^2} \bigg) \mathrm{Im} \Pi^{(1+0)}(s) - 2\frac{s}{m_N^2} \mathrm{Im} \Pi^{(0)}(s)  \bigg] \\
& = 6 \pi i \oint\limits_{s=m_N^2} d s \, \bigg(1 - \frac{s}{m_N^2} \bigg)^2 \bigg[ \bigg( 1+2 \frac{s}{m_N^2} \bigg) \Pi^{(1+0)}(s) - 2\frac{s}{m_N^2}  \Pi^{(0)}(s)  \bigg],
\end{align*}
the integral only depends on $s$ at the value of the mass of the decaying particle and so low energy effects are already accounted for in the decay width. Indeed in this way the inclusive decay width correctly averages over all possible resonances induced by the mesons.

\section{Easy-to-use Formulae for Phase-Space Integrals}
\label{app:intfit}
\begin{figure}[t]
\centering
\includegraphics[width=0.65\textwidth]{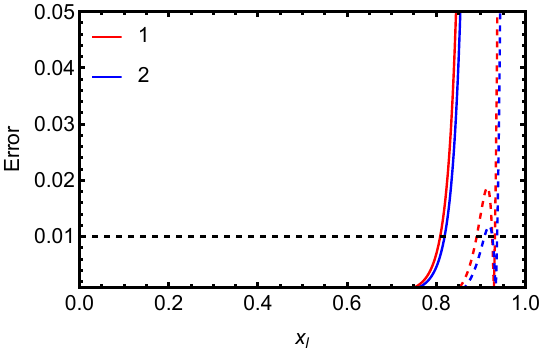}
\caption{The absolute error $|1- I_i^\mathrm{approx. ,\, T/P}/I_i| $ of the approximations
in \eqsto{eq:tay}{eq:pol}. The solid red and blue curves correspond to the Taylor expansion and the dashed blue and red curves are the polynomial fit.  The dashed black line indicates the $1\%$ level. For $x_\ell \lesssim 0.8$ both approximations have  uncertainties 
$|1- I_i^\mathrm{approx. ,\, T/P}/I_i| < 1\%$.}
\label{fig:error}
\end{figure}

For easier use we have expanded Eqs. (\ref{eq:I1}), (\ref{eq:I2}) as
\begin{align}
I_1^\mathrm{approx. ,\, T} &= \frac{2}{525}x_{\ell}^{14} + \frac{1}{60} x_{\ell}^{12}+\frac{2}{15} x_{\ell}^{10} -\frac{43}{24} x_{\ell}^8 +\frac{34}{3} x_{\ell}^6 +\left(\frac{15}{2}-2 \pi ^2\right) x_{\ell}^4 \label{eq:tay}\\
& +\frac{10 }{3}x_{\ell}^2 -\frac{19}{24} +\left(x_{\ell}^4-8x_{\ell}^2-6\right) x_{\ell}^4 \log (x_{\ell}) + \mathcal{O}(x_\ell^{16}) \nonumber\\
I_2^\mathrm{approx. ,\, T} &= -\frac{787 }{110250} x_{\ell}^{14}+\frac{1}{1800} x_{\ell}^{12}+\frac{217}{450} x_{\ell}^{10}+\frac{1}{6} \left(2 \pi ^2-5\right) x_{\ell}^8  \\
& +\frac{2}{3} \left(37-4 \pi ^2\right) x_{\ell}^6 -2  \left(-12 \zeta (3)+3+\pi ^2\right) x_{\ell}^4 -\frac{56 }{9} x_{\ell}^2 +\frac{265}{144} \nonumber\\
&-\frac{\left(8 x_{\ell}^8+35 x_{\ell}^6+280 x_{\ell}^4-2100 x_{\ell}^2+16800\right) x_{\ell}^6 \log (x_{\ell})}{1050} + \mathcal{O}(x_\ell^{16}). \nonumber
\end{align}
Additionally we have also fitted Eq. (\ref{eq:I1}), (\ref{eq:I2}) to \kt{polynomials (see Fig.~\ref{fig:error})}
\begin{align}
I_1^\mathrm{approx. ,\, P} =& 1.27683 x_{\ell}^9-1.11042 x_\ell^8-10.3142 x_{\ell}^7+21.4421 x_{\ell}^6-9.05503 x_{\ell}^5 \\
&-6.19752 x_{\ell}^4+1.5042 x_{\ell}^3+3.24256 x_{\ell}^2+0.00309017 x_{\ell}-0.7917 \nonumber \\
I_2^\mathrm{approx., \, P} =& -4.83878 x_{\ell}^9+29.1877 x_{\ell}^8-53.4361 x_{\ell}^7+34.6139 x_{\ell}^6-7.13307 x_{\ell}^5 \label{eq:pol}\\
&+6.63816 x_{\ell}^4-0.720101 x_{\ell}^3-6.14877 x_{\ell}^2-0.00323628 x_{\ell}+1.84032 
\nonumber
\end{align}

\section{Integrated Spectrum}
\label{app:integratedspectrum}
Here we present the analytical solutions of the integral
\begin{equation}
I_k(x_\ell^2,x_\mathrm{max}) = \int\limits_0^{x_\mathrm{max}} d x \, \bigg((1+x_\ell^2 -x)  (1+2x+x_\ell^2) - 4 x_\ell^2 \bigg) \sqrt{\lambda(1,x,x_\ell^2)} \ln^k( x), \label{eq:integratedspec}
\end{equation}
for $k<3$.
\begin{figure}[t]
\centering
\includegraphics[width=0.7
\textwidth]{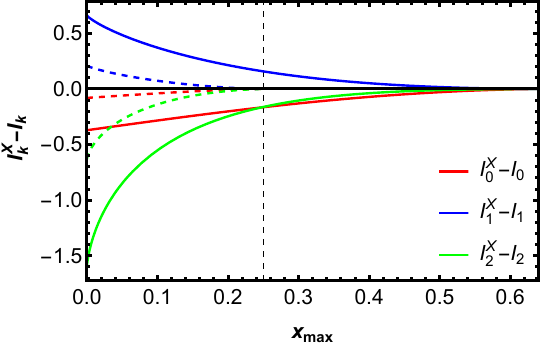}
\caption{Here $I_k^X-I_k$ is plotted for two different values of $x_\ell=0.2$ (closed line) and $x_\ell=0.5$ (dashed line). The vertically dashed line lies at the end of the spectrum of the dashed curves i.e. $(1-0.5)^2 = 0.25 $. At the end of the spectrum the $I_i^X-I_i = 0$}
\label{fig:spectrum}
\end{figure}
In Fig. \ref{fig:spectrum} the $I_k(x_\ell^2,x_\mathrm{max})-I_k(x_\ell^2,(1-x_\ell)^2)$ are shown.

The $I_k(x_\ell^2,x_\mathrm{max})-I_k(x_\ell^2,(1-x_\ell)^2)$, which we denote as $I_k^X-I_k$ and use $x_\mathrm{max} = \hat{x}$ as a shorthand notation, take the following form
\kt{
\begin{align}
I_0^{X}-I_0 =& \frac{1}{2} \bigg(-\bigg(\hat{x}^2+7\bigg) x_{\ell}^2-(\hat{x}+7) x_{\ell}^4 + x_{\ell}^6 \\
&\hspace{1cm} +(\hat{x}-1)^2 (\hat{x}+1)\bigg) \bigg(\hat{x}-(x_{\ell}+1)^2\bigg) X_r -24 x_{\ell}^4 \mathrm{ArcTanh}(X_r) \nonumber\\
I_1^{X}-I_1 =& 12 x_{\ell}^4 \bigg(\Delta \mathrm{Li}_2\bigg(-\frac{X_{-}}{2
   x_{\ell}}\bigg)-\Delta \mathrm{Li}_2\bigg(\frac{X_{-}}{2}\bigg)+\Delta \mathrm{Li}_2\bigg(\frac{1-X_r}{2}\bigg)\bigg) \\
&+\frac{1}{24} X_r \bigg(\hat{x}-(x_{\ell}+1)^2\bigg) \nonumber\\ 
& \times\bigg(-3 \hat{x}^3+12 \big(-(\hat{x}^2+7) x_{\ell}^2-(\hat{x}+7) x_{\ell}^4 +x_{\ell}^6 +(\hat{x}-1)^2 (\hat{x}+1)\big) \ln (\hat{x}) \nonumber\\
&\hspace{1cm} +5 (\hat{x}^2+\hat{x})+(\hat{x} (5 \hat{x}+4)+85) x_{\ell}^2 +5 (\hat{x}+17) x_{\ell}^4-19 x_{\ell}^6-19\bigg) \nonumber\\
&+\bigg(-8 x_{\ell}^2 -6 x_{\ell}^4 \bigg(4 \ln (\hat{x})+2 \ln\bigg(1-X_r^2\bigg) -4\ln (2)+2) \bigg) \nonumber\\
&\hspace{1cm} -8 x_{\ell}^6+x_{\ell}^8+1\bigg) \mathrm{ArcTanh}(X_r) \nonumber\\
&-\bigg(8 x_{\ell}^2 +24 x_{\ell}^4 \ln
   (x_{\ell})-8 x_{\ell}^6 + x_{\ell}^8-1\bigg) \mathrm{ArcTanh}\bigg(-X_r \frac{1+x_{\ell}}{1-x_{\ell}}\bigg) \nonumber
\end{align}
\begingroup
\allowdisplaybreaks
\begin{align}
I_2^{X}-I_2 =& X_r \bigg[\bigg(\big(\hat{x}^2+7\big) x_\ell^3-(\hat{x}-1) \big(\hat{x}^2+3\big) x_{\ell}^2+(\hat{x}+3) x_{\ell}^6+(\hat{x}+7) x_{\ell}^5 \\
&+(7-3 \hat{x}) x_{\ell}^4-(\hat{x}-1)^2 (\hat{x}+1) x_{\ell}+\frac{1}{2} (\hat{x}-1)^3 (\hat{x}+1)-\frac{x_{\ell}^8}{2}-x_{\ell}^7\bigg) \ln^2(\hat{x})\nonumber\\
&+\bigg(\frac{1}{12} \big(-3 \hat{x}^4+8 \hat{x}^3-24 \hat{x}+19\big)+\bigg(-2 \hat{x}-\frac{11}{2}\bigg) x_{\ell}^6-\frac{5}{6} (\hat{x}+17) x_{\ell}^5 \nonumber\\
&+\frac{1}{6} (38 \hat{x}-85) x_{\ell}^4+\frac{1}{6} (-\hat{x} (5\hat{x}+4)-85) x_{\ell}^3 \nonumber\\
&+\frac{1}{6} (\hat{x} (\hat{x} (4 \hat{x}-3)+38)-33) x_{\ell}^2 +\frac{1}{6} (\hat{x} (\hat{x} (3 \hat{x}-5)-5)+19) x_{\ell}\nonumber\\
&+\frac{19 x_{\ell}^8}{12}+\frac{19 x_{\ell}^7}{6}\bigg) \ln (\hat{x}) +\frac{1}{144} \bigg(9 \hat{x}^4-32\hat{x}^3+288 \hat{x}-265\bigg) \nonumber\\
&+\bigg(2 \hat{x}+\frac{113}{24}\bigg) x_{\ell}^6+\frac{23}{72} (\hat{x}+41) x_{\ell}^5 +\frac{1}{72} (943-446 \hat{x}) x_{\ell}^4\nonumber\\
&+\frac{1}{72} (\hat{x} (23 \hat{x}+28)+943) x_{\ell}^3 +\frac{1}{72} (\hat{x} ((9-16\hat{x}) \hat{x}-446)+339) x_{\ell}^2\nonumber\\
&+\frac{1}{72} (\hat{x} ((23-9 \hat{x}) \hat{x}+23)-265) x_{\ell} -\frac{265 x_{\ell}^8}{144}-\frac{265 x_{\ell}^7}{72}\bigg]\nonumber\\
&+ x_{\ell}^4 \bigg[-16 \mathcal{P}_{(0,7,5,0)} \mathrm{ArcTanh}^2(X_r) \nonumber\\
&+24 \ln \bigg(1-X_r^2\bigg) \bigg(\text{Li}_2\bigg(-\frac{X_{-}
   x_{\ell}}{X_{+}}\bigg)-\text{Li}_2\bigg(-\frac{X_{-}}{X_{+} x_{\ell}}\bigg)\bigg) \nonumber\\
&+24 \text{Li}_3\bigg(-\frac{X_{-}}{X_{+} x_{\ell}}\bigg)-24
   \text{Li}_3\bigg(-\frac{X_{-} x_{\ell}}{X_{+}}\bigg)+48 \text{Li}_3\bigg(-\frac{X_{-}}{X_r-1}\bigg) \nonumber\\
&+48\text{Li}_3\bigg(\frac{X_r+1}{(X_r-1) x_{\ell}}+1\bigg)+24 \text{Li}_3\bigg(-\frac{(X_r+1) x_{\ell}}{X_r-1}\bigg) \nonumber\\
& +24 \text{Li}_3\bigg(\frac{x_{\ell}-X_r x_{\ell}}{X_r+1}\bigg)+30 \mathrm{ArcTanh}(X_r)-48 \zeta (3)\bigg]\nonumber\\
&+\frac{1}{6} x_{\ell}^8 \mathcal{P}_{(0,19,0,-3)}+x_{\ell}^6
   \mathcal{P}_{(0,-\frac{46}{3},-2,4)}+x_{\ell}^2 \mathcal{P}_{(0,2,\frac{46}{3},-4)} \nonumber\\
&+ \ln (x_\ell) \bigg[x_\ell^8 \mathcal{P}_{(0,1,3,0)}-8 x_\ell^6 \mathcal{P}_{(0,1,3,0)}-12 x_\ell^4 \mathcal{P}_{(\pi ^2,0,2,-1)} \nonumber\\
&+\ln (\hat{x}) \bigg(-12 x_{\ell}^4 \bigg(4 \mathrm{ArcTanh}(X_r)+1\bigg)+2 x_{\ell}^8-16 x_{\ell}^6\bigg) \nonumber\\
&-12 x_{\ell}^4 \ln ^2(\hat{x})+48 x_{\ell}^4 \ln (X_{-}) \mathrm{ArcTanh}(X_r) \nonumber\\
&+\ln (X_{+}) \bigg(-24 x_{\ell}^4 \bigg(2 \ln
   \bigg(1-X_r^2\bigg)+2 \mathrm{ArcTanh}(X_r)-1\bigg)-4 x_{\ell}^8+32 x_{\ell}^6\bigg)\nonumber\\
&+48 x_{\ell}^4 \ln ^2(X_{+})+96 x_{\ell}^4 \mathrm{ArcTanh}(X_r) \ln (1-x_{\ell}) \nonumber\\
&  -16 x_{\ell}^2 \mathrm{ArcTanh}(X_r)+2 \mathrm{ArcTanh}(X_r)\bigg] \nonumber\\
&+ \ln^2(x_{\ell}) \bigg(12 x_{\ell}^4 (2 \ln (X_{+})-\mathcal{P}_{(0,2,0,0)})+\frac{p(x_\ell)}{2}\bigg)-12 x_{\ell}^4 \ln ^3(x_{\ell})  \nonumber\\
&+\ln (\hat{x}) \bigg[x_{\ell}^8 \mathcal{P}_{(\frac{19}{12},-2,0,0)}+x_{\ell}^6 \mathcal{P}_{(-\frac{26}{3},16,0,0)}+x_{\ell}^2
   \mathcal{P}_{(\frac{26}{3},0,-16,0)}+\mathcal{P}_{(-\frac{19}{12},0,2,0)} \nonumber\\
&-24 x_{\ell}^4 \bigg(\text{Li}_2\bigg(-\frac{X_{-}}{X_{+}
   x_{\ell}}\bigg)-\text{Li}_2\bigg(-\frac{X_{-} x_{\ell}}{X_{+}}\bigg) \nonumber\\
&+2 \mathrm{ArcTanh}(X_r)^2+\mathrm{ArcTanh}(X_r)\bigg) -2 p(x_\ell)( \ln(X_{+})- \ln (x_{\ell}+1))\bigg] \nonumber\\
&+\frac{1}{2} \ln^2(\hat{x}) \bigg(p(x_\ell)-48 x_\ell^4 \mathrm{ArcTanh}(X_r)\bigg) + 2 \ln^2(X_{+}) p(x_\ell) \nonumber\\
&+\ln (X_{+}) \bigg[\bigg(x_{\ell}^8-1\bigg) \mathcal{P}_{(-\frac{19}{6},2,2,0)}-\bigg(x_{\ell}^4-1\bigg) x_{\ell}^2
   \mathcal{P}_{(-\frac{52}{3},16,16,0)} \nonumber\\
 &  +48 x_{\ell}^4 \bigg(\text{Li}_2\bigg(-\frac{X_{-}}{X_{+} x_{\ell}}\bigg)-\text{Li}_2\bigg(-\frac{X_{-}
   x_{\ell}}{X_{+}}\bigg)\bigg) \nonumber\\
&-4 p(x_{\ell}) \ln (x_{\ell}+1)+144 x_{\ell}^4 \mathrm{ArcTanh}^2(X_r)\bigg] \nonumber\\
&+48 x_{\ell}^4 \mathrm{ArcTanh}^2(X_r) \ln(X_{-}) \nonumber\\
&+ \ln(1-x_{\ell}) \bigg[8 x_{\ell}^4 \bigg(6 \Delta\mathrm{Li}_2\bigg(\frac{X_{-}}{1-X_r}\bigg)+6 \Delta\mathrm{Li}_2\bigg(-\frac{X_{-}}{2
   x_{\ell}}\bigg) \nonumber\\
& -6 \Delta\mathrm{Li}_2\bigg(-\frac{1}{2} (X_r-1) (x_{\ell}+1)\bigg) \nonumber\\
& -12 \text{Li}_2\bigg(\frac{X_r+1}{2}\bigg)+6 \ln(X_{-}) \bigg(\ln (X_r+1)+\ln \bigg(\frac{x_{\ell}+1}{2}\bigg)\bigg) \nonumber\\
&-6 \ln (X_{+}) \bigg(\ln \bigg(\frac{1-X_r}{2}\bigg)+\ln(x_{\ell}+1)\bigg)+6\ln (2) \ln \bigg(1-X_r^2\bigg) \nonumber\\
&-6 \ln ^2(X_r+1) +12 \mathrm{ArcTanh}^2(X_r)+\pi^2-6 \ln ^2(2)\bigg) \nonumber\\
&+2 p(x_\ell)
   \bigg(\ln (\hat{x})-2 \ln (X_{+})+\ln \bigg(1-X_r^2\bigg)\bigg)\bigg] \nonumber\\
&+ \frac{1}{6} \bigg(\mathcal{P}_{(0,0,-19,3)}+p(x_{\ell})\bigg(6 \text{Li}_2\bigg(-\frac{X_{-}}{X_{+} x_{\ell}}\bigg)+6 \text{Li}_2\bigg(-\frac{X_{-}x_{\ell}}{X_{+}}\bigg) \nonumber\\
&+12 \ln \bigg(1-X_r^2\bigg) \ln (x_{\ell}+1)+\pi ^2\bigg)\bigg) \nonumber\\
&+\mathrm{Li}_2\bigg(\frac{X_{-}}{1-X_r}\bigg) q_{-}(x_{\ell}) - \mathrm{Li}_2\bigg(\frac{X_{+}}{1+X_r}\bigg) q_{+}(x_{\ell}) \nonumber
\end{align}
\endgroup
}
where the $I_i$ are given in Eq.~(\ref{eq:I0}-\ref{eq:I2}) and $X_r=\sqrt{(\hat{x}-(1-x_\ell)^2)/(\hat{x}-(1+x_\ell)^2)}$. We defined the following abbreviations
\begin{align}
X_{\pm} &= (1-x_\ell) \pm X_r(1+x_\ell), \\
\mathcal{P}_{a,b,c,d} &= a + b \ln(1-X_r) + c \ln(1+X_r) +d \ln^2(1-X_r^2) \\
\Delta\mathrm{Li}_n(f(X_r)) &= \mathrm{Li}_n(f(X_r)) - \mathrm{Li}_n(f(-X_r)) \\
p(x_\ell)&=1 - 8 x_{\ell}^2 + 8 x_{\ell}^6 - x_{\ell}^8 \\
q_{\pm}(x_{\ell})&= 1 - 8 x_{\ell}^2 - 12 x_{\ell}^4 - 8 x_{\ell}^6 + x_{\ell}^8 \\
& \hspace{2cm}- 24 x_{\ell}^4 \ln(\hat{x}) \pm 24 x_{\ell}^4 \ln\bigg(\frac{\hat{x}(1-X_r^2)}{X_{+}^2}\bigg) \nonumber
\end{align}
where $f(X_r)$ is an arbitrary function depending on $X_r$. 

\begin{figure}[t]
\centering
\includegraphics[width=0.7
\textwidth]{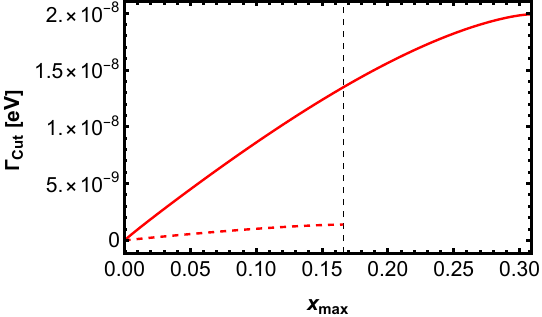}
\caption{Integrated spectrum $\Gamma_N^X(x_\mathrm{max})$. The width is plotted with a mixing angle of $V_{N\ell} = 10^{-3}$ and the lepton is a tau $\ell = \tau$. In the full curve the mass of the HSN is $m_N = \SI{4}{\giga\electronvolt}$ whereas in the dashed curve it is $m_N = \SI{3}{\giga\electronvolt}$; the dashed vertical line marks the endpoint of the integrated spectrum for the dashed red curve.}
\label{fig:spectrum2}
\end{figure}

Using Eq.~(\ref{eq:integratedspec}) we plot the integrated decay width $\Gamma_\mathrm{cut}$ defined as 
\begin{align}
\Gamma_\mathrm{cut}(N \rightarrow \ell X) =& N_c \frac{G_F^2 m_N^5 |V_{N\ell}|^2 }{192 \pi^3} \\
& \times 2 (|V_{ud}|^2 + |V_{us}|^2) \bigg[ I_0 c_{0,1} + a_{m_N}  c_{1,1} I_0 \nonumber \\
&+a_{m_N}^2 \big( c_{2,1} I_0 +2 c_{2,2} I_1 \big) \nonumber\\
&+a_{m_N}^3 \big(  c_{3,1} I_0 +2 c_{3,2} I_1 - (\pi^2 I_0-3  I_2 )c_{3,3} \big) \nonumber\\
&+a_{m_N}^4 \big(  c_{4,1} I_0^Q+2 c_{4,2} I_1 \nonumber\\
&\hphantom{a_\mu^4 }
 -(\pi^2 I_0-3 I_2)c_{4,3} - (4\pi^2 I_1 -4 I_3 )c_{4,4} \big) 
\bigg] \nonumber
\end{align}
\kt{where $I_k = I_k(x_\ell^2,x_\mathrm{max})$, see Fig.~\ref{fig:spectrum2}}.

\newpage
\bibliographystyle{JHEP}
\bibliography{bibQCDmasslesscorr}

@article{Bondarenko_2018,
   title="{Phenomenology of GeV-scale heavy neutral leptons}",
   volume={2018},
   ISSN={1029-8479},
   url={http://dx.doi.org/10.1007/JHEP11(2018)032},
   DOI={10.1007/jhep11(2018)032},
   number={11},
   journal={Journal of High Energy Physics},
   publisher={Springer Science and Business Media LLC},
   author={Bondarenko, Kyrylo and Boyarsky, Alexey and Gorbunov, Dmitry and Ruchayskiy, Oleg},
   year={2018},
   month=nov }

@article{Beneke_2008,
   title={$\alpha_s$ and the $\tau$ hadronic width: fixed-order, contour- improved and higher-order perturbation theory},
   volume={2008},
   ISSN={1029-8479},
   url={http://dx.doi.org/10.1088/1126-6708/2008/09/044},
   DOI={10.1088/1126-6708/2008/09/044},
   number={09},
   journal={Journal of High Energy Physics},
   publisher={Springer Science and Business Media LLC},
   author={Beneke, Martin and Jamin, Matthias},
   year={2008},
   month=sep, pages={044–044} }

@article{Chetyrkin_1998,
   title={Determining the strange quark mass in Cabibbo-suppressed tau lepton decays},
   volume={533},
   ISSN={0550-3213},
   url={http://dx.doi.org/10.1016/S0550-3213(98)00511-2},
   DOI={10.1016/s0550-3213(98)00511-2},
   number={1–3},
   journal={Nuclear Physics B},
   publisher={Elsevier BV},
   author={Chetyrkin, K.G. and Kühn, J.H. and Pivovarov, A.A.},
   year={1998},
   month=nov, pages={473–493} }

@article{Baikov:2008jh,
    author = "Baikov, P. A. and Chetyrkin, K. G. and Kuhn, Johann H.",
    title = "{Order $\alpha^4(s)$ QCD Corrections to $Z$ and $\tau$ Decays}",
    eprint = "0801.1821",
    archivePrefix = "arXiv",
    primaryClass = "hep-ph",
    reportNumber = "SFB-CPP-08-04, TTP08-01",
    doi = "10.1103/PhysRevLett.101.012002",
    journal = "Phys. Rev. Lett.",
    volume = "101",
    pages = "012002",
    year = "2008"
}

@article{PhysRevLett.108.222003,
  title = "{Complete $\mathcal{O}(\alpha_{s}^{4})$ QCD Corrections to Hadronic $Z$ Decays}",
  author = {Baikov, P. A. and Chetyrkin, K. G. and K\"uhn, J. H. and Rittinger, J.},
  journal = {Phys. Rev. Lett.},
  volume = {108},
  issue = {22},
  pages = {222003},
  numpages = {4},
  year = {2012},
  month = May,
  publisher = {American Physical Society},
  doi = {10.1103/PhysRevLett.108.222003},
  url = {https://link.aps.org/doi/10.1103/PhysRevLett.108.222003}
}

@article{Pich_2021,
   title="{Precision physics with inclusive QCD processes}",
   volume={117},
   ISSN={0146-6410},
   url={http://dx.doi.org/10.1016/j.ppnp.2020.103846},
   DOI={10.1016/j.ppnp.2020.103846},
   journal={Progress in Particle and Nuclear Physics},
   publisher={Elsevier BV},
   author={Pich, Antonio},
   year={2021},
   month=mar, pages={103846} }

@article{BRAATEN1992581,
title = "{QCD analysis of the tau hadronic width}",
journal = {Nuclear Physics B},
volume = {373},
number = {3},
pages = {581-612},
year = {1992},
issn = {0550-3213},
doi = {https://doi.org/10.1016/0550-3213(92)90267-F},
url = {https://www.sciencedirect.com/science/article/pii/055032139290267F},
author = {E. Braaten and S. Narison and A. Pich}
}

@article{Czakon_2005,
   title={The four-loop QCD $\beta$-function and anomalous dimensions},
   volume={710},
   ISSN={0550-3213},
   url={http://dx.doi.org/10.1016/j.nuclphysb.2005.01.012},
   DOI={10.1016/j.nuclphysb.2005.01.012},
   number={1–2},
   journal={Nuclear Physics B},
   publisher={Elsevier BV},
   author={Czakon, M.},
   year={2005},
   month=mar, pages={485–498} }

@article{Chetyrkin:2016uhw,
    author = {Chetyrkin, Konstantin and Baikov, Pavel and K\"uhn, Johann},
    title = "{The $\beta$-function of Quantum Chromodynamics and the effective Higgs-gluon-gluon coupling in five-loop order}",
    doi = "10.22323/1.260.0010",
    journal = "PoS",
    volume = "LL2016",
    pages = "010",
    year = "2016"
}

@article{PhysRevLett.61.1815,
  title = {Electroweak radiative corrections to \ensuremath{\tau} decay},
  author = {Marciano, William J. and Sirlin, A.},
  journal = {Phys. Rev. Lett.},
  volume = {61},
  issue = {16},
  pages = {1815--1818},
  numpages = {0},
  year = {1988},
  month = Oct,
  publisher = {American Physical Society},
  doi = {10.1103/PhysRevLett.61.1815},
  url = {https://link.aps.org/doi/10.1103/PhysRevLett.61.1815}
}

@article{Adler:1974gd,
    author = "Adler, Stephen L.",
    title = "{Some Simple Vacuum Polarization Phenomenology: $e^+ e^- \to \mathrm{Hadrons}$: The muonic-atom x-Ray Discrepancy and $g_{\mu}-2$}",
    reportNumber = "FNAL-PUB-74-063-THY, FERMILAB-PUB-74-063-T",
    doi = "10.1103/PhysRevD.10.3714",
    journal = "Phys. Rev. D",
    volume = "10",
    pages = "3714",
    year = "1974"
}

@article{Herren:2017osy,
    author = "Herren, Florian and Steinhauser, Matthias",
    title = "{Version 3 of RunDec and CRunDec}",
    eprint = "1703.03751",
    archivePrefix = "arXiv",
    primaryClass = "hep-ph",
    reportNumber = "TTP17-011",
    doi = "10.1016/j.cpc.2017.11.014",
    journal = "Comput. Phys. Commun.",
    volume = "224",
    pages = "333--345",
    year = "2018"
}

@article{Chetyrkin:2000yt,
    author = "Chetyrkin, K. G. and Kuhn, Johann H. and Steinhauser, M.",
    title = "{RunDec: A Mathematica package for running and decoupling of the strong coupling and quark masses}",
    eprint = "hep-ph/0004189",
    archivePrefix = "arXiv",
    reportNumber = "DESY-00-034, TTP-00-05",
    doi = "10.1016/S0010-4655(00)00155-7",
    journal = "Comput. Phys. Commun.",
    volume = "133",
    pages = "43--65",
    year = "2000"
}

@article{Surguladze:1990tg,
    author = "Surguladze, Levan R. and Samuel, Mark A.",
    title = "{Total hadronic cross-section in $e^+ e^-$ annihilation at the four loop level of perturbative QCD}",
    reportNumber = "OSU-RN-250",
    doi = "10.1103/PhysRevLett.66.560",
    journal = "Phys. Rev. Lett.",
    volume = "66",
    pages = "560--563",
    year = "1991",
    note = "[Erratum: Phys.Rev.Lett. 66, 2416 (1991)]"
}

@article{Gorishnii:1990vf,
    author = "Gorishnii, S. G. and Kataev, A. L. and Larin, S. A.",
    title = "{The $O(\alpha^{3}_{s})$-corrections to $\sigma_{tot}(e^{+}e^{-}\rightarrow \mathrm{hadrons})$ and $\Gamma(\tau^{-} \rightarrow \nu_{\tau} + \mathrm{hadrons})$ in QCD}",
    reportNumber = "UM-TH-91-01",
    doi = "10.1016/0370-2693(91)90149-K",
    journal = "Phys. Lett. B",
    volume = "259",
    pages = "144--150",
    year = "1991"
}

@article{PhysRevD.94.053001,
  title = "{Rare decays of $B$ mesons via on-shell sterile neutrinos}",
  author = {Cveti\v{c}, Gorazd and Kim, C. S.},
  journal = {Phys. Rev. D},
  volume = {94},
  issue = {5},
  pages = {053001},
  numpages = {28},
  year = {2016},
  month = Sep,
  publisher = {American Physical Society},
  doi = {10.1103/PhysRevD.94.053001},
  url = {https://link.aps.org/doi/10.1103/PhysRevD.94.053001}
}

@article{Johnson_1997,
   title={Extending sensitivity for low-mass neutral heavy lepton searches},
   volume={56},
   ISSN={1089-4918},
   url={http://dx.doi.org/10.1103/PhysRevD.56.2970},
   DOI={10.1103/physrevd.56.2970},
   number={5},
   journal={Physical Review D},
   publisher={American Physical Society (APS)},
   author={Johnson, Loretta M. and McKay, Douglas W. and Bolton, Tim},
   year={1997},
   month=sep, pages={2970–2981} }

@article{Gribanov_2001,
   title={Sterile neutrinos in tau lepton decays},
   volume={607},
   ISSN={0550-3213},
   url={http://dx.doi.org/10.1016/S0550-3213(01)00169-9},
   DOI={10.1016/s0550-3213(01)00169-9},
   number={1–2},
   journal={Nuclear Physics B},
   publisher={Elsevier BV},
   author={Gribanov, Vladimir and Kovalenko, Sergey and Schmidt, Ivan},
   year={2001},
   month=jul, pages={355–368} }

@article{Gorbunov_2007,
   title="{How to find neutral leptons of the $\nu$ MSM?}",
   volume={2007},
   ISSN={1029-8479},
   url={http://dx.doi.org/10.1088/1126-6708/2007/10/015},
   DOI={10.1088/1126-6708/2007/10/015},
   number={10},
   journal={Journal of High Energy Physics},
   publisher={Springer Science and Business Media LLC},
   author={Gorbunov, Dmitry and Shaposhnikov, Mikhail},
   year={2007},
   month=oct, pages={015–015} }

@article{Atre_2009,
   title={The search for heavy Majorana neutrinos},
   volume={2009},
   ISSN={1029-8479},
   url={http://dx.doi.org/10.1088/1126-6708/2009/05/030},
   DOI={10.1088/1126-6708/2009/05/030},
   number={05},
   journal={Journal of High Energy Physics},
   publisher={Springer Science and Business Media LLC},
   author={Atre, Anupama and Han, Tao and Pascoli, Silvia and Zhang, Bin},
   year={2009},
   month=may, pages={030–030} }

@article{Helo_2011,
   title={Sterile neutrinos in lepton number and lepton flavor violating decays},
   volume={853},
   ISSN={0550-3213},
   url={http://dx.doi.org/10.1016/j.nuclphysb.2011.07.020},
   DOI={10.1016/j.nuclphysb.2011.07.020},
   number={1},
   journal={Nuclear Physics B},
   publisher={Elsevier BV},
   author={Helo, Juan Carlos and Kovalenko, Sergey and Schmidt, Ivan},
   year={2011},
   month=dec, pages={80–104} }

@article{PhysRevD.15.755,
  title = {Determination of the quark-gluon coupling constant},
  author = {Shankar, R.},
  journal = {Phys. Rev. D},
  volume = {15},
  issue = {3},
  pages = {755--758},
  numpages = {0},
  year = {1977},
  month = Feb,
  publisher = {American Physical Society},
  doi = {10.1103/PhysRevD.15.755},
  url = {https://link.aps.org/doi/10.1103/PhysRevD.15.755}
}

@article{PhysRevLett.60.1606,
  title = "{QCD predictions for the decay of the \ensuremath{\tau} lepton}",
  author = {Braaten, Eric},
  journal = {Phys. Rev. Lett.},
  volume = {60},
  issue = {16},
  pages = {1606--1609},
  numpages = {0},
  year = {1988},
  month = Apr,
  publisher = {American Physical Society},
  doi = {10.1103/PhysRevLett.60.1606},
  url = {https://link.aps.org/doi/10.1103/PhysRevLett.60.1606}
}

@article{PhysRevD.16.703,
  title = {Decays of a heavy lepton and an intermediate weak boson in quantum chromodynamics},
  author = {Lam, C. S. and Yan, T. M.},
  journal = {Phys. Rev. D},
  volume = {16},
  issue = {3},
  pages = {703--706},
  numpages = {0},
  year = {1977},
  month = Aug,
  publisher = {American Physical Society},
  doi = {10.1103/PhysRevD.16.703},
  url = {https://link.aps.org/doi/10.1103/PhysRevD.16.703}
}

@article{Becchi:1980vz,
    author = "Becchi, C. and Narison, Stephan and de Rafael, E. and Yndurain, F. J.",
    title = "{Light Quark Masses in Quantum Chromodynamics and Chiral Symmetry Breaking}",
    reportNumber = "CERN-TH-2920",
    doi = "10.1007/BF01546328",
    journal = "Z. Phys. C",
    volume = "8",
    pages = "335",
    year = "1981"
}

@article{Chetyrkin:1979bj,
    author = "Chetyrkin, K. G. and Kataev, A. L. and Tkachov, F. V.",
    title = "{Higher Order Corrections to $\sigma_{\mathrm{tot}} (e^+ e^- \rightarrow \mathrm{Hadrons})$ in Quantum Chromodynamics}",
    reportNumber = "IYaI-P-0126",
    doi = "10.1016/0370-2693(79)90596-3",
    journal = "Phys. Lett. B",
    volume = "85",
    pages = "277--279",
    year = "1979"
}

@article{Poggio:1975af,
    author = "Poggio, E. C. and Quinn, Helen R. and Weinberg, Steven",
    title = "{Smearing the Quark Model}",
    reportNumber = "Print-75-1056 (HARVARD)",
    doi = "10.1103/PhysRevD.13.1958",
    journal = "Phys. Rev. D",
    volume = "13",
    pages = "1958",
    year = "1976"
}

@article{Robinson:2018gza,
        archiveprefix = {arXiv},
        author = {Robinson, Dean J. and Shakya, Bibhushan and Zupan, Jure},
        doi = {10.1007/JHEP02(2019)119},
        eprint = {1807.04753},
        journal = {JHEP},
        pages = {119},
        primaryclass = {hep-ph},
        reportnumber = {LCTP-18-19},
        title = {{Right-handed neutrinos and R(D$^{(\ast)}$)}},
        volume = {02},
        year = {2019}}

@article{Bernlochner:2024xiz,
    author = "Bernlochner, Florian U. and Fedele, Marco and Kretz, Tim and Nierste, Ulrich and Prim, Markus T.",
    title = "{Model independent bounds on heavy sterile neutrinos from
                  the angular distribution of $B\to D^{*}\ell \nu$ decays}",
    eprint = "2410.11945",
    archivePrefix = "arXiv",
    primaryClass = "hep-ph",
    reportNumber = "TTP24-041",
    doi = "10.1007/JHEP01(2025)040",
    journal = "JHEP",
    volume = "01",
    pages = "040",
    year = "2025"
}

@article{Asaka_2005,
   title="{The $\nu$MSM, dark matter and baryon asymmetry of the universe}",
   volume={620},
   ISSN={0370-2693},
   url={http://dx.doi.org/10.1016/j.physletb.2005.06.020},
   DOI={10.1016/j.physletb.2005.06.020},
   number={1–2},
   journal={Physics Letters B},
   publisher={Elsevier BV},
   author={Asaka, Takehiko and Shaposhnikov, Mikhail},
   year={2005},
   month=jul, pages={17–26} }

@article{Asaka_2005_2,
   title="{The $\nu$MSM, dark matter and neutrino masses}",
   volume={631},
   ISSN={0370-2693},
   url={http://dx.doi.org/10.1016/j.physletb.2005.09.070},
   DOI={10.1016/j.physletb.2005.09.070},
   number={4},
   journal={Physics Letters B},
   publisher={Elsevier BV},
   author={Asaka, Takehiko and Blanchet, Steve and Shaposhnikov, Mikhail},
   year={2005},
   month=dec, pages={151–156} }

@article{Yanagida:1979as,
    author = "Yanagida, Tsutomu",
    editor = "Sawada, Osamu and Sugamoto, Akio",
    title = "{Horizontal gauge symmetry and masses of neutrinos}",
    reportNumber = "KEK-79-18-95",
    journal = "Conf. Proc. C",
    volume = "7902131",
    pages = "95--99",
    year = "1979"
}

@article{Gell-Mann:1979vob,
    author = "Gell-Mann, Murray and Ramond, Pierre and Slansky, Richard",
    title = "{Complex Spinors and Unified Theories}",
    eprint = "1306.4669",
    archivePrefix = "arXiv",
    primaryClass = "hep-th",
    reportNumber = "PRINT-80-0576",
    journal = "Conf. Proc. C",
    volume = "790927",
    pages = "315--321",
    year = "1979"
}

@article{Glashow:1979nm,
    author = "Glashow, S. L.",
    editor = "L{\'e}vy, Maurice and Basdevant, Jean-Louis and Speiser, David and Weyers, Jacques and Gastmans, Raymond and Jacob, Maurice",
    title = "{The Future of Elementary Particle Physics}",
    reportNumber = "HUTP-79-A059",
    doi = "10.1007/978-1-4684-7197-7_15",
    journal = "NATO Sci. Ser. B",
    volume = "61",
    pages = "687",
    year = "1980"
}

@article{Mohapatra:1979ia,
    author = "Mohapatra, Rabindra N. and Senjanovic, Goran",
    title = "{Neutrino Mass and Spontaneous Parity Nonconservation}",
    reportNumber = "MDDP-TR-80-060, MDDP-PP-80-105, CCNY-HEP-79-10",
    doi = "10.1103/PhysRevLett.44.912",
    journal = "Phys. Rev. Lett.",
    volume = "44",
    pages = "912",
    year = "1980"
}

@article{Fukugita:1986hr,
    author = "Fukugita, M. and Yanagida, T.",
    title = "{Baryogenesis Without Grand Unification}",
    reportNumber = "RIFP-641",
    doi = "10.1016/0370-2693(86)91126-3",
    journal = "Phys. Lett. B",
    volume = "174",
    pages = "45--47",
    year = "1986"
}

@article{Davidson:2002qv,
    author = "Davidson, Sacha and Ibarra, Alejandro",
    title = "{A Lower bound on the right-handed neutrino mass from leptogenesis}",
    eprint = "hep-ph/0202239",
    archivePrefix = "arXiv",
    reportNumber = "OUTP-02-10P, IPPP-02-16, DCPT-02-32",
    doi = "10.1016/S0370-2693(02)01735-5",
    journal = "Phys. Lett. B",
    volume = "535",
    pages = "25--32",
    year = "2002"
}

@article{Minkowski:1977sc,
    author = "Minkowski, Peter",
    title = "{$\mu \to e\gamma$ at a Rate of One Out of $10^{9}$ Muon Decays?}",
    reportNumber = "Print-77-0182 (BERN)",
    doi = "10.1016/0370-2693(77)90435-X",
    journal = "Phys. Lett. B",
    volume = "67",
    pages = "421--428",
    year = "1977"
}

@article{PhysRevLett.43.668,
  title = {Higher-Order Quantum Chromodynamic Corrections in ${e}^{+}{e}^{\ensuremath{-}}$ Annihilation},
  author = {Dine, Michael and Sapirstein, Jonathan},
  journal = {Phys. Rev. Lett.},
  volume = {43},
  issue = {10},
  pages = {668--671},
  numpages = {0},
  year = {1979},
  month = {Sep},
  publisher = {American Physical Society},
  doi = {10.1103/PhysRevLett.43.668},
  url = {https://link.aps.org/doi/10.1103/PhysRevLett.43.668}
}

@article{PhysRevLett.44.560,
  title = {Analytic Calculation of Higher-Order Quantum-Chromodynamic Corrections in ${e}^{+}{e}^{\ensuremath{-}}$ Annihilation},
  author = {Celmaster, William and Gonsalves, Richard J.},
  journal = {Phys. Rev. Lett.},
  volume = {44},
  issue = {9},
  pages = {560--564},
  numpages = {0},
  year = {1980},
  month = {Mar},
  publisher = {American Physical Society},
  doi = {10.1103/PhysRevLett.44.560},
  url = {https://link.aps.org/doi/10.1103/PhysRevLett.44.560}
}

@article{Gorishnii:1988bc,
    author = "Gorishnii, S. G. and Kataev, A. L. and Larin, S. A.",
    title = "{Next-To-Leading $\mathcal{O}(\alpha_s^3)$ QCD Correction to $\sigma_{\mathrm{tot}}(e^+ e^- \rightarrow \mathrm{Hadrons})$: Analytical Calculation and Estimation of the Parameter Lambda (MS)}",
    reportNumber = "JINR-E2-88-254",
    doi = "10.1016/0370-2693(88)90532-1",
    journal = "Phys. Lett. B",
    volume = "212",
    pages = "238--244",
    year = "1988"
}

@article{HeavyFlavorAveragingGroupHFLAV:2024ctg,
    author = "Banerjee, Swagato and others",
    collaboration = "Heavy Flavor Averaging Group (HFLAV)",
    title = "{Averages of $b$-hadron, $c$-hadron, and $\tau$-lepton properties as of 2023}",
    eprint = "2411.18639",
    archivePrefix = "arXiv",
    primaryClass = "hep-ex",
    month = "11",
    year = "2024"
}

@article{ParticleDataGroup:2024cfk,
    author = "Navas, S. and others",
    collaboration = "Particle Data Group",
    title = "{Review of particle physics}",
    doi = "10.1103/PhysRevD.110.030001",
    journal = "Phys. Rev. D",
    volume = "110",
    number = "3",
    pages = "030001",
    year = "2024"
}

@article{Erler:2002bu,
    author = "Erler, Jens and Luo, Mingxing",
    title = "{Precision determination of heavy quark masses and the strong coupling constant}",
    eprint = "hep-ph/0207114",
    archivePrefix = "arXiv",
    reportNumber = "UPR-1003-T, FT-2002-04",
    doi = "10.1016/S0370-2693(03)00276-4",
    journal = "Phys. Lett. B",
    volume = "558",
    pages = "125--131",
    year = "2003"
}

@article{Lacker:2010zz,
    author = "Lacker, H. and Menzel, A.",
    title = "{Simultaneous Extraction of the Fermi constant and PMNS matrix elements in the presence of a fourth generation}",
    eprint = "1003.4532",
    archivePrefix = "arXiv",
    primaryClass = "hep-ph",
    reportNumber = "HU-EP-10-10",
    doi = "10.1007/JHEP07(2010)006",
    journal = "JHEP",
    volume = "07",
    pages = "006",
    year = "2010"
}

@article{2024CMS,
   title={Search for long-lived heavy neutral leptons with lepton flavour conserving or violating decays to a jet and a charged lepton},
   volume={2024},
   ISSN={1029-8479},
   url={http://dx.doi.org/10.1007/JHEP03(2024)105},
   DOI={10.1007/jhep03(2024)105},
   number={3},
   journal={Journal of High Energy Physics},
   publisher={Springer Science and Business Media LLC},
   author={Hayrapetyan, A. and Tumasyan, A. and Adam, W. and Andrejkovic, J. W. and Bergauer, T. and Chatterjee, S. and Damanakis, K. and Dragicevic, M. and Hussain, P. S. and Jeitler, M. and Krammer, N. and Li, A. and Liko, D. and Mikulec, I. and Schieck, J. and Schöfbeck, R. and Schwarz, D. and Sonawane, M. and Templ, S. and Waltenberger, W. and Wulz, C.-E. and Darwish, M. R. and Janssen, T. and Van Mechelen, P. and Bols, E. S. and D’Hondt, J. and Dansana, S. and De Moor, A. and Delcourt, M. and El Faham, H. and Lowette, S. and Makarenko, I. and Müller, D. and Sahasransu, A. R. and Tavernier, S. and Tytgat, M. and Van Onsem, G. P. and Van Putte, S. and Vannerom, D. and Clerbaux, B. and Das, A. K. and De Lentdecker, G. and Favart, L. and Gianneios, P. and Hohov, D. and Jaramillo, J. and Khalilzadeh, A. and Lee, K. and Mahdavikhorrami, M. and Malara, A. and Paredes, S. and Pétré, L. and Postiau, N. and Thomas, L. and Vanden Bemden, M. and Vander Velde, C. and Vanlaer, P. and De Coen, M. and Dobur, D. and Hong, Y. and Knolle, J. and Lambrecht, L. and Mestdach, G. and Mota Amarilo, K. and Rendón, C. and Samalan, A. and Skovpen, K. and Van Den Bossche, N. and van der Linden, J. and Wezenbeek, L. and Benecke, A. and Bethani, A. and Bruno, G. and Caputo, C. and Delaere, C. and Donertas, I. S. and Giammanco, A. and Jaffel, K. and Jain, Sa. and Lemaitre, V. and Lidrych, J. and Mastrapasqua, P. and Mondal, K. and Tran, T. T. and Wertz, S. and Alves, G. A. and Coelho, E. and Hensel, C. and Menezes De Oliveira, T. and Moraes, A. and Rebello Teles, P. and Soeiro, M. and Aldá Júnior, W. L. and Alves Gallo Pereira, M. and Barroso Ferreira Filho, M. and Brandao Malbouisson, H. and Carvalho, W. and Chinellato, J. and Da Costa, E. M. and Da Silveira, G. G. and De Jesus Damiao, D. and Fonseca De Souza, S. and Gomes De Souza, R. and Martins, J. and Mora Herrera, C. and Mundim, L. and Nogima, H. and Pinheiro, J. P. and Santoro, A. and Sznajder, A. and Thiel, M. and Vilela Pereira, A. and Bernardes, C. A. and Calligaris, L. and Fernandez Perez Tomei, T. R. and Gregores, E. M. and Mercadante, P. G. and Novaes, S. F. and Orzari, B. and Padula, Sandra S. and Aleksandrov, A. and Antchev, G. and Hadjiiska, R. and Iaydjiev, P. and Misheva, M. and Shopova, M. and Sultanov, G. and Dimitrov, A. and Litov, L. and Pavlov, B. and Petkov, P. and Petrov, A. and Shumka, E. and Keshri, S. and Thakur, S. and Cheng, T. and Javaid, T. and Yuan, L. and Hu, Z. and Liu, J. and Yi, K. and Chen, G. M. and Chen, H. S. and Chen, M. and Iemmi, F. and Jiang, C. H. and Kapoor, A. and Liao, H. and Liu, Z.-A. and Sharma, R. and Song, J. N. and Tao, J. and Wang, C. and Wang, J. and Wang, Z. and Zhang, H. and Agapitos, A. and Ban, Y. and Levin, A. and Li, C. and Li, Q. and Mao, Y. and Qian, S. J. and Sun, X. and Wang, D. and Yang, H. and Zhang, L. and Zhou, C. and You, Z. and Lu, N. and Bauer, G. and Gao, X. and Leggat, D. and Okawa, H. and Lin, Z. and Lu, C. and Xiao, M. and Avila, C. and Barbosa Trujillo, D. A. and Cabrera, A. and Florez, C. and Fraga, J. and Reyes Vega, J. A. and Mejia Guisao, J. and Ramirez, F. and Rodriguez, M. and Ruiz Alvarez, J. D. and Giljanovic, D. and Godinovic, N. and Lelas, D. and Sculac, A. and Kovac, M. and Sculac, T. and Bargassa, P. and Brigljevic, V. and Chitroda, B. K. and Ferencek, D. and Mishra, S. and Starodumov, A. and Susa, T. and Attikis, A. and Christoforou, K. and Konstantinou, S. and Mousa, J. and Nicolaou, C. and Ptochos, F. and Razis, P. A. and Rykaczewski, H. and Saka, H. and Stepennov, A. and Finger, M. and Finger, M. and Kveton, A. and Ayala, E. and Carrera Jarrin, E. and Abdalla, H. and Assran, Y. and Mahmoud, M. A. and Mohammed, Y. and Ehataht, K. and Kadastik, M. and Lange, T. and Nandan, S. and Nielsen, C. and Pata, J. and Raidal, M. and Tani, L. and Veelken, C. and Kirschenmann, H. and Osterberg, K. and Voutilainen, M. and Bharthuar, S. and Brücken, E. and Garcia, F. and Kallonen, K. T. S. and Kinnunen, R. and Lampén, T. and Lassila-Perini, K. and Lehti, S. and Lindén, T. and Martikainen, L. and Myllymäki, M. and Rantanen, M. m. and Siikonen, H. and Tuominen, E. and Tuominiemi, J. and Luukka, P. and Petrow, H. and Besancon, M. and Couderc, F. and Dejardin, M. and Denegri, D. and Faure, J. L. and Ferri, F. and Ganjour, S. and Gras, P. and Hamel de Monchenault, G. and Lohezic, V. and Malcles, J. and Rander, J. and Rosowsky, A. and Sahin, M. Ö. and Savoy-Navarro, A. and Simkina, P. and Titov, M. and Tornago, M. and Baldenegro Barrera, C. and Beaudette, F. and Buchot Perraguin, A. and Busson, P. and Cappati, A. and Charlot, C. and Chiusi, M. and Damas, F. and Davignon, O. and De Wit, A. and Fontana Santos Alves, B. A. and Ghosh, S. and Gilbert, A. and Granier de Cassagnac, R. and Hakimi, A. and Harikrishnan, B. and Kalipoliti, L. and Liu, G. and Motta, J. and Nguyen, M. and Ochando, C. and Portales, L. and Salerno, R. and Sauvan, J. B. and Sirois, Y. and Tarabini, A. and Vernazza, E. and Zabi, A. and Zghiche, A. and Agram, J.-L. and Andrea, J. and Apparu, D. and Bloch, D. and Brom, J.-M. and Chabert, E. C. and Collard, C. and Falke, S. and Goerlach, U. and Grimault, C. and Haeberle, R. and Le Bihan, A.-C. and Meena, M. and Saha, G. and Sessini, M. A. and Van Hove, P. and Beauceron, S. and Blancon, B. and Boudoul, G. and Chanon, N. and Choi, J. and Contardo, D. and Depasse, P. and Dozen, C. and El Mamouni, H. and Fay, J. and Gascon, S. and Gouzevitch, M. and Greenberg, C. and Grenier, G. and Ille, B. and Laktineh, I. B. and Lethuillier, M. and Mirabito, L. and Perries, S. and Purohit, A. and Vander Donckt, M. and Verdier, P. and Xiao, J. and Bagaturia, I. and Lomidze, I. and Tsamalaidze, Z. and Botta, V. and Feld, L. and Klein, K. and Lipinski, M. and Meuser, D. and Pauls, A. and Röwert, N. and Teroerde, M. and Diekmann, S. and Dodonova, A. and Eich, N. and Eliseev, D. and Engelke, F. and Erdmann, J. and Erdmann, M. and Fackeldey, P. and Fischer, B. and Hebbeker, T. and Hoepfner, K. and Ivone, F. and Jung, A. and Lee, M. y. and Mastrolorenzo, L. and Mausolf, F. and Merschmeyer, M. and Meyer, A. and Mukherjee, S. and Noll, D. and Novak, A. and Nowotny, F. and Pozdnyakov, A. and Rath, Y. and Redjeb, W. and Rehm, F. and Reithler, H. and Sarkar, U. and Sarkisovi, V. and Schmidt, A. and Sharma, A. and Spah, J. L. and Stein, A. and Torres Da Silva De Araujo, F. and Vigilante, L. and Wiedenbeck, S. and Zaleski, S. and Dziwok, C. and Flügge, G. and Haj Ahmad, W. and Kress, T. and Nowack, A. and Pooth, O. and Stahl, A. and Ziemons, T. and Zotz, A. and Aarup Petersen, H. and Aldaya Martin, M. and Alimena, J. and Amoroso, S. and An, Y. and Baxter, S. and Bayatmakou, M. and Becerril Gonzalez, H. and Behnke, O. and Belvedere, A. and Bhattacharya, S. and Blekman, F. and Borras, K. and Campbell, A. and Cardini, A. and Cheng, C. and Colombina, F. and Consuegra Rodríguez, S. and Correia Silva, G. and De Silva, M. and Eckerlin, G. and Eckstein, D. and Estevez Banos, L. I. and Filatov, O. and Gallo, E. and Geiser, A. and Giraldi, A. and Greau, G. and Guglielmi, V. and Guthoff, M. and Hinzmann, A. and Jafari, A. and Jeppe, L. and Jomhari, N. Z. and Kaech, B. and Kasemann, M. and Kleinwort, C. and Kogler, R. and Komm, M. and Krücker, D. and Lange, W. and Leyva Pernia, D. and Lipka, K. and Lohmann, W. and Mankel, R. and Melzer-Pellmann, I.-A. and Mendizabal Morentin, M. and Meyer, A. B. and Milella, G. and Mussgiller, A. and Nair, L. P. and Nürnberg, A. and Otarid, Y. and Park, J. and Pérez Adán, D. and Ranken, E. and Raspereza, A. and Ribeiro Lopes, B. and Rübenach, J. and Saggio, A. and Scham, M. and Schnake, S. and Schütze, P. and Schwanenberger, C. and Selivanova, D. and Sharko, K. and Shchedrolosiev, M. and Sosa Ricardo, R. E. and Stafford, D. and Vazzoler, F. and Ventura Barroso, A. and Walsh, R. and Wang, Q. and Wen, Y. and Wichmann, K. and Wiens, L. and Wissing, C. and Yang, Y. and Zimermmane Castro Santos, A. and Albrecht, A. and Albrecht, S. and Antonello, M. and Bein, S. and Benato, L. and Bonanomi, M. and Connor, P. and Eich, M. and El Morabit, K. and Fischer, Y. and Fröhlich, A. and Garbers, C. and Garutti, E. and Grohsjean, A. and Hajheidari, M. and Haller, J. and Jabusch, H. R. and Kasieczka, G. and Keicher, P. and Klanner, R. and Korcari, W. and Kramer, T. and Kutzner, V. and Labe, F. and Lange, J. and Lobanov, A. and Matthies, C. and Mehta, A. and Moureaux, L. and Mrowietz, M. and Nigamova, A. and Nissan, Y. and Paasch, A. and Pena Rodriguez, K. J. and Quadfasel, T. and Raciti, B. and Rieger, M. and Savoiu, D. and Schindler, J. and Schleper, P. and Schröder, M. and Schwandt, J. and Sommerhalder, M. and Stadie, H. and Steinbrück, G. and Tews, A. and Wolf, M. and Brommer, S. and Burkart, M. and Butz, E. and Chwalek, T. and Dierlamm, A. and Droll, A. and Faltermann, N. and Giffels, M. and Gottmann, A. and Hartmann, F. and Hofsaess, R. and Horzela, M. and Husemann, U. and Kieseler, J. and Klute, M. and Koppenhöfer, R. and Lawhorn, J. M. and Link, M. and Lintuluoto, A. and Maier, S. and Mitra, S. and Mormile, M. and Müller, Th. and Neukum, M. and Oh, M. and Presilla, M. and Quast, G. and Rabbertz, K. and Regnery, B. and Shadskiy, N. and Shvetsov, I. and Simonis, H. J. and Toms, M. and Trevisani, N. and Ulrich, R. and Von Cube, R. F. and Wassmer, M. and Wieland, S. and Wittig, F. and Wolf, R. and Zuo, X. and Anagnostou, G. and Daskalakis, G. and Kyriakis, A. and Papadopoulos, A. and Stakia, A. and Kontaxakis, P. and Melachroinos, G. and Panagiotou, A. and Papavergou, I. and Paraskevas, I. and Saoulidou, N. and Theofilatos, K. and Tziaferi, E. and Vellidis, K. and Zisopoulos, I. and Bakas, G. and Chatzistavrou, T. and Karapostoli, G. and Kousouris, K. and Papakrivopoulos, I. and Siamarkou, E. and Tsipolitis, G. and Zacharopoulou, A. and Adamidis, K. and Bestintzanos, I. and Evangelou, I. and Foudas, C. and Kamtsikis, C. and Katsoulis, P. and Kokkas, P. and Kosmoglou Kioseoglou, P. G. and Manthos, N. and Papadopoulos, I. and Strologas, J. and Bartók, M. and Hajdu, C. and Horvath, D. and Sikler, F. and Veszpremi, V. and Csanád, M. and Farkas, K. and Gadallah, M. M. A. and Kadlecsik, Á. and Major, P. and Mandal, K. and Pásztor, G. and Rádl, A. J. and Veres, G. I. and Raics, P. and Ujvari, B. and Zilizi, G. and Bencze, G. and Czellar, S. and Molnar, J. and Szillasi, Z. and Csorgo, T. and Nemes, F. and Novak, T. and Babbar, J. and Bansal, S. and Beri, S. B. and Bhatnagar, V. and Chaudhary, G. and Chauhan, S. and Dhingra, N. and Kaur, A. and Kaur, A. and Kaur, H. and Kaur, M. and Kumar, S. and Sandeep, K. and Sheokand, T. and Singh, J. B. and Singla, A. and Ahmed, A. and Bhardwaj, A. and Chhetri, A. and Choudhary, B. C. and Kumar, A. and Kumar, A. and Naimuddin, M. and Ranjan, K. and Saumya, S. and Baradia, S. and Barman, S. and Bhattacharya, S. and Dutta, S. and Dutta, S. and Sarkar, S. and Ameen, M. M. and Behera, P. K. and Behera, S. C. and Chatterjee, S. and Jana, P. and Kalbhor, P. and Komaragiri, J. R. and Kumar, D. and Panwar, L. and Pujahari, P. R. and Saha, N. R. and Sharma, A. and Sikdar, A. K. and Verma, S. and Dugad, S. and Kumar, M. and Mohanty, G. B. and Suryadevara, P. and Bala, A. and Banerjee, S. and Chatterjee, R. M. and Dewanjee, R. K. and Guchait, M. and Jain, Sh. and Karmakar, S. and Kumar, S. and Majumder, G. and Mazumdar, K. and Parolia, S. and Thachayath, A. and Bahinipati, S. and Kar, C. and Maity, D. and Mal, P. and Mishra, T. and Muraleedharan Nair Bindhu, V. K. and Naskar, K. and Nayak, A. and Sadangi, P. and Saha, P. and Swain, S. K. and Varghese, S. and Vats, D. and Acharya, S. and Alpana, A. and Dube, S. and Gomber, B. and Kansal, B. and Laha, A. and Sahu, B. and Sharma, S. and Bakhshiansohi, H. and Khazaie, E. and Zeinali, M. and Chenarani, S. and Etesami, S. M. and Khakzad, M. and Mohammadi Najafabadi, M. and Grunewald, M. and Abbrescia, M. and Aly, R. and Colaleo, A. and Creanza, D. and D’Anzi, B. and De Filippis, N. and De Palma, M. and Di Florio, A. and Elmetenawee, W. and Fiore, L. and Iaselli, G. and Louka, M. and Maggi, G. and Maggi, M. and Margjeka, I. and Mastrapasqua, V. and My, S. and Nuzzo, S. and Pellecchia, A. and Pompili, A. and Pugliese, G. and Radogna, R. and Ramirez-Sanchez, G. and Ramos, D. and Ranieri, A. and Silvestris, L. and Simone, F. M. and Sözbilir, Ü. and Stamerra, A. and Venditti, R. and Verwilligen, P. and Zaza, A. and Abbiendi, G. and Battilana, C. and Bonacorsi, D. and Borgonovi, L. and Campanini, R. and Capiluppi, P. and Cavallo, F. R. and Cuffiani, M. and Dallavalle, G. M. and Diotalevi, T. and Fabbri, F. and Fanfani, A. and Fasanella, D. and Giacomelli, P. and Giommi, L. and Grandi, C. and Guiducci, L. and Lo Meo, S. and Lunerti, L. and Marcellini, S. and Masetti, G. and Navarria, F. L. and Perrotta, A. and Primavera, F. and Rossi, A. M. and Rovelli, T. and Siroli, G. P. and Costa, S. and Di Mattia, A. and Potenza, R. and Tricomi, A. and Tuve, C. and Assiouras, P. and Barbagli, G. and Bardelli, G. and Camaiani, B. and Cassese, A. and Ceccarelli, R. and Ciulli, V. and Civinini, C. and D’Alessandro, R. and Focardi, E. and Kello, T. and Latino, G. and Lenzi, P. and Lizzo, M. and Meschini, M. and Paoletti, S. and Papanastassiou, A. and Sguazzoni, G. and Viliani, L. and Benussi, L. and Bianco, S. and Meola, S. and Piccolo, D. and Chatagnon, P. and Ferro, F. and Robutti, E. and Tosi, S. and Benaglia, A. and Boldrini, G. and Brivio, F. and Cetorelli, F. and De Guio, F. and Dinardo, M. E. and Dini, P. and Gennai, S. and Gerosa, R. and Ghezzi, A. and Govoni, P. and Guzzi, L. and Lucchini, M. T. and Malberti, M. and Malvezzi, S. and Massironi, A. and Menasce, D. and Moroni, L. and Paganoni, M. and Pedrini, D. and Pinolini, B. S. and Ragazzi, S. and Tabarelli de Fatis, T. and Zuolo, D. and Buontempo, S. and Cagnotta, A. and Carnevali, F. and Cavallo, N. and De Iorio, A. and Fabozzi, F. and Iorio, A. O. M. and Lista, L. and Paolucci, P. and Rossi, B. and Sciacca, C. and Ardino, R. and Azzi, P. and Bacchetta, N. and Bisello, D. and Bortignon, P. and Bragagnolo, A. and Carlin, R. and Checchia, P. and Dorigo, T. and Fantinel, S. and Fanzago, F. and Gasparini, U. and Lusiani, E. and Margoni, M. and Marini, F. and Meneguzzo, A. T. and Migliorini, M. and Pazzini, J. and Ronchese, P. and Rossin, R. and Strong, G. and Tosi, M. and Triossi, A. and Ventura, S. and Yarar, H. and Zanetti, M. and Zotto, P. and Zucchetta, A. and Zumerle, G. and Abu Zeid, S. and Aimè, C. and Braghieri, A. and Calzaferri, S. and Fiorina, D. and Montagna, P. and Re, V. and Riccardi, C. and Salvini, P. and Vai, I. and Vitulo, P. and Ajmal, S. and Bilei, G. M. and Ciangottini, D. and Fanò, L. and Magherini, M. and Mantovani, G. and Mariani, V. and Menichelli, M. and Moscatelli, F. and Rossi, A. and Santocchia, A. and Spiga, D. and Tedeschi, T. and Asenov, P. and Azzurri, P. and Bagliesi, G. and Bhattacharya, R. and Bianchini, L. and Boccali, T. and Bossini, E. and Bruschini, D. and Castaldi, R. and Ciocci, M. A. and Cipriani, M. and D’Amante, V. and Dell’Orso, R. and Donato, S. and Giassi, A. and Ligabue, F. and Matos Figueiredo, D. and Messineo, A. and Musich, M. and Palla, F. and Rizzi, A. and Rolandi, G. and Roy Chowdhury, S. and Sarkar, T. and Scribano, A. and Spagnolo, P. and Tenchini, R. and Tonelli, G. and Turini, N. and Venturi, A. and Verdini, P. G. and Barria, P. and Campana, M. and Cavallari, F. and Cunqueiro Mendez, L. and Del Re, D. and Di Marco, E. and Diemoz, M. and Errico, F. and Longo, E. and Meridiani, P. and Mijuskovic, J. and Organtini, G. and Pandolfi, F. and Paramatti, R. and Quaranta, C. and Rahatlou, S. and Rovelli, C. and Santanastasio, F. and Soffi, L. and Amapane, N. and Arcidiacono, R. and Argiro, S. and Arneodo, M. and Bartosik, N. and Bellan, R. and Bellora, A. and Biino, C. and Borca, C. and Cartiglia, N. and Costa, M. and Covarelli, R. and Demaria, N. and Finco, L. and Grippo, M. and Kiani, B. and Legger, F. and Luongo, F. and Mariotti, C. and Markovic, L. and Maselli, S. and Mecca, A. and Migliore, E. and Monteno, M. and Mulargia, R. and Obertino, M. M. and Ortona, G. and Pacher, L. and Pastrone, N. and Pelliccioni, M. and Ruspa, M. and Siviero, F. and Sola, V. and Solano, A. and Staiano, A. and Tarricone, C. and Trocino, D. and Umoret, G. and Vlasov, E. and Belforte, S. and Candelise, V. and Casarsa, M. and Cossutti, F. and De Leo, K. and Della Ricca, G. and Dogra, S. and Hong, J. and Huh, C. and Kim, B. and Kim, D. H. and Kim, J. and Lee, H. and Lee, S. W. and Moon, C. S. and Oh, Y. D. and Ryu, M. S. and Sekmen, S. and Yang, Y. C. and Kim, M. S. and Bak, G. and Gwak, P. and Kim, H. and Moon, D. H. and Asilar, E. and Kim, D. and Kim, T. J. and Merlin, J. A. and Choi, S. and Han, S. and Hong, B. and Lee, K. and Lee, K. S. and Lee, S. and Park, J. and Park, S. K. and Yoo, J. and Goh, J. and Yang, S. and Kim, H. S. and Kim, Y. and Lee, S. and Almond, J. and Bhyun, J. H. and Choi, J. and Jun, W. and Kim, J. and Ko, S. and Kwon, H. and Lee, H. and Lee, J. and Lee, J. and Oh, B. H. and Oh, S. B. and Seo, H. and Yang, U. K. and Yoon, I. and Jang, W. and Kang, D. Y. and Kang, Y. and Kim, S. and Ko, B. and Lee, J. S. H. and Lee, Y. and Park, I. C. and Roh, Y. and Watson, I. J. and Ha, S. and Yoo, H. D. and Choi, M. and Kim, M. R. and Lee, H. and Lee, Y. and Yu, I. and Beyrouthy, T. and Maghrbi, Y. and Dreimanis, K. and Gaile, A. and Pikurs, G. and Potrebko, A. and Seidel, M. and Veckalns, V. and Strautnieks, N. R. and Ambrozas, M. and Juodagalvis, A. and Rinkevicius, A. and Tamulaitis, G. and Bin Norjoharuddeen, N. and Yusuff, I. and Zolkapli, Z. and Benitez, J. F. and Castaneda Hernandez, A. and Encinas Acosta, H. A. and Gallegos Maríñez, L. G. and León Coello, M. and Murillo Quijada, J. A. and Sehrawat, A. and Valencia Palomo, L. and Ayala, G. and Castilla-Valdez, H. and Crotte Ledesma, H. and De La Cruz-Burelo, E. and Heredia-De La Cruz, I. and Lopez-Fernandez, R. and Mondragon Herrera, C. A. and Sánchez Hernández, A. and Oropeza Barrera, C. and Ramírez García, M. and Bautista, I. and Pedraza, I. and Salazar Ibarguen, H. A. and Uribe Estrada, C. and Bubanja, I. and Raicevic, N. and Butler, P. H. and Ahmad, A. and Asghar, M. I. and Awais, A. and Awan, M. I. M. and Hoorani, H. R. and Khan, W. A. and Avati, V. and Grzanka, L. and Malawski, M. and Bialkowska, H. and Bluj, M. and Boimska, B. and Górski, M. and Kazana, M. and Szleper, M. and Zalewski, P. and Bunkowski, K. and Doroba, K. and Kalinowski, A. and Konecki, M. and Krolikowski, J. and Muhammad, A. and Pozniak, K. and Zabolotny, W. and Araujo, M. and Bastos, D. and Beirão Da Cruz E Silva, C. and Boletti, A. and Bozzo, M. and Camporesi, T. and Da Molin, G. and Faccioli, P. and Gallinaro, M. and Hollar, J. and Leonardo, N. and Niknejad, T. and Petrilli, A. and Pisano, M. and Seixas, J. and Varela, J. and Wulff, J. W. and Adzic, P. and Milenovic, P. and Dordevic, M. and Milosevic, J. and Rekovic, V. and Aguilar-Benitez, M. and Alcaraz Maestre, J. and Bedoya, Cristina F. and Cepeda, M. and Cerrada, M. and Colino, N. and De La Cruz, B. and Delgado Peris, A. and Escalante Del Valle, A. and Fernández Del Val, D. and Fernández Ramos, J. P. and Flix, J. and Fouz, M. C. and Gonzalez Lopez, O. and Goy Lopez, S. and Hernandez, J. M. and Josa, M. I. and Moran, D. and Morcillo Perez, C. M. and Navarro Tobar, Á. and Perez Dengra, C. and Pérez-Calero Yzquierdo, A. and Puerta Pelayo, J. and Redondo, I. and Redondo Ferrero, D. D. and Romero, L. and Sánchez Navas, S. and Urda Gómez, L. and Vazquez Escobar, J. and Willmott, C. and de Trocóniz, J. F. and Alvarez Gonzalez, B. and Cuevas, J. and Fernandez Menendez, J. and Folgueras, S. and Gonzalez Caballero, I. and González Fernández, J. R. and Palencia Cortezon, E. and Ramón Álvarez, C. and Rodríguez Bouza, V. and Soto Rodríguez, A. and Trapote, A. and Vico Villalba, C. and Vischia, P. and Bhowmik, S. and Blanco Fernández, S. and Brochero Cifuentes, J. A. and Cabrillo, I. J. and Calderon, A. and Duarte Campderros, J. and Fernandez, M. and Gomez, G. and Lasaosa García, C. and Martinez Rivero, C. and Martinez Ruiz del Arbol, P. and Matorras, F. and Matorras Cuevas, P. and Navarrete Ramos, E. and Piedra Gomez, J. and Scodellaro, L. and Vila, I. and Vizan Garcia, J. M. and Jayananda, M. K. and Kailasapathy, B. and Sonnadara, D. U. J. and Wickramarathna, D. D. C. and Dharmaratna, W. G. D. and Liyanage, K. and Perera, N. and Wickramage, N. and Abbaneo, D. and Amendola, C. and Auffray, E. and Auzinger, G. and Baechler, J. and Barney, D. and Bermúdez Martínez, A. and Bianco, M. and Bilin, B. and Bin Anuar, A. A. and Bocci, A. and Botta, C. and Brondolin, E. and Caillol, C. and Cerminara, G. and Chernyavskaya, N. and d’Enterria, D. and Dabrowski, A. and David, A. and De Roeck, A. and Defranchis, M. M. and Deile, M. and Dobson, M. and Forthomme, L. and Franzoni, G. and Funk, W. and Giani, S. and Gigi, D. and Gill, K. and Glege, F. and Gouskos, L. and Haranko, M. and Hegeman, J. and Huber, B. and Innocente, V. and James, T. and Janot, P. and Laurila, S. and Lecoq, P. and Leutgeb, E. and Lourenço, C. and Maier, B. and Malgeri, L. and Mannelli, M. and Marini, A. C. and Matthewman, M. and Meijers, F. and Mersi, S. and Meschi, E. and Milosevic, V. and Monti, F. and Moortgat, F. and Mulders, M. and Neutelings, I. and Orfanelli, S. and Pantaleo, F. and Petrucciani, G. and Pfeiffer, A. and Pierini, M. and Piparo, D. and Qu, H. and Rabady, D. and Reales Gutiérrez, G. and Rovere, M. and Sakulin, H. and Scarfi, S. and Schwick, C. and Selvaggi, M. and Sharma, A. and Shchelina, K. and Silva, P. and Sphicas, P. and Stahl Leiton, A. G. and Steen, A. and Summers, S. and Treille, D. and Tropea, P. and Tsirou, A. and Walter, D. and Wanczyk, J. and Wang, J. and Wuchterl, S. and Zehetner, P. and Zejdl, P. and Zeuner, W. D. and Bevilacqua, T. and Caminada, L. and Ebrahimi, A. and Erdmann, W. and Horisberger, R. and Ingram, Q. and Kaestli, H. C. and Kotlinski, D. and Lange, C. and Missiroli, M. and Noehte, L. and Rohe, T. and Aarrestad, T. K. and Androsov, K. and Backhaus, M. and Calandri, A. and Cazzaniga, C. and Datta, K. and De Cosa, A. and Dissertori, G. and Dittmar, M. and Donegà, M. and Eble, F. and Galli, M. and Gedia, K. and Glessgen, F. and Grab, C. and Hits, D. and Lustermann, W. and Lyon, A.-M. and Manzoni, R. A. and Marchegiani, M. and Marchese, L. and Martin Perez, C. and Mascellani, A. and Nessi-Tedaldi, F. and Pauss, F. and Perovic, V. and Pigazzini, S. and Reissel, C. and Reitenspiess, T. and Ristic, B. and Riti, F. and Ruini, D. and Seidita, R. and Steggemann, J. and Valsecchi, D. and Wallny, R. and Amsler, C. and Bärtschi, P. and Brzhechko, D. and Canelli, M. F. and Cormier, K. and Heikkilä, J. K. and Huwiler, M. and Jin, W. and Jofrehei, A. and Kilminster, B. and Leontsinis, S. and Liechti, S. P. and Macchiolo, A. and Meiring, P. and Molinatti, U. and Reimers, A. and Robmann, P. and Sanchez Cruz, S. and Senger, M. and Takahashi, Y. and Tramontano, R. and Adloff, C. and Bhowmik, D. and Kuo, C. M. and Lin, W. and Rout, P. K. and Tiwari, P. C. and Yu, S. S. and Ceard, L. and Chao, Y. and Chen, K. F. and Chen, P. s. and Chen, Z. g. and Hou, W.-S. and Hsu, T. h. and Kao, Y. w. and Khurana, R. and Kole, G. and Li, Y. y. and Lu, R.-S. and Paganis, E. and Su, X. f. and Thomas-Wilsker, J. and Tsai, L. s. and Wu, H. y. and Yazgan, E. and Asawatangtrakuldee, C. and Srimanobhas, N. and Wachirapusitanand, V. and Agyel, D. and Boran, F. and Demiroglu, Z. S. and Dolek, F. and Dumanoglu, I. and Eskut, E. and Guler, Y. and Gurpinar Guler, E. and Isik, C. and Kara, O. and Kayis Topaksu, A. and Kiminsu, U. and Onengut, G. and Ozdemir, K. and Polatoz, A. and Tali, B. and Tok, U. G. and Turkcapar, S. and Uslan, E. and Zorbakir, I. S. and Yalvac, M. and Akgun, B. and Atakisi, I. O. and Gülmez, E. and Kaya, M. and Kaya, O. and Tekten, S. and Cakir, A. and Cankocak, K. and Komurcu, Y. and Sen, S. and Aydilek, O. and Cerci, S. and Epshteyn, V. and Hacisahinoglu, B. and Hos, I. and Kaynak, B. and Ozkorucuklu, S. and Potok, O. and Sert, H. and Simsek, C. and Zorbilmez, C. and Isildak, B. and Sunar Cerci, D. and Boyaryntsev, A. and Grynyov, B. and Levchuk, L. and Anthony, D. and Brooke, J. J. and Bundock, A. and Bury, F. and Clement, E. and Cussans, D. and Flacher, H. and Glowacki, M. and Goldstein, J. and Heath, H. F. and Kreczko, L. and Paramesvaran, S. and Seif El Nasr-Storey, S. and Smith, V. J. and Stylianou, N. and Walkingshaw Pass, K. and White, R. and Ball, A. H. and Bell, K. W. and Belyaev, A. and Brew, C. and Brown, R. M. and Cockerill, D. J. A. and Cooke, C. and Ellis, K. V. and Harder, K. and Harper, S. and Holmberg, M.-L. and Linacre, J. and Manolopoulos, K. and Newbold, D. M. and Olaiya, E. and Petyt, D. and Reis, T. and Salvi, G. and Schuh, T. and Shepherd-Themistocleous, C. H. and Tomalin, I. R. and Williams, T. and Bainbridge, R. and Bloch, P. and Brown, C. E. and Buchmuller, O. and Cacchio, V. and Carrillo Montoya, C. A. and Cepaitis, V. and Chahal, G. S. and Colling, D. and Dancu, J. S. and Das, I. and Dauncey, P. and Davies, G. and Davies, J. and Della Negra, M. and Fayer, S. and Fedi, G. and Hall, G. and Hassanshahi, M. H. and Howard, A. and Iles, G. and Knight, M. and Langford, J. and León Holgado, J. and Lyons, L. and Magnan, A.-M. and Malik, S. and Mieskolainen, M. and Nash, J. and Pesaresi, M. and Radburn-Smith, B. C. and Richards, A. and Rose, A. and Savva, K. and Seez, C. and Shukla, R. and Tapper, A. and Uchida, K. and Uttley, G. P. and Vage, L. H. and Virdee, T. and Vojinovic, M. and Wardle, N. and Winterbottom, D. and Coldham, K. and Cole, J. E. and Khan, A. and Kyberd, P. and Reid, I. D. and Abdullin, S. and Brinkerhoff, A. and Caraway, B. and Dittmann, J. and Hatakeyama, K. and Hiltbrand, J. and McMaster, B. and Saunders, M. and Sawant, S. and Sutantawibul, C. and Wilson, J. and Bartek, R. and Dominguez, A. and Huerta Escamilla, C. and Simsek, A. E. and Uniyal, R. and Vargas Hernandez, A. M. and Bam, B. and Chudasama, R. and Cooper, S. I. and Gleyzer, S. V. and Perez, C. U. and Rumerio, P. and Usai, E. and Yi, R. and Akpinar, A. and Arcaro, D. and Cosby, C. and Demiragli, Z. and Erice, C. and Fangmeier, C. and Fernandez Madrazo, C. and Fontanesi, E. and Gastler, D. and Golf, F. and Jeon, S. and Reed, I. and Rohlf, J. and Salyer, K. and Sperka, D. and Spitzbart, D. and Suarez, I. and Tsatsos, A. and Yuan, S. and Zecchinelli, A. G. and Benelli, G. and Coubez, X. and Cutts, D. and Hadley, M. and Heintz, U. and Hogan, J. M. and Kwon, T. and Landsberg, G. and Lau, K. T. and Li, D. and Luo, J. and Mondal, S. and Narain, M. and Pervan, N. and Sagir, S. and Simpson, F. and Stamenkovic, M. and Wong, W. Y. and Yan, X. and Zhang, W. and Abbott, S. and Bonilla, J. and Brainerd, C. and Breedon, R. and Calderon De La Barca Sanchez, M. and Chertok, M. and Citron, M. and Conway, J. and Cox, P. T. and Erbacher, R. and Jensen, F. and Kukral, O. and Mocellin, G. and Mulhearn, M. and Pellett, D. and Wei, W. and Yao, Y. and Zhang, F. and Bachtis, M. and Cousins, R. and Datta, A. and Flores Avila, G. and Hauser, J. and Ignatenko, M. and Iqbal, M. A. and Lam, T. and Manca, E. and Nunez Del Prado, A. and Saltzberg, D. and Valuev, V. and Clare, R. and Gary, J. W. and Gordon, M. and Hanson, G. and Si, W. and Wimpenny, S. and Branson, J. G. and Cittolin, S. and Cooperstein, S. and Diaz, D. and Duarte, J. and Giannini, L. and Guiang, J. and Kansal, R. and Krutelyov, V. and Lee, R. and Letts, J. and Masciovecchio, M. and Mokhtar, F. and Mukherjee, S. and Pieri, M. and Quinnan, M. and Sathia Narayanan, B. V. and Sharma, V. and Tadel, M. and Vourliotis, E. and Würthwein, F. and Xiang, Y. and Yagil, A. and Barzdukas, A. and Brennan, L. and Campagnari, C. and Dorsett, A. and Incandela, J. and Kim, J. and Li, A. J. and Masterson, P. and Mei, H. and Richman, J. and Sarica, U. and Schmitz, R. and Setti, F. and Sheplock, J. and Stuart, D. and Vámi, T. Á. and Wang, S. and Bornheim, A. and Cerri, O. and Latorre, A. and Mao, J. and Newman, H. B. and Spiropulu, M. and Vlimant, J. R. and Wang, C. and Xie, S. and Zhu, R. Y. and Alison, J. and An, S. and Andrews, M. B. and Bryant, P. and Cremonesi, M. and Dutta, V. and Ferguson, T. and Harilal, A. and Liu, C. and Mudholkar, T. and Murthy, S. and Palit, P. and Paulini, M. and Roberts, A. and Sanchez, A. and Terrill, W. and Cumalat, J. P. and Ford, W. T. and Hassani, A. and Karathanasis, G. and MacDonald, E. and Manganelli, N. and Perloff, A. and Savard, C. and Schonbeck, N. and Stenson, K. and Ulmer, K. A. and Wagner, S. R. and Zipper, N. and Alexander, J. and Bright-Thonney, S. and Chen, X. and Cranshaw, D. J. and Fan, J. and Fan, X. and Gadkari, D. and Hogan, S. and Kotamnives, P. and Monroy, J. and Oshiro, M. and Patterson, J. R. and Reichert, J. and Reid, M. and Ryd, A. and Thom, J. and Wittich, P. and Zou, R. and Albrow, M. and Alyari, M. and Amram, O. and Apollinari, G. and Apresyan, A. and Bauerdick, L. A. T. and Berry, D. and Berryhill, J. and Bhat, P. C. and Burkett, K. and Butler, J. N. and Canepa, A. and Cerati, G. B. and Cheung, H. W. K. and Chlebana, F. and Cummings, G. and Dickinson, J. and Dutta, I. and Elvira, V. D. and Feng, Y. and Freeman, J. and Gandrakota, A. and Gecse, Z. and Gray, L. and Green, D. and Grummer, A. and Grünendahl, S. and Guerrero, D. and Gutsche, O. and Harris, R. M. and Heller, R. and Herwig, T. C. and Hirschauer, J. and Horyn, L. and Jayatilaka, B. and Jindariani, S. and Johnson, M. and Joshi, U. and Klijnsma, T. and Klima, B. and Kwok, K. H. M. and Lammel, S. and Lincoln, D. and Lipton, R. and Liu, T. and Madrid, C. and Maeshima, K. and Mantilla, C. and Mason, D. and McBride, P. and Merkel, P. and Mrenna, S. and Nahn, S. and Ngadiuba, J. and Noonan, D. and Papadimitriou, V. and Pastika, N. and Pedro, K. and Pena, C. and Ravera, F. and Reinsvold Hall, A. and Ristori, L. and Sexton-Kennedy, E. and Smith, N. and Soha, A. and Spiegel, L. and Stoynev, S. and Strait, J. and Taylor, L. and Tkaczyk, S. and Tran, N. V. and Uplegger, L. and Vaandering, E. W. and Zoi, I. and Aruta, C. and Avery, P. and Bourilkov, D. and Cadamuro, L. and Chang, P. and Cherepanov, V. and Field, R. D. and Koenig, E. and Kolosova, M. and Konigsberg, J. and Korytov, A. and Lo, K. H. and Matchev, K. and Menendez, N. and Mitselmakher, G. and Mohrman, K. and Muthirakalayil Madhu, A. and Rawal, N. and Rosenzweig, D. and Rosenzweig, S. and Shi, K. and Wang, J. and Adams, T. and Al Kadhim, A. and Askew, A. and Bower, S. and Habibullah, R. and Hagopian, V. and Hashmi, R. and Kim, R. S. and Kim, S. and Kolberg, T. and Martinez, G. and Prosper, H. and Prova, P. R. and Wulansatiti, M. and Yohay, R. and Zhang, J. and Alsufyani, B. and Baarmand, M. M. and Butalla, S. and Elkafrawy, T. and Hohlmann, M. and Kumar Verma, R. and Rahmani, M. and Yanes, E. and Adams, M. R. and Baty, A. and Bennett, C. and Cavanaugh, R. and Escobar Franco, R. and Evdokimov, O. and Gerber, C. E. and Hofman, D. J. and Lee, J. h. and Lemos, D. S. and Merrit, A. H. and Mills, C. and Nanda, S. and Oh, G. and Ozek, B. and Pilipovic, D. and Pradhan, R. and Roy, T. and Rudrabhatla, S. and Tonjes, M. B. and Varelas, N. and Ye, Z. and Yoo, J. and Alhusseini, M. and Blend, D. and Dilsiz, K. and Emediato, L. and Karaman, G. and Köseyan, O. K. and Merlo, J.-P. and Mestvirishvili, A. and Nachtman, J. and Neogi, O. and Ogul, H. and Onel, Y. and Penzo, A. and Snyder, C. and Tiras, E. and Blumenfeld, B. and Corcodilos, L. and Davis, J. and Gritsan, A. V. and Kang, L. and Kyriacou, S. and Maksimovic, P. and Roguljic, M. and Roskes, J. and Sekhar, S. and Swartz, M. and Abreu, A. and Alcerro Alcerro, L. F. and Anguiano, J. and Baringer, P. and Bean, A. and Flowers, Z. and Grove, D. and King, J. and Krintiras, G. and Lazarovits, M. and Le Mahieu, C. and Lindsey, C. and Marquez, J. and Minafra, N. and Murray, M. and Nickel, M. and Pitt, M. and Popescu, S. and Rogan, C. and Royon, C. and Salvatico, R. and Sanders, S. and Smith, C. and Wang, Q. and Wilson, G. and Allmond, B. and Ivanov, A. and Kaadze, K. and Kalogeropoulos, A. and Kim, D. and Maravin, Y. and Nam, K. and Natoli, J. and Roy, D. and Sorrentino, G. and Rebassoo, F. and Wright, D. and Baden, A. and Belloni, A. and Chen, Y. M. and Eno, S. C. and Hadley, N. J. and Jabeen, S. and Kellogg, R. G. and Koeth, T. and Lai, Y. and Lascio, S. and Mignerey, A. C. and Nabili, S. and Palmer, C. and Papageorgakis, C. and Paranjpe, M. M. and Wang, L. and Bendavid, J. and Busza, W. and Cali, I. A. and D’Alfonso, M. and Eysermans, J. and Freer, C. and Gomez-Ceballos, G. and Goncharov, M. and Grosso, G. and Harris, P. and Hoang, D. and Kovalskyi, D. and Krupa, J. and Lavezzo, L. and Lee, Y.-J. and Long, K. and Mironov, C. and Paus, C. and Rankin, D. and Roland, C. and Roland, G. and Rothman, S. and Stephans, G. S. F. and Wang, Z. and Wyslouch, B. and Yang, T. J. and Crossman, B. and Joshi, B. M. and Kapsiak, C. and Krohn, M. and Mahon, D. and Mans, J. and Marzocchi, B. and Pandey, S. and Revering, M. and Rusack, R. and Saradhy, R. and Schroeder, N. and Strobbe, N. and Wadud, M. A. and Cremaldi, L. M. and Bloom, K. and Claes, D. R. and Haza, G. and Hossain, J. and Joo, C. and Kravchenko, I. and Siado, J. E. and Tabb, W. and Vagnerini, A. and Wightman, A. and Yan, F. and Yu, D. and Bandyopadhyay, H. and Hay, L. and Iashvili, I. and Kharchilava, A. and Morris, M. and Nguyen, D. and Rappoccio, S. and Rejeb Sfar, H. and Williams, A. and Alverson, G. and Barberis, E. and Dervan, J. and Haddad, Y. and Han, Y. and Krishna, A. and Li, J. and Lu, M. and Madigan, G. and Mccarthy, R. and Morse, D. M. and Nguyen, V. and Orimoto, T. and Parker, A. and Skinnari, L. and Tishelman-Charny, A. and Wang, B. and Wood, D. and Bhattacharya, S. and Bueghly, J. and Chen, Z. and Dittmer, S. and Hahn, K. A. and Liu, Y. and Miao, Y. and Monk, D. G. and Schmitt, M. H. and Taliercio, A. and Velasco, M. and Agarwal, G. and Band, R. and Bucci, R. and Castells, S. and Das, A. and Goldouzian, R. and Hildreth, M. and Ho, K. W. and Hurtado Anampa, K. and Ivanov, T. and Jessop, C. and Lannon, K. and Lawrence, J. and Loukas, N. and Lutton, L. and Mariano, J. and Marinelli, N. and Mcalister, I. and McCauley, T. and Mcgrady, C. and Moore, C. and Musienko, Y. and Nelson, H. and Osherson, M. and Piccinelli, A. and Ruchti, R. and Townsend, A. and Wan, Y. and Wayne, M. and Yockey, H. and Zarucki, M. and Zygala, L. and Basnet, A. and Bylsma, B. and Carrigan, M. and Durkin, L. S. and Hill, C. and Joyce, M. and Nunez Ornelas, M. and Wei, K. and Winer, B. L. and Yates, B. R. and Addesa, F. M. and Bouchamaoui, H. and Das, P. and Dezoort, G. and Elmer, P. and Frankenthal, A. and Greenberg, B. and Haubrich, N. and Kopp, G. and Kwan, S. and Lange, D. and Loeliger, A. and Marlow, D. and Ojalvo, I. and Olsen, J. and Shevelev, A. and Stickland, D. and Tully, C. and Malik, S. and Bakshi, A. S. and Barnes, V. E. and Chandra, S. and Chawla, R. and Das, S. and Gu, A. and Gutay, L. and Jones, M. and Jung, A. W. and Kondratyev, D. and Koshy, A. M. and Liu, M. and Negro, G. and Neumeister, N. and Paspalaki, G. and Piperov, S. and Scheurer, V. and Schulte, J. F. and Stojanovic, M. and Thieman, J. and Virdi, A. K. and Wang, F. and Xie, W. and Dolen, J. and Parashar, N. and Pathak, A. and Acosta, D. and Carnahan, T. and Ecklund, K. M. and Fernández Manteca, P. J. and Freed, S. and Gardner, P. and Geurts, F. J. M. and Li, W. and Miguel Colin, O. and Padley, B. P. and Redjimi, R. and Rotter, J. and Yigitbasi, E. and Zhang, Y. and Bodek, A. and de Barbaro, P. and Demina, R. and Dulemba, J. L. and Garcia-Bellido, A. and Hindrichs, O. and Khukhunaishvili, A. and Parmar, N. and Parygin, P. and Popova, E. and Taus, R. and Goulianos, K. and Chiarito, B. and Chou, J. P. and Gershtein, Y. and Halkiadakis, E. and Hart, A. and Heindl, M. and Jaroslawski, D. and Karacheban, O. and Laflotte, I. and Lath, A. and Montalvo, R. and Nash, K. and Routray, H. and Salur, S. and Schnetzer, S. and Somalwar, S. and Stone, R. and Thayil, S. A. and Thomas, S. and Vora, J. and Wang, H. and Acharya, H. and Ally, D. and Delannoy, A. G. and Fiorendi, S. and Higginbotham, S. and Holmes, T. and Kanuganti, A. R. and Karunarathna, N. and Lee, L. and Nibigira, E. and Spanier, S. and Aebi, D. and Ahmad, M. and Bouhali, O. and Eusebi, R. and Gilmore, J. and Huang, T. and Kamon, T. and Kim, H. and Luo, S. and Mueller, R. and Overton, D. and Rathjens, D. and Safonov, A. and Akchurin, N. and Damgov, J. and Hegde, V. and Hussain, A. and Kazhykarim, Y. and Lamichhane, K. and Lee, S. W. and Mankel, A. and Peltola, T. and Volobouev, I. and Whitbeck, A. and Appelt, E. and Chen, Y. and Greene, S. and Gurrola, A. and Johns, W. and Kunnawalkam Elayavalli, R. and Melo, A. and Romeo, F. and Sheldon, P. and Tuo, S. and Velkovska, J. and Viinikainen, J. and Cardwell, B. and Cox, B. and Hakala, J. and Hirosky, R. and Ledovskoy, A. and Neu, C. and Perez Lara, C. E. and Karchin, P. E. and Aravind, A. and Banerjee, S. and Black, K. and Bose, T. and Dasu, S. and De Bruyn, I. and Everaerts, P. and Galloni, C. and He, H. and Herndon, M. and Herve, A. and Koraka, C. K. and Lanaro, A. and Loveless, R. and Madhusudanan Sreekala, J. and Mallampalli, A. and Mohammadi, A. and Mondal, S. and Parida, G. and Pinna, D. and Savin, A. and Shang, V. and Sharma, V. and Smith, W. H. and Teague, D. and Tsoi, H. F. and Vetens, W. and Warden, A. and Afanasiev, S. and Andreev, V. and Andreev, Yu. and Aushev, T. and Azarkin, M. and Babaev, A. and Belyaev, A. and Blinov, V. and Boos, E. and Borshch, V. and Budkouski, D. and Bunichev, V. and Chadeeva, M. and Chekhovsky, V. and Chistov, R. and Dermenev, A. and Dimova, T. and Druzhkin, D. and Dubinin, M. and Dudko, L. and Ershov, A. and Gavrilov, G. and Gavrilov, V. and Gninenko, S. and Golovtcov, V. and Golubev, N. and Golutvin, I. and Gorbunov, I. and Gribushin, A. and Ivanov, Y. and Kachanov, V. and Karjavine, V. and Karneyeu, A. and Kim, V. and Kirakosyan, M. and Kirpichnikov, D. and Kirsanov, M. and Klyukhin, V. and Kodolova, O. and Korenkov, V. and Kozyrev, A. and Krasnikov, N. and Lanev, A. and Levchenko, P. and Lychkovskaya, N. and Makarenko, V. and Malakhov, A. and Matveev, V. and Murzin, V. and Nikitenko, A. and Obraztsov, S. and Oreshkin, V. and Palichik, V. and Perelygin, V. and Perfilov, M. and Petrushanko, S. and Polikarpov, S. and Popov, V. and Radchenko, O. and Savina, M. and Savrin, V. and Shalaev, V. and Shmatov, S. and Shulha, S. and Skovpen, Y. and Slabospitskii, S. and Smirnov, V. and Sosnov, D. and Sulimov, V. and Tcherniaev, E. and Terkulov, A. and Teryaev, O. and Tlisova, I. and Toropin, A. and Uvarov, L. and Uzunian, A. and Vorobyev, A. and Voytishin, N. and Yuldashev, B. S. and Zarubin, A. and Zhizhin, I. and Zhokin, A.},
   year={2024},
   month=mar }

@article{CMS:2024ake,
    author = "Hayrapetyan, Aram and others",
    collaboration = "CMS",
    title = "{Search for long-lived heavy neutral leptons decaying in the CMS muon detectors in proton-proton collisions at s=13{\,}{\,}TeV}",
    eprint = "2402.18658",
    archivePrefix = "arXiv",
    primaryClass = "hep-ex",
    reportNumber = "CMS-EXO-22-017, CERN-EP-2024-022",
    doi = "10.1103/PhysRevD.110.012004",
    journal = "Phys. Rev. D",
    volume = "110",
    pages = "012004",
    year = "2024"
}

@article{CMS:2024xdq,
    author = "Hayrapetyan, Aram and others",
    collaboration = "CMS",
    title = "{Search for heavy neutral leptons in final states with electrons, muons, and hadronically decaying tau leptons in proton-proton collisions at $ \sqrt{s} $ = 13 TeV}",
    eprint = "2403.00100",
    archivePrefix = "arXiv",
    primaryClass = "hep-ex",
    reportNumber = "CMS-EXO-22-011, CERN-EP-2024-032",
    doi = "10.1007/JHEP06(2024)123",
    journal = "JHEP",
    volume = "06",
    pages = "123",
    year = "2024"
}

@article{ATLAS:2024fcs,
    author = "Aad, Georges and others",
    collaboration = "ATLAS",
    title = "{Search for heavy right-handed Majorana neutrinos in the decay of top quarks produced in proton-proton collisions at s=13{\,}{\,}TeV with the ATLAS detector}",
    eprint = "2408.05000",
    archivePrefix = "arXiv",
    primaryClass = "hep-ex",
    reportNumber = "CERN-EP-2024-154",
    doi = "10.1103/PhysRevD.110.112004",
    journal = "Phys. Rev. D",
    volume = "110",
    number = "11",
    pages = "112004",
    year = "2024"
}

@article{ATLAS:2025uah,
    author = "Aad, Georges and others",
    collaboration = "ATLAS",
    title = "{Search for heavy neutral leptons in decays of W bosons using leptonic and semi-leptonic displaced vertices in $ \sqrt{s} $ = 13 TeV pp collisions with the ATLAS detector}",
    eprint = "2503.16213",
    archivePrefix = "arXiv",
    primaryClass = "hep-ex",
    reportNumber = "CERN-EP-2025-052",
    doi = "10.1007/JHEP07(2025)196",
    journal = "JHEP",
    volume = "07",
    pages = "196",
    year = "2025"
}

@article{Ellis:2016jkw,
    author = "Ellis, Joshua",
    title = "{TikZ-Feynman: Feynman diagrams with TikZ}",
    eprint = "1601.05437",
    archivePrefix = "arXiv",
    primaryClass = "hep-ph",
    doi = "10.1016/j.cpc.2016.08.019",
    journal = "Comput. Phys. Commun.",
    volume = "210",
    pages = "103--123",
    year = "2017"
}

@article{Cirigliano:2021yto,
    author = "Cirigliano, Vincenzo and D{\'\i}az-Calder{\'o}n, David and Falkowski, Adam and Gonz{\'a}lez-Alonso, Mart{\'\i}n and Rodr{\'\i}guez-S{\'a}nchez, Antonio",
    title = "{Semileptonic tau decays beyond the Standard Model}",
    eprint = "2112.02087",
    archivePrefix = "arXiv",
    primaryClass = "hep-ph",
    doi = "10.1007/JHEP04(2022)152",
    journal = "JHEP",
    volume = "04",
    pages = "152",
    year = "2022"
}

@article{Kinoshita:1958ru,
    author = "Kinoshita, Toichiro and Sirlin, Alberto",
    title = "{Radiative corrections to Fermi interactions}",
    doi = "10.1103/PhysRev.113.1652",
    journal = "Phys. Rev.",
    volume = "113",
    pages = "1652--1660",
    year = "1959"
}

@article{vanRitbergen:1999fi,
    author = "van Ritbergen, Timo and Stuart, Robin G.",
    title = "{On the precise determination of the Fermi coupling constant from the muon lifetime}",
    eprint = "hep-ph/9904240",
    archivePrefix = "arXiv",
    reportNumber = "TTP-99-18, UM-TH-99-04",
    doi = "10.1016/S0550-3213(99)00572-6",
    journal = "Nucl. Phys. B",
    volume = "564",
    pages = "343--390",
    year = "2000"
}

@article{Steinhauser:1999bx,
    author = "Steinhauser, M. and Seidensticker, T.",
    title = "{Second order corrections to the muon lifetime and the semileptonic B decay}",
    eprint = "hep-ph/9909436",
    archivePrefix = "arXiv",
    reportNumber = "BUTP-99-17, TTP-99-38",
    doi = "10.1016/S0370-2693(99)01168-5",
    journal = "Phys. Lett. B",
    volume = "467",
    pages = "271--278",
    year = "1999"
}

@article{Nir:1989rm,
    author = "Nir, Yosef",
    title = "{The Mass Ratio $m_c/m_b$ in Semileptonic $b$ Decays}",
    reportNumber = "SLAC-PUB-4847",
    doi = "10.1016/0370-2693(89)91495-0",
    journal = "Phys. Lett. B",
    volume = "221",
    pages = "184--190",
    year = "1989"
}

@article{Pak:2008qt,
    author = "Pak, Alexey and Czarnecki, Andrzej",
    title = "{Mass effects in muon and semileptonic $b \rightarrow c$ decays}",
    eprint = "0803.0960",
    archivePrefix = "arXiv",
    primaryClass = "hep-ph",
    reportNumber = "ALBERTA-THY-12-08",
    doi = "10.1103/PhysRevLett.100.241807",
    journal = "Phys. Rev. Lett.",
    volume = "100",
    pages = "241807",
    year = "2008"
}

@article{FlavourLatticeAveragingGroupFLAG:2024oxs,
    author = "Aoki, Y. and others",
    collaboration = "Flavour Lattice Averaging Group (FLAG)",
    title = "{FLAG Review 2024}",
    eprint = "2411.04268",
    archivePrefix = "arXiv",
    primaryClass = "hep-lat",
    reportNumber = "CERN-TH-2024-192, FERMILAB-PUB-24-0785-T",
    month = "11",
    year = "2024"
}

@article{Dowdall:2013rya,
    author = "Dowdall, R. J. and Davies, C. T. H. and Lepage, G. P. and McNeile, C.",
    title = "{$V_{us}$ from $\pi$ and $K$ decay constants in full lattice QCD with physical $u$, $d$, $s$ and $c$ quarks}",
    eprint = "1303.1670",
    archivePrefix = "arXiv",
    primaryClass = "hep-lat",
    doi = "10.1103/PhysRevD.88.074504",
    journal = "Phys. Rev. D",
    volume = "88",
    pages = "074504",
    year = "2013"
}

@article{Carrasco:2014poa,
    author = "Carrasco, N. and others",
    title = "{Leptonic decay constants $f_{K},f_{D},$ and $f_{{D}_{s}}$ with $N_{f} = 2+1+1$ twisted-mass lattice QCD}",
    eprint = "1411.7908",
    archivePrefix = "arXiv",
    primaryClass = "hep-lat",
    doi = "10.1103/PhysRevD.91.054507",
    journal = "Phys. Rev. D",
    volume = "91",
    number = "5",
    pages = "054507",
    year = "2015"
}

@article{Bazavov:2014wgs,
      author         = "{[FNAL/MILC 14A] A. Bazavov} and others",
      title          = "{Charmed and light pseudoscalar meson decay constants
                        from four-flavor lattice QCD with physical light quarks}",
      journal        = "Phys.Rev.",
      number         = "7",
      volume         = "D90",
      pages          = "074509",
      doi            = "10.1103/PhysRevD.90.074509",
      year           = "2014",
      eprint         = "1407.3772",
      archivePrefix  = "arXiv",
      primaryClass   = "hep-lat",
      reportNumber   = "FERMILAB-PUB-14-230-T",
      SLACcitation   = "%%CITATION = ARXIV:1407.3772;%%",
}

@article{ExtendedTwistedMass:2021qui,
    author = "Alexandrou, C. and others",
    collaboration = "Extended Twisted Mass",
    title = "{Ratio of kaon and pion leptonic decay constants with $N_f=2+1+1$ Wilson-clover twisted-mass fermions}",
    eprint = "2104.06747",
    archivePrefix = "arXiv",
    primaryClass = "hep-lat",
    doi = "10.1103/PhysRevD.104.074520",
    journal = "Phys. Rev. D",
    volume = "104",
    number = "7",
    pages = "074520",
    year = "2021"
}

@article{Bazavov:2017lyh,
    author = "Bazavov, A. and others",
    title = "{$B$- and $D$-meson leptonic decay constants from four-flavor lattice QCD}",
    eprint = "1712.09262",
    archivePrefix = "arXiv",
    primaryClass = "hep-lat",
    reportNumber = "FERMILAB-PUB-17/491-T, FERMILAB-PUB-17-491-T",
    doi = "10.1103/PhysRevD.98.074512",
    journal = "Phys. Rev. D",
    volume = "98",
    number = "7",
    pages = "074512",
    year = "2018"
}

@article{Miller:2020xhy,
    author = "Miller, Nolan and others",
    title = {{$F_K / F_\pi$ from M\"obius Domain-Wall fermions solved on gradient-flowed HISQ ensembles}},
    eprint = "2005.04795",
    archivePrefix = "arXiv",
    primaryClass = "hep-lat",
    reportNumber = "LLNL-JRNL-809712, RIKEN-iTHEMS-Report-20, JLAB-THY-20-3192",
    doi = "10.1103/PhysRevD.102.034507",
    journal = "Phys. Rev. D",
    volume = "102",
    number = "3",
    pages = "034507",
    year = "2020"
}

@article{Erler:2002mv,
    author = "Erler, Jens",
    title = "{Electroweak radiative corrections to semileptonic tau decays}",
    eprint = "hep-ph/0211345",
    archivePrefix = "arXiv",
    journal = "Rev. Mex. Fis.",
    volume = "50",
    pages = "200--202",
    year = "2004"
}

@article{Erler:1998sy,
    author = "Erler, Jens",
    title = "{Calculation of the QED coupling $\hat{\alpha}(M_Z)$ in the modified minimal subtraction scheme}",
    eprint = "hep-ph/9803453",
    archivePrefix = "arXiv",
    reportNumber = "UPR-796T",
    doi = "10.1103/PhysRevD.59.054008",
    journal = "Phys. Rev. D",
    volume = "59",
    pages = "054008",
    year = "1999"
}

@article{Tireli:2025pno,
    author = "Tireli, Edis D. and Klausen, Rikke S. and Ruchayskiy, Oleg",
    title = "{Constraining Heavy Neutral Leptons Coupled to the Tau-Neutrino Flavor at the Large Hadron Collider}",
    eprint = "2510.12248",
    archivePrefix = "arXiv",
    primaryClass = "hep-ph",
    month = "10",
    year = "2025"
}

@article{Gelmini:1980re,
    author = "Gelmini, G. B. and Roncadelli, M.",
    title = "{Left-Handed Neutrino Mass Scale and Spontaneously Broken Lepton Number}",
    reportNumber = "MPI-PAE-PTH-50-80",
    doi = "10.1016/0370-2693(81)90559-1",
    journal = "Phys. Lett. B",
    volume = "99",
    pages = "411--415",
    year = "1981"
}

@article{Barenboim:2020dmg,
    author = "Barenboim, Gabriela and Nierste, Ulrich",
    title = "{Modified majoron model for cosmological anomalies}",
    eprint = "2005.13280",
    archivePrefix = "arXiv",
    primaryClass = "hep-ph",
    reportNumber = "IFIC/20-23, TTP20-023, P3H-20-022",
    doi = "10.1103/PhysRevD.104.023013",
    journal = "Phys. Rev. D",
    volume = "104",
    number = "2",
    pages = "023013",
    year = "2021"
}

@article{Chikashige:1980ui,
    author = "Chikashige, Y. and Mohapatra, Rabindra N. and Peccei, R. D.",
    title = "{Are There Real Goldstone Bosons Associated with Broken Lepton Number?}",
    reportNumber = "MPI-PAE-PTH-36-80",
    doi = "10.1016/0370-2693(81)90011-3",
    journal = "Phys. Lett. B",
    volume = "98",
    pages = "265--268",
    year = "1981"
}

@article{Rosner:2015wva,
    author = "Rosner, Jonathan L. and Stone, Sheldon and Van de Water, Ruth S.",
    title = "{Leptonic Decays of Charged Pseudoscalar Mesons - 2015}",
    eprint = "1509.02220",
    archivePrefix = "arXiv",
    primaryClass = "hep-ph",
    reportNumber = "EFI-15-21, FERMILAB-PUB-15-384-T",
    month = "9",
    year = "2015"
}

@article{Arroyo-Urena:2021nil,
    author = "Arroyo-Ure{\~n}a, M. A. and Hern{\'a}ndez-Tom{\'e}, G. and L{\'o}pez-Castro, G. and Roig, P. and Rosell, I.",
    title = "{Radiative corrections to {\ensuremath{\tau}}{\textrightarrow}{\ensuremath{\pi}}(K){\ensuremath{\nu}}{\ensuremath{\tau}}[{\ensuremath{\gamma}}]: A reliable new physics test}",
    eprint = "2107.04603",
    archivePrefix = "arXiv",
    primaryClass = "hep-ph",
    doi = "10.1103/PhysRevD.104.L091502",
    journal = "Phys. Rev. D",
    volume = "104",
    number = "9",
    pages = "L091502",
    year = "2021"
}

@article{Arroyo-Urena:2021dfe,
    author = "Arroyo-Ure{\~n}a, M. A. and Hern{\'a}ndez-Tom{\'e}, G. and L{\'o}pez-Castro, G. and Roig, P. and Rosell, I.",
    title = "{One-loop determination of {\ensuremath{\tau}} {\textrightarrow} {\ensuremath{\pi}}(K){\ensuremath{\nu}}$_{\tau}$ [{\ensuremath{\gamma}}] branching ratios and new physics tests}",
    eprint = "2112.01859",
    archivePrefix = "arXiv",
    primaryClass = "hep-ph",
    doi = "10.1007/JHEP02(2022)173",
    journal = "JHEP",
    volume = "02",
    pages = "173",
    year = "2022"
}

@article{Cirigliano:2007xi,
    author = "Cirigliano, Vincenzo and Rosell, Ignasi",
    title = "{The Standard Model prediction for $R_{e/\mu}^{(\pi,K)}$}",
    eprint = "0707.3439",
    archivePrefix = "arXiv",
    primaryClass = "hep-ph",
    reportNumber = "LAUR-07-3194",
    doi = "10.1103/PhysRevLett.99.231801",
    journal = "Phys. Rev. Lett.",
    volume = "99",
    pages = "231801",
    year = "2007"
}

@article{DELPHI:1996qcc,
    author = "Abreu, P. and others",
    collaboration = "DELPHI",
    title = "{Search for neutral heavy leptons produced in Z decays}",
    reportNumber = "CERN-PPE-96-195",
    doi = "10.1007/s002880050370",
    journal = "Z. Phys. C",
    volume = "74",
    pages = "57--71",
    year = "1997",
    note = "[Erratum: Z.Phys.C 75, 580 (1997)]"
}

@article{BaBar:2022cqj,
    author = "Lees, J. P. and others",
    collaboration = "BaBar",
    title = "{Search for heavy neutral leptons using tau lepton decays at BaBaR}",
    eprint = "2207.09575",
    archivePrefix = "arXiv",
    primaryClass = "hep-ex",
    reportNumber = "BABAR-PUB-22/002, SLAC-PUB-17695",
    doi = "10.1103/PhysRevD.107.052009",
    journal = "Phys. Rev. D",
    volume = "107",
    number = "5",
    pages = "052009",
    year = "2023"
}

@article{Belle:2024wyk,
    author = "Nayak, M. and others",
    collaboration = "Belle",
    title = "{Search for a heavy neutral lepton that mixes predominantly with the tau neutrino}",
    eprint = "2402.02580",
    archivePrefix = "arXiv",
    primaryClass = "hep-ex",
    reportNumber = "Belle Preprint 2023-22, KEK Preprint 2023-44",
    doi = "10.1103/PhysRevD.109.L111102",
    journal = "Phys. Rev. D",
    volume = "109",
    number = "11",
    pages = "L111102",
    year = "2024"
}

@article{ArgoNeuT:2021clc,
    author = "Acciarri, R. and others",
    collaboration = "ArgoNeuT",
    title = "{New Constraints on Tau-Coupled Heavy Neutral Leptons with Masses mN=280{\textendash}970{\,}{\,}MeV}",
    eprint = "2106.13684",
    archivePrefix = "arXiv",
    primaryClass = "hep-ex",
    reportNumber = "FERMILAB-PUB-21-296-ND-T",
    doi = "10.1103/PhysRevLett.127.121801",
    journal = "Phys. Rev. Lett.",
    volume = "127",
    number = "12",
    pages = "121801",
    year = "2021"
}

@article{Boiarska:2021yho,
    author = "Boiarska, Iryna and Boyarsky, Alexey and Mikulenko, Oleksii and Ovchynnikov, Maksym",
    title = "{Constraints from the CHARM experiment on heavy neutral leptons with tau mixing}",
    eprint = "2107.14685",
    archivePrefix = "arXiv",
    primaryClass = "hep-ph",
    doi = "10.1103/PhysRevD.104.095019",
    journal = "Phys. Rev. D",
    volume = "104",
    number = "9",
    pages = "095019",
    year = "2021"
}

@article{Barouki:2022bkt,
    author = "Barouki, Ryan and Marocco, Giacomo and Sarkar, Subir",
    title = "{Blast from the past II: Constraints on heavy neutral leptons from the BEBC WA66 beam dump experiment}",
    eprint = "2208.00416",
    archivePrefix = "arXiv",
    primaryClass = "hep-ph",
    doi = "10.21468/SciPostPhys.13.5.118",
    journal = "SciPost Phys.",
    volume = "13",
    pages = "118",
    year = "2022"
}

@article{Kobach:2014hea,
    author = "Kobach, Andrew and Dobbs, Sean",
    title = "{Heavy Neutrinos and the Kinematics of Tau Decays}",
    eprint = "1412.4785",
    archivePrefix = "arXiv",
    primaryClass = "hep-ph",
    reportNumber = "NUHEP-TH-14-09",
    doi = "10.1103/PhysRevD.91.053006",
    journal = "Phys. Rev. D",
    volume = "91",
    number = "5",
    pages = "053006",
    year = "2015"
}
\end{document}